\IfFileExists{figcommons-local.tex}{\input{figcommons-local}}{}
\PassOptionsToPackage{dvipsnames}{xcolor}
\documentclass[letterpaper,11pt]{article}
\usepackage[margin=1in]{geometry}

\providecommand\apxparameter{appendix=append}
\usepackage[\apxparameter]{apxproof}
\newcommand\apxproofmode[1]{#1}
\newcommand\apxproofmodesubsection[1]{#1}

\newcommand\NODECORATION[1]{}
\newcommand\SHORTER[1]{}

\usepackage{version}
\excludeversion{SIDE}
\excludeversion{PASFINALISE}
\excludeversion{LONGUER}
\excludeversion{AVOIR}

\newcommand\myparagraph[1]{\medskip\noindent\textbf{#1}.}

\providecommand\nolncs[1]{#1}	

\providecommand\BIBFILES{@@reference-biblio}

\newcommand\affiliationofficielleolivier{LIX, CNRS, École polytechnique, Institut Polytechnique de Paris, Palaiseau, France}

\providecommand\apxproofmode[1]{}
	
\providecommand\apxproofmodesubsection[1]{}

\providecommand\lipicsmode{}
		
\newcommand\nolipics[1]{#1}
\lipicsmode{\renewcommand\nolipics[1]{}}

\providecommand\NODECORATION[1]{\ifnotSLIDE{#1}}
\lipicsmode{\renewcommand\NODECORATION[1]{}}

\providecommand\PASSIAPXPROOF[1]{#1}
\apxproofmode{\renewcommand\PASSIAPXPROOF[1]{}}

\providecommand\ifnotTDEXAM[1]{#1}
\providecommand\ifnotSLIDE[1]{#1}

\providecommand\PREUVESENAPPENDIX[1]{}

\NODECORATION{
\usepackage{fourier}
}

\providecommand\danger{}

\usepackage[framemethod=tikz]{mdframed}

\newcommand\ENVCOLOR[1]{#1}
\newcommand\BEGINTEMPENVCOLOR[1]{\let\envcolorwas\ENVCOLOR\renewcommand\ENVCOLOR[1]{#1}}
\newcommand\ENDTEMPENVCOLOR{\let\ENVCOLOR\envcolorwas}

\NODECORATION{
\surroundwithmdframed[hidealllines=true,backgroundcolor=\ENVCOLOR{blue!10},roundcorner=8pt]{exercice}
\surroundwithmdframed[hidealllines=true,backgroundcolor=\ENVCOLOR{blue!10},roundcorner=8pt]{exercise}
\surroundwithmdframed[hidealllines=true,backgroundcolor=\ENVCOLOR{blue!10},roundcorner=8pt]{exercicee}
\surroundwithmdframed[hidealllines=true,backgroundcolor=\ENVCOLOR{blue!10},roundcorner=8pt]{exerciced}
\surroundwithmdframed[hidealllines=true,backgroundcolor=\ENVCOLOR{blue!10},roundcorner=8pt]{exercicec}
\surroundwithmdframed[hidealllines=true,backgroundcolor=\ENVCOLOR{blue!10},roundcorner=8pt]{exercicedc}
\surroundwithmdframed[hidealllines=true,backgroundcolor=\ENVCOLOR{orange!20},roundcorner=8pt]{example}
\surroundwithmdframed[hidealllines=true,backgroundcolor=\ENVCOLOR{orange!20},roundcorner=8pt]{exemple}
\surroundwithmdframed[hidealllines=true,backgroundcolor=\ENVCOLOR{orange!10},roundcorner=8pt]{remark}
\ifnotSLIDE{\surroundwithmdframed[backgroundcolor=\ENVCOLOR{green!15},roundcorner=8pt]{definition}}
\surroundwithmdframed[hidealllines=true,backgroundcolor=\ENVCOLOR{gray!35}]{proposition}
\ifnotSLIDE{\surroundwithmdframed[backgroundcolor=\ENVCOLOR{gray!35},roundcorner=8pt]{theorem}}
\surroundwithmdframed[hidealllines=true,backgroundcolor=\ENVCOLOR{gray!35}]{lemma}
\surroundwithmdframed[hidealllines=true,backgroundcolor=\ENVCOLOR{gray!35}]{corollary}

\surroundwithmdframed[backgroundcolor=red!35,roundcorner=8pt]{theoremaprouver}
\surroundwithmdframed[backgroundcolor=red!35,roundcorner=8pt]{conjectureaprouver}
}

\usepackage{tcolorbox}
\tcbuselibrary{skins}

\newtcolorbox{maboiteconfiance}[2][]{colback=red!5!white, colframe=red!75!black,fonttitle=\bfseries, colbacktitle=red!85!black,enhanced,
attach boxed title to top center={yshift=-2mm},
  title={#2},#1}
  
\newtcolorbox{maboiteresume}[2][]{colback=red!5!white, colframe=red!75!black,fonttitle=\bfseries, colbacktitle=red!85!black,enhanced,
attach boxed title to top center={yshift=-2mm},
  title={#2},#1}
  \newtcolorbox{maboiteresumeperso}[2][]{colback=red!5!white, colframe=red!75!black,fonttitle=\bfseries, colbacktitle=red!85!black,enhanced,
attach boxed title to top center={yshift=-2mm},
  title={#2},#1}
  \newtcolorbox{maboitecommentaire}[2][]{colback=red!5!white, colframe=red!75!black,fonttitle=\bfseries, colbacktitle=red!85!black,enhanced,
attach boxed title to top center={yshift=-2mm},
  title={#2},#1}
\newtcolorbox{maboitetruc}[2][]{colback=red!5!white, colframe=red!75!black,fonttitle=\bfseries, colbacktitle=red!85!black,enhanced,
attach boxed title to top left={yshift=-2mm},
  title={#2},#1}

\ifdefined\CESTDUFRANCAIS
	\providecommand\LANGUE[2]{#2}
	\usepackage[french]{babel}
\else
	\providecommand\LANGUE[2]{#1}
\fi
\providecommand\LANGUEinv[2]{\LANGUE{#2}{#1}}

\providecommand\ifnotmodeDIFFUSIONFINALE[1]{
\ifdefined\SAFEMODE
\else
#1
\fi
}
\providecommand\ifmodeDIFFUSIONFINALE[1]{
\ifdefined\SAFEMODE
#1
\else
\fi
}

\usepackage{xr-hyper}
\usepackage{hyperref}

\usepackage{versions}
\usepackage{amsmath,amssymb,mathrsfs,amsfonts}
\usepackage{listings}
\usepackage{url}
\usepackage{versions}
\usepackage{color}
\usepackage[colorinlistoftodos]{todonotes}
\usepackage{proof}
\usepackage{xstring}

\usepackage{makeidx}
\makeindex
\providecommand\DOINDEXPRINT{}
\providecommand\SIGNALEMOTNOUV[1]{}
\ifdefined\INDEXPRINT
	\ifnotSAFEMODE{
	\renewcommand\DOINDEXPRINT{
	      	\let\oldindex\index
		\renewcommand{\index}[1]{\oldindex{##1}\marginpar{IDXLA: \tiny##1}}
		\renewcommand\SIGNALEMOTNOUV[1]{\marginpar{MNV: \small\textbf{##1}}}
	}
	}
\fi

\usepackage[utf8]{inputenc}
\usepackage[T1]{fontenc}

\usepackage{cleveref}

\usepackage{bbm}

\DeclareUnicodeCharacter{2212}{-}
\DeclareUnicodeCharacter{2297}{\ensuremath{\otimes}}       
\DeclareUnicodeCharacter{2A33}{\ensuremath{\smashtimes}}
\DeclareUnicodeCharacter{1D56F}{\ensuremath{\mathfrak{D}}}  
\DeclareUnicodeCharacter{1D40E}{\ensuremath{\mathbf{O}}}  
\DeclareUnicodeCharacter{1D427}{\ensuremath{\mathbf{n}}}  
\DeclareUnicodeCharacter{3008}{\ensuremath{\langle}} 
\DeclareUnicodeCharacter{3009}{\ensuremath{\rangle}} 

\DeclareUnicodeCharacter{00AC}{\ensuremath{\neg}}                     
\DeclareUnicodeCharacter{00B1}{\ensuremath{\pm}}                      
\DeclareUnicodeCharacter{00B2}{\ensuremath{^2}}                       
\DeclareUnicodeCharacter{00B3}{\ensuremath{^3}}                       
\DeclareUnicodeCharacter{00B7}{\ensuremath{\cdot}}                    
\DeclareUnicodeCharacter{00B9}{\ensuremath{^1}}                       
\DeclareUnicodeCharacter{00D7}{\ensuremath{\times}}                   
\DeclareUnicodeCharacter{02B0}{\ensuremath{^h}}                       
\DeclareUnicodeCharacter{02B3}{\ensuremath{^r}}                       
\DeclareUnicodeCharacter{02E2}{\ensuremath{^s}}                       
\DeclareUnicodeCharacter{0302}{\ensuremath{\hat{\phantom{x}}}}        
\DeclareUnicodeCharacter{0332}{\ensuremath{\underline{\phantom{x}}}}  
\DeclareUnicodeCharacter{0308}{\ensuremath{\ddot{\phantom{x}}}}       
\DeclareUnicodeCharacter{0393}{\ensuremath{\Gamma}}                   
\DeclareUnicodeCharacter{0394}{\ensuremath{\Delta}}                   
\DeclareUnicodeCharacter{0398}{\ensuremath{\Theta}}                   
\DeclareUnicodeCharacter{039B}{\ensuremath{\Lambda}}                  
\DeclareUnicodeCharacter{039E}{\ensuremath{\Xi}}                      
\DeclareUnicodeCharacter{03A0}{\ensuremath{\Pi}}                      
\DeclareUnicodeCharacter{03A3}{\ensuremath{\Sigma}}                   

\DeclareUnicodeCharacter{03A4}{\ensuremath{\mathrm{T}}}  

\DeclareUnicodeCharacter{03A5}{\ensuremath{\Upsilon}}   
\DeclareUnicodeCharacter{03A6}{\ensuremath{\Phi}}       
\DeclareUnicodeCharacter{03A8}{\ensuremath{\Psi}}       
\DeclareUnicodeCharacter{03A9}{\ensuremath{\Omega}}     
\DeclareUnicodeCharacter{03B1}{\ensuremath{\alpha}}     
\DeclareUnicodeCharacter{03B2}{\ensuremath{\beta}}      
\DeclareUnicodeCharacter{03B3}{\ensuremath{\gamma}}     
\DeclareUnicodeCharacter{03B4}{\ensuremath{\delta}}     
\DeclareUnicodeCharacter{03B5}{\ensuremath{\epsilon}}   
\DeclareUnicodeCharacter{03B6}{\ensuremath{\zeta}}      
\DeclareUnicodeCharacter{03B7}{\ensuremath{\eta}}       
\DeclareUnicodeCharacter{03B8}{\ensuremath{\theta}}     
\DeclareUnicodeCharacter{03B9}{\ensuremath{\iota}}      
\DeclareUnicodeCharacter{03BA}{\ensuremath{\kappa}}     
\DeclareUnicodeCharacter{03BB}{\ensuremath{\lambda}}    
\DeclareUnicodeCharacter{03BC}{\ensuremath{\mu}}        
\DeclareUnicodeCharacter{03BD}{\ensuremath{\nu}}        
\DeclareUnicodeCharacter{03BE}{\ensuremath{\xi}}        
\DeclareUnicodeCharacter{03C0}{\ensuremath{\pi}}        
\DeclareUnicodeCharacter{03C1}{\ensuremath{\rho}}       
\DeclareUnicodeCharacter{03C2}{\ensuremath{\varsigma}}  
\DeclareUnicodeCharacter{03C3}{\ensuremath{\sigma}}     
\DeclareUnicodeCharacter{03C4}{\ensuremath{\tau}}       
\DeclareUnicodeCharacter{03C5}{\ensuremath{\upsilon}}   
\DeclareUnicodeCharacter{03C6}{\ensuremath{\varphi}}    
\DeclareUnicodeCharacter{03C7}{\ensuremath{\chi}}       
\DeclareUnicodeCharacter{03C8}{\ensuremath{\psi}}       
\DeclareUnicodeCharacter{03C9}{\ensuremath{\omega}}     
\DeclareUnicodeCharacter{03D1}{\ensuremath{\vartheta}}  

\DeclareUnicodeCharacter{0432}{\ensuremath{\mathrm{PDF\;TODO}}}  
\DeclareUnicodeCharacter{0435}{\ensuremath{\mathrm{PDF\;TODO}}}  
\DeclareUnicodeCharacter{0438}{\ensuremath{\mathrm{PDF\;TODO}}}  
\DeclareUnicodeCharacter{043C}{\ensuremath{\mathrm{PDF\;TODO}}}  
\DeclareUnicodeCharacter{0440}{\ensuremath{\mathrm{PDF\;TODO}}}  
\DeclareUnicodeCharacter{0442}{\ensuremath{\mathrm{PDF\;TODO}}}  
\DeclareUnicodeCharacter{041F}{\ensuremath{\mathrm{PDF\;TODO}}}  
\DeclareUnicodeCharacter{0BA8}{\ensuremath{\mathrm{PDF\;TODO}}}  
\DeclareUnicodeCharacter{0BBF}{\ensuremath{\mathrm{PDF\;TODO}}}  
\DeclareUnicodeCharacter{1100}{\ensuremath{\mathrm{PDF\;TODO}}}  
\DeclareUnicodeCharacter{11F9}{\ensuremath{\mathrm{PDF\;TODO}}}  

\DeclareUnicodeCharacter{1D48}{\ensuremath{^d}}  
\DeclareUnicodeCharacter{1D50}{\ensuremath{^m}}  
\DeclareUnicodeCharacter{1D56}{\ensuremath{^p}}  
\DeclareUnicodeCharacter{1D62}{\ensuremath{_i}}  
\DeclareUnicodeCharacter{1D63}{\ensuremath{_r}}  
\DeclareUnicodeCharacter{1D64}{\ensuremath{_u}}  
\DeclareUnicodeCharacter{1D9C}{\ensuremath{^c}}  

\DeclareUnicodeCharacter{2032}{\ensuremath{\text{\textasciiacute}}}  

\DeclareUnicodeCharacter{2070}{\ensuremath{^0}}  
\DeclareUnicodeCharacter{2074}{\ensuremath{^4}}  
\DeclareUnicodeCharacter{2075}{\ensuremath{^5}}  
\DeclareUnicodeCharacter{2076}{\ensuremath{^6}}  
\DeclareUnicodeCharacter{2077}{\ensuremath{^7}}  
\DeclareUnicodeCharacter{2078}{\ensuremath{^8}}  
\DeclareUnicodeCharacter{2079}{\ensuremath{^9}}  
\DeclareUnicodeCharacter{207A}{\ensuremath{^+}}  
\DeclareUnicodeCharacter{207B}{\ensuremath{^-}}  
\DeclareUnicodeCharacter{207F}{\ensuremath{^n}}  
\DeclareUnicodeCharacter{2080}{\ensuremath{_0}}  
\DeclareUnicodeCharacter{2081}{\ensuremath{_1}}  
\DeclareUnicodeCharacter{2082}{\ensuremath{_2}}  
\DeclareUnicodeCharacter{2083}{\ensuremath{_3}}  
\DeclareUnicodeCharacter{2084}{\ensuremath{_4}}  
\DeclareUnicodeCharacter{2085}{\ensuremath{_5}}  
\DeclareUnicodeCharacter{2086}{\ensuremath{_6}}  
\DeclareUnicodeCharacter{2087}{\ensuremath{_7}}  
\DeclareUnicodeCharacter{2088}{\ensuremath{_8}}  
\DeclareUnicodeCharacter{2089}{\ensuremath{_9}}  
\DeclareUnicodeCharacter{208A}{\ensuremath{_+}}  
\DeclareUnicodeCharacter{208B}{\ensuremath{_-}}  
\DeclareUnicodeCharacter{2091}{\ensuremath{_e}}  
\DeclareUnicodeCharacter{2095}{\ensuremath{_h}}  
\DeclareUnicodeCharacter{2096}{\ensuremath{_k}}  
\DeclareUnicodeCharacter{2097}{\ensuremath{_l}}  
\DeclareUnicodeCharacter{2098}{\ensuremath{_m}}  
\DeclareUnicodeCharacter{2099}{\ensuremath{_n}}  
\DeclareUnicodeCharacter{209A}{\ensuremath{_p}}  
\DeclareUnicodeCharacter{209B}{\ensuremath{_s}}  
\DeclareUnicodeCharacter{209C}{\ensuremath{_t}}  

\DeclareUnicodeCharacter{2113}{\ensuremath{\mathpzc{l}}} 

\DeclareUnicodeCharacter{2115}{\ensuremath{\mathbbm{N}}}  
\DeclareUnicodeCharacter{211A}{\ensuremath{\mathbbm{Q}}}  
\DeclareUnicodeCharacter{211D}{\ensuremath{\mathbbm{R}}}  
\DeclareUnicodeCharacter{2124}{\ensuremath{\mathbbm{Z}}}  

\DeclareUnicodeCharacter{2135}{\ensuremath{\aleph}}           
\DeclareUnicodeCharacter{2190}{\ensuremath{\leftarrow}}       
\DeclareUnicodeCharacter{2192}{\ensuremath{\rightarrow}}      
\DeclareUnicodeCharacter{2194}{\ensuremath{\leftrightarrow}}  

\DeclareUnicodeCharacter{21A0}{\ensuremath{\twoheadrightarrow}}  

\DeclareUnicodeCharacter{21A6}{\ensuremath{\mapsto}}  
\DeclareUnicodeCharacter{21A6}{\ensuremath{\mapsto}}  

\DeclareUnicodeCharacter{21BE}{\ensuremath{\restriction}}  

\DeclareUnicodeCharacter{21D0}{\ensuremath{\Leftarrow}}       
\DeclareUnicodeCharacter{21D2}{\ensuremath{\Rightarrow}}      
\DeclareUnicodeCharacter{21D4}{\ensuremath{\Leftrightarrow}}  

\DeclareUnicodeCharacter{2200}{\ensuremath{\forall}}      
\DeclareUnicodeCharacter{2203}{\ensuremath{\exists}}      
\DeclareUnicodeCharacter{2204}{\ensuremath{\not\exists}}  
\DeclareUnicodeCharacter{2205}{\ensuremath{\emptyset}}    
\DeclareUnicodeCharacter{2208}{\ensuremath{\in}}          
\DeclareUnicodeCharacter{2209}{\ensuremath{\not\in}}      
\DeclareUnicodeCharacter{220B}{\ensuremath{\ni}}          
\DeclareUnicodeCharacter{220C}{\ensuremath{\not\ni}}      

\DeclareUnicodeCharacter{220E}{{\tiny \ensuremath{\blacksquare}}} 

\DeclareUnicodeCharacter{220F}{{\tiny \ensuremath{\prod}}}  
\DeclareUnicodeCharacter{2211}{\ensuremath{\sum}}           
\DeclareUnicodeCharacter{2218}{\ensuremath{\circ}}          
\DeclareUnicodeCharacter{2219}{\ensuremath{\bullet}}        
\DeclareUnicodeCharacter{221E}{\ensuremath{\infty}}         
\DeclareUnicodeCharacter{2223}{\ensuremath{\mid}}           
\DeclareUnicodeCharacter{2225}{\ensuremath{\parallel}}      
\DeclareUnicodeCharacter{2227}{\ensuremath{\wedge}}         
\DeclareUnicodeCharacter{2228}{\ensuremath{\vee}}           
\DeclareUnicodeCharacter{2229}{\ensuremath{\cap}}           
\DeclareUnicodeCharacter{222A}{\ensuremath{\cup}}           
\DeclareUnicodeCharacter{222B}{\ensuremath{\int}}           
\DeclareUnicodeCharacter{2234}{\ensuremath{\therefore}}     

\DeclareUnicodeCharacter{2237}{\ensuremath{\dblcolon}} 

\newcommand{\dotminus}{\mathop{\mbox{$-^{\hspace{-.5em}\cdot}\,$}}}
\DeclareUnicodeCharacter{2238}{\ensuremath{\dotminus}}  

\DeclareUnicodeCharacter{223C}{\ensuremath{\sim}}       
\DeclareUnicodeCharacter{2243}{\ensuremath{\simeq}}     
\DeclareUnicodeCharacter{2245}{\ensuremath{\cong}}      
\DeclareUnicodeCharacter{2248}{\ensuremath{\approx}}    
\DeclareUnicodeCharacter{224D}{\ensuremath{\asymp}}       
\DeclareUnicodeCharacter{2250}{\ensuremath{\doteq}}     

\DeclareUnicodeCharacter{2254}{\ensuremath{\coloneqq}}  

\newcommand{\eqq}{\stackrel{\text{\tiny ?}}{=}}
\DeclareUnicodeCharacter{225F}{\ensuremath{\eqq}}  

\DeclareUnicodeCharacter{2260}{\ensuremath{\neq}}         
\DeclareUnicodeCharacter{2261}{\ensuremath{\equiv}}       
\DeclareUnicodeCharacter{2262}{\ensuremath{\not\equiv}}   
\DeclareUnicodeCharacter{2264}{\ensuremath{\le}}          
\DeclareUnicodeCharacter{2265}{\ensuremath{\ge}}          
\DeclareUnicodeCharacter{226E}{\ensuremath{\nless}}       
\DeclareUnicodeCharacter{226F}{\ensuremath{\ngtr}}        
\DeclareUnicodeCharacter{2270}{\ensuremath{\nleq}}        
\DeclareUnicodeCharacter{2271}{\ensuremath{\ngeq}}        
\DeclareUnicodeCharacter{227A}{\ensuremath{\prec}}        
\DeclareUnicodeCharacter{227C}{\ensuremath{\preceq}}      
\DeclareUnicodeCharacter{2282}{\ensuremath{\subset}}      
\DeclareUnicodeCharacter{2284}{\ensuremath{\not\subset}} 
\DeclareUnicodeCharacter{2283}{\ensuremath{\supset}}      
\DeclareUnicodeCharacter{2286}{\ensuremath{\subseteq}}    
\DeclareUnicodeCharacter{228E}{\ensuremath{\uplus}}       
\DeclareUnicodeCharacter{2291}{\ensuremath{\sqsubseteq}}  
\DeclareUnicodeCharacter{2293}{\ensuremath{\sqcap}}       
\DeclareUnicodeCharacter{2294}{\ensuremath{\sqcup}}       
\DeclareUnicodeCharacter{2295}{\ensuremath{\oplus}}       
\DeclareUnicodeCharacter{22A2}{\ensuremath{\vdash}}       
\DeclareUnicodeCharacter{22A4}{\ensuremath{\top}}         
\DeclareUnicodeCharacter{22A5}{\ensuremath{\bot}}         
\DeclareUnicodeCharacter{22C0}{\ensuremath{\bigwedge}}    
\DeclareUnicodeCharacter{22C2}{\ensuremath{\bigcap}}      
\DeclareUnicodeCharacter{22C3}{\ensuremath{\bigcup}}      
\DeclareUnicodeCharacter{22EE}{\ensuremath{\vdots}}       
\DeclareUnicodeCharacter{22EF}{\ensuremath{\cdots}}       

\DeclareUnicodeCharacter{22F0}{\ensuremath{\iddots}} 

\DeclareUnicodeCharacter{230A}{\ensuremath{\lfloor}}  
\DeclareUnicodeCharacter{230B}{\ensuremath{\rfloor}}  
\DeclareUnicodeCharacter{2308}{\ensuremath{\lceil}}   
\DeclareUnicodeCharacter{2309}{\ensuremath{\rceil}}   
\DeclareUnicodeCharacter{2329}{\ensuremath{\langle}}  
\DeclareUnicodeCharacter{232A}{\ensuremath{\rangle}}  

\DeclareUnicodeCharacter{23CE}{\ensuremath{\Return}}  

\DeclareUnicodeCharacter{2460}{\ensuremath{\text{\ding{172}}}} 
\DeclareUnicodeCharacter{2461}{\ensuremath{\text{\ding{173}}}} 
\DeclareUnicodeCharacter{2462}{\ensuremath{\text{\ding{174}}}} 
\DeclareUnicodeCharacter{2463}{\ensuremath{\text{\ding{175}}}} 
\DeclareUnicodeCharacter{2464}{\ensuremath{\text{\ding{176}}}} 
\DeclareUnicodeCharacter{2465}{\ensuremath{\text{\ding{177}}}} 
\DeclareUnicodeCharacter{2466}{\ensuremath{\text{\ding{178}}}} 
\DeclareUnicodeCharacter{2467}{\ensuremath{\text{\ding{179}}}} 
\DeclareUnicodeCharacter{2468}{\ensuremath{\text{\ding{180}}}} 

\DeclareUnicodeCharacter{25C5}{\ensuremath{\triangleleft}}  

\DeclareUnicodeCharacter{2660}{\ensuremath{\spadesuit}} 
\DeclareUnicodeCharacter{266D}{\ensuremath{\flat}}      
\DeclareUnicodeCharacter{266F}{\ensuremath{\sharp}}     

\DeclareUnicodeCharacter{2713}{\ensuremath{\checkmark}}  

\DeclareUnicodeCharacter{27E6}{\ensuremath{\llbracket}}  
\DeclareUnicodeCharacter{27E7}{\ensuremath{\rrbracket}}  

\DeclareUnicodeCharacter{27E8}{\ensuremath{\langle}}          
\DeclareUnicodeCharacter{27E9}{\ensuremath{\rangle}}          
\DeclareUnicodeCharacter{27F6}{\ensuremath{\longrightarrow}}  

\DeclareUnicodeCharacter{2983}{\ensuremath{\mathrm{PDF\;TODO}}}  
\DeclareUnicodeCharacter{2984}{\ensuremath{\mathrm{PDF\;TODO}}}  
\DeclareUnicodeCharacter{2985}{\ensuremath{\mathrm{PDF\;TODO}}}  
\DeclareUnicodeCharacter{2986}{\ensuremath{\mathrm{PDF\;TODO}}}  

\DeclareUnicodeCharacter{2987}{\ensuremath{\llparenthesis}}  
\DeclareUnicodeCharacter{2988}{\ensuremath{\rrparenthesis}}  

\DeclareUnicodeCharacter{29F5}{\ensuremath{\setminus}}   

\DeclareUnicodeCharacter{2A06}{\ensuremath{\bigcup}}  
\DeclareUnicodeCharacter{2C7C}{\ensuremath{_j}}       

\DeclareUnicodeCharacter{D7B0}{\ensuremath{\mathrm{PDF\;TODO}}} 

\DeclareUnicodeCharacter{1D7D9}{\ensuremath{\mathbbm{1}}}  

\DeclareUnicodeCharacter{22AC}{\ensuremath{\neg\vdash}}
\DeclareUnicodeCharacter{279C}{\ensuremath{\to}}
\DeclareUnicodeCharacter{1D442}{ }
\DeclareUnicodeCharacter{2000}{ }
\DeclareUnicodeCharacter{1D440}{M}
\DeclareUnicodeCharacter{1D442}{O}
\DeclareUnicodeCharacter{20E3}{2}
\DeclareUnicodeCharacter{1D45A}{m}
\DeclareUnicodeCharacter{1D45B}{n}

\usepackage{pifont}
\DeclareUnicodeCharacter{2714}{\ding{52}}
\DeclareUnicodeCharacter{2705}{\ding{52}}
\DeclareUnicodeCharacter{2717}{\ding{55}}
\DeclareUnicodeCharacter{274C}{\ding{53}}

\DeclareUnicodeCharacter{1F9E9}{}	
\DeclareUnicodeCharacter{2699}{}
\DeclareUnicodeCharacter{FE0F}{}
\DeclareUnicodeCharacter{2080}{\ensuremath{_{0}}}
\DeclareUnicodeCharacter{1F4DA}{}
\DeclareUnicodeCharacter{1F9EE}{}
\DeclareUnicodeCharacter{1F9E0}{}
\DeclareUnicodeCharacter{27A1}{0}
\DeclareUnicodeCharacter{1F9ED}{} 
\DeclareUnicodeCharacter{1F501}{}
\DeclareUnicodeCharacter{1F50D}{}
\DeclareUnicodeCharacter{1F4A1}{}
\DeclareUnicodeCharacter{1F4D8}{}
\DeclareUnicodeCharacter{1F447}{}
\DeclareUnicodeCharacter{1F4BB}{}
\DeclareUnicodeCharacter{1F4AC}{}
\DeclareUnicodeCharacter{1F449}{}
\DeclareUnicodeCharacter{1F4C9}{}
\DeclareUnicodeCharacter{1F522}{}
\DeclareUnicodeCharacter{1F52C}{}
\DeclareUnicodeCharacter{1F51A}{}
\DeclareUnicodeCharacter{1F33F}{}
\DeclareUnicodeCharacter{1F539}{}
\DeclareUnicodeCharacter{1F538}{}
\DeclareUnicodeCharacter{1F44C}{}
\DeclareUnicodeCharacter{1F7E6}{}
\DeclareUnicodeCharacter{1F44F}{}
\DeclareUnicodeCharacter{26A0}{}
\DeclareUnicodeCharacter{1F517}{}
\DeclareUnicodeCharacter{1F52D}{}
\DeclareUnicodeCharacter{1F44D}{}
\DeclareUnicodeCharacter{202F}{}
\DeclareUnicodeCharacter{1F3AF}{}
\DeclareUnicodeCharacter{1F4CF}{}

}

\usepackage{stmaryrd}
\usepackage{ifthen}
\usepackage{mathtools}
\usepackage{dsfont}
\PASSIAPXPROOF{\usepackage{thmtools}}
\usepackage{longtable}
\usepackage{mathdots}
\usepackage{booktabs}

\usepackage{tikz}
\usetikzlibrary{calc}
\usepgflibrary{arrows}
\usetikzlibrary{fit}
\usetikzlibrary{patterns}
\usetikzlibrary{decorations.pathreplacing}
\usetikzlibrary{shapes.multipart}
\usetikzlibrary{shapes.geometric}
\usetikzlibrary{shapes.misc}
\usetikzlibrary{positioning}
\usetikzlibrary{graphs}
\usetikzlibrary{topaths}
\usetikzlibrary{arrows.meta}
\usetikzlibrary{decorations.pathreplacing}
\usetikzlibrary{decorations.markings}
\usetikzlibrary{decorations.pathmorphing}
 \usetikzlibrary{calc}
 \usetikzlibrary{automata}
\usetikzlibrary{chains}
\usetikzlibrary{scopes}
 \usetikzlibrary{positioning}
 \usetikzlibrary{through,backgrounds}
\usepackage{circuitikz}

\providecommand\ifnotPOLYCHAPTERMODE[1]{#1}
\providecommand\ifPOLYCHAPTERMODE[1]{}

\ifnotPOLYCHAPTERMODE{
\ifdefined\thechapter
\providecommand\spnewtheorem[4]{\newtheorem{#1}{#2}[chapter]}
\providecommand\spnewtheoremi[5]{\newtheorem{#1}[#5]{#2}}
\else
\providecommand\spnewtheorem[4]{\newtheorem{#1}{#2}}
\providecommand\spnewtheoremi[5]{\newtheorem{#1}[#5]{#2}}
\fi
}
\ifPOLYCHAPTERMODE{
\providecommand\spnewtheorem[4]{\newtheorem{#1}{#2}}
\providecommand\spnewtheoremi[5]{\newtheorem{#1}[#5]{#2}}
}
\providecommand\nolipics[1]{#1}

\includeversion{metdate}

\newcommand\XPAGEDEGARDE[2]{
\thispagestyle{empty}
\title{{#1} \\[2cm] 
{\large #2}
}
\author{
 Olivier Bournez
\\[0.5cm] 
bournez@lix.polytechnique.fr
}
\begin{metdate}
\date{
\vspace{2cm}
Version \LANGUE{of}{du} \today \\[1cm] 
\includegraphics[height=3.5cm]{/Users/bournez/lib/LaTeX/Perso/FIG-COMMONS/logo_X_new}
}
\end{metdate}
\maketitle
}

\newcommand\vectorl[1]{{\mathbf#1}}

\newcommand\Q{\mathbb{Q}}
\newcommand\R{\mathbb{R}}
\newcommand\Z{\mathbb{Z}}

\newcommand\N{\mathbb{N}}

\newcommand{\cp}[1]{\operatorname{#1}}

\LANGUE{

}
{
}

\newcommand{\Ptime}{\cp{P}}      

\newcommand\NP{\cp{NP}}

\newcommand\PSPACE{\cp{ PSPACE}}

\LANGUEinv{

}{

}

\LANGUEinv{}
{}

\LANGUEinv{}
{}

\LANGUEinv{}
{}
\LANGUEinv{}{
}
\LANGUEinv{
}
{}

\LANGUEinv{
}
{}
\LANGUEinv{
}
{}
\LANGUEinv{
}
{}
\LANGUEinv{
}
{}
\LANGUEinv{
}
{}
\LANGUEinv{
}
{}

\LANGUEinv{
}
{}
\LANGUEinv{
}
{}

\LANGUEinv{
}
{}

\LANGUEinv{
}
{}

\newcommand\semspace{\kern -.25ex}

\renewcommand{\restriction}{\mathord{\upharpoonright}}

\newcommand{\myop}[1]{\operatorname{#1}}

\newcommand{\myclass}[1]{\operatorname{#1}}
 \newcommand{\gval}[2][]{\ensuremath{\myclass{GVAL}_{#1}\ifthenelse{\equal{#2}{}}{}{[#2]}}}

 \newcommand{\gpc}[1][]{\ensuremath{\myclass{ATSP}}}
\newcommand{\cgc}[1][]{\ensuremath{\myclass{ATS}}}
\newcommand{\gexpc}[1][]{\ensuremath{\myclass{AEXP}}}

\newcommand{\gplc}[1][]{\ensuremath{\myclass{ALP}}}
\newcommand{\cglc}[1][]{\ensuremath{\myclass{AL}}}

\newenvironment{dessinp}[1]{
\begin{center}\begin{tikzpicture}[baseline=(current bounding
      box.north),#1]
}
{\end{tikzpicture}
\end{center}}

\newcommand{\MAT}[3]{M_{#1\ifthenelse{\equal{#2}{}}{}{,#2}}\ifthenelse{\equal{#3}{}}{}{\left(#3\right)}}

\newcommand{\FP}{\ensuremath{\operatorname{FP}}}

\ifnotSLIDE{
\spnewtheorem{openproblem}{Open Problem}{\bfseries}{\rmfamily}
\spnewtheorem{remarkolivier}{Commentaire d'Olivier: }{\bfseries}{\rmfamily}
\spnewtheorem{postulate}{Postulate}{\bfseries}{\rmfamily}
}

\ifnotSLIDE{
\nolipics{
\nolncs{\spnewtheorem{theorem}{\LANGUE{Theorem}{Théorème}}{\bfseries}{\rmfamily}}
\nolncs{\spnewtheoremi{definition}{\LANGUE{Definition}{Définition}}{\bfseries}{\rmfamily}{theorem}}
\nolncs{\spnewtheoremi{lemma}{\LANGUE{Lemma}{Lemme}}{\bfseries}{\rmfamily}{theorem}}
\nolncs{\spnewtheoremi{corollary}{\LANGUE{Corollary}{Corollaire}}{\bfseries}{\rmfamily}{theorem}}
\nolncs{\spnewtheoremi{proposition}{Proposition}{\bfseries}{\rmfamily}{theorem}}
\nolncs{\spnewtheoremi{example}{\LANGUE{Example}{Exemple}}{\bfseries}{\rmfamily}{theorem}}
\nolncs{\spnewtheoremi{remark}{\LANGUE{Remark}{Remarque}}{\bfseries}{\rmfamily}{theorem}}
\nolncs{\spnewtheorem{exercise}{\LANGUE{Exercise}{Exercice}}{\bfseries}{\rmfamily}}
\spnewtheorem{exercisee}{\LANGUE{*Exercise}{*Exercice}}{\bfseries}{\rmfamily}
\spnewtheorem{summary}{\LANGUE{Summary}{Résumé}}{\bfseries}{\rmfamily}
\nolncs{\spnewtheorem{conjecture}{\LANGUE{Conjecture}{Conjecture}}{\bfseries}{\rmfamily}}
 \spnewtheorem{theoremaprouver}{\LANGUE{Theorem TO BE PROVED}{Théorème (A PROUVER)}}{\bfseries}{\rmfamily}
  \spnewtheorem{conjectureaprouver}{\LANGUE{Conjecture TO BE VERIFIED}{Conjecture (A VERIFIER)}}{\bfseries}{\rmfamily}
 }}
            
 \makeatletter
 \ifnotSLIDE{
  \ifnotTDEXAM{
\@ifundefined{proof}{
	\newenvironment{proof}{{\bf \LANGUE{Proof}{Démonstration}}:}{\hfill $\Box$ 
	}{}
	}}}
\makeatother

\newcommand\zero{\vectorl{0}}
\newcommand\un{\vectorl{1}}

\providecommand\citep[1]{\cite{#1}}

\renewcommand\Pr{\cp{Pr}}

\tikzstyle{case} = []

\RequirePackage{listings}
\def\beginlisting{\begin{lstlisting}}
\def\endlisting{\end{lstlisting}}

\newenvironment{algo}[1]{
  \lstset{escapechar=\@}
  \def\input{$#1$}

}{}

\newenvironment{FRANCAIS}{
\small
\color{orange}
}
{\normalsize
\color{black}
}

\newcommand\nl{\ \\ }

\usepackage{csquotes}

\definecolor{deepgreen}{RGB}{0,160,40}
\definecolor{forestgreen}{RGB}{0,150,70}
\definecolor{mossgreen}{RGB}{0,140,90}
\definecolor{greenwave}{RGB}{0,150,90}
\definecolor{forestmist}{RGB}{0,160,90}
\definecolor{darkteal}{RGB}{0,130,70}

\definecolor{seagreen}{RGB}{0,130,100}
\definecolor{freshteal}{RGB}{0,140,130}
\definecolor{greenteal}{RGB}{0,140,120}
\definecolor{blueteal}{RGB}{0,150,120}      
\definecolor{lagoon}{RGB}{0,180,120}
\definecolor{greenmist}{RGB}{0,170,110}

\definecolor{springgreen}{RGB}{0,190,100}
\definecolor{mintgreen}{RGB}{10,200,150}
\definecolor{oceanfoam}{RGB}{0,200,150}
\definecolor{aquagreen}{RGB}{30,210,170}
\definecolor{coastalaqua}{RGB}{40,210,180}
\definecolor{softaqua}{RGB}{40,210,190}

\definecolor{emeraldteal}{RGB}{0,180,140}
\definecolor{deepcyan}{RGB}{0,180,160}
\definecolor{lightcyan}{RGB}{0,200,160}
\definecolor{aquamarine}{RGB}{0,200,160}
\definecolor{lightaqua}{RGB}{60,230,200}
\definecolor{mistblue}{RGB}{80,240,210}

\definecolor{paleaqua}{RGB}{80,200,190}
\definecolor{glacier}{RGB}{90,150,200}
\definecolor{skyblueA}{RGB}{60,150,220}      
\definecolor{lightsky}{RGB}{90,170,230}
\definecolor{iceblue}{RGB}{120,190,240}
\definecolor{frost}{RGB}{150,210,250}

\definecolor{steelblue}{RGB}{30,100,160}
\definecolor{cadetblue}{RGB}{70,120,180}
\definecolor{coolblue}{RGB}{110,170,220}
\definecolor{morningblue}{RGB}{140,200,240}
\definecolor{blueiceA}{RGB}{40,150,210}       
\definecolor{blueair}{RGB}{60,170,225}

\definecolor{tealblue}{RGB}{0,110,130}
\definecolor{deepbluecyan}{RGB}{0,110,170}
\definecolor{azur}{RGB}{0,70,180}
\definecolor{brightazure}{RGB}{20,130,200}
\definecolor{lightazure}{RGB}{80,180,235}
\definecolor{skylight}{RGB}{100,200,245}

\definecolor{nightblue}{RGB}{40,0,110}
\definecolor{slateblue}{RGB}{20,40,120}
\definecolor{violetA}{RGB}{60,50,160}         
\definecolor{blueviolet}{RGB}{80,60,180}
\definecolor{purpleA}{RGB}{100,70,200}        
\definecolor{lightpurple}{RGB}{130,90,210}

\definecolor{lilac}{RGB}{160,110,220}
\definecolor{indigoA}{RGB}{50,0,130}          
\definecolor{deepindigo}{RGB}{70,10,150}
\definecolor{royalpurple}{RGB}{90,20,170}

\definecolor{coolmint}{RGB}{0,130,70}
\definecolor{deepsea}{RGB}{0,130,70}

\definecolor{crimson}{RGB}{200,30,60}
\definecolor{brickred}{RGB}{180,40,40}
\definecolor{coral}{RGB}{220,90,70}
\definecolor{salmon}{RGB}{240,130,120}

\definecolor{magenta}{RGB}{200,40,150}
\definecolor{fuchsia}{RGB}{220,60,170}
\definecolor{violetB}{RGB}{140,70,190}        
\definecolor{indigoB}{RGB}{90,50,180}         

\definecolor{royalblueA}{RGB}{50,80,200}      
\definecolor{azure}{RGB}{30,120,230}
\definecolor{skyblueB}{RGB}{80,160,230}       
\definecolor{tealB}{RGB}{0,150,170}           

\definecolor{marine}{RGB}{0,80,140}

\newcommand{\ProtaColorPair}[1]{%
  \colorlet{#1fill}{#1}%
  \colorlet{#1text}{#1!80!black}%
}

	\foreach \i/\name in {
	  1/deepgreen,
	  2/greenwave,
	  3/lagoon,
	  4/oceanfoam,
	  5/softaqua,
	  6/mistblue,
 	 7/lightsky,
	  8/blueiceA,   
	  9/deepcyan,
	  10/azur,
	  11/blueviolet,
	  12/royalpurple}
	  {\ProtaColorPair{\name}}

\newcommand{\DarkenForText}[2]{%
  \colorlet{#1text}{black!#2!#1}%
}

\DarkenForText{deepgreen}{20}
\DarkenForText{forestgreen}{20}
\DarkenForText{mossgreen}{20}
\DarkenForText{greenwave}{20}
\DarkenForText{forestmist}{20}
\DarkenForText{darkteal}{20}

\DarkenForText{seagreen}{20}
\DarkenForText{freshteal}{20}
\DarkenForText{greenteal}{20}
\DarkenForText{blueteal}{20}
\DarkenForText{lagoon}{20}
\DarkenForText{greenmist}{20}

\DarkenForText{springgreen}{20}
\DarkenForText{mintgreen}{20}
\DarkenForText{oceanfoam}{20}
\DarkenForText{aquagreen}{20}
\DarkenForText{coastalaqua}{20}
\DarkenForText{softaqua}{20}

\DarkenForText{emeraldteal}{20}
\DarkenForText{deepcyan}{20}
\DarkenForText{lightcyan}{20}
\DarkenForText{aquamarine}{20}
\DarkenForText{lightaqua}{20}
\DarkenForText{mistblue}{30}

\DarkenForText{paleaqua}{20}
\DarkenForText{glacier}{20}
\DarkenForText{skyblueA}{20}
\DarkenForText{lightsky}{20}
\DarkenForText{iceblue}{20}
\DarkenForText{frost}{20}

\DarkenForText{steelblue}{20}
\DarkenForText{cadetblue}{20}
\DarkenForText{coolblue}{20}
\DarkenForText{morningblue}{20}
\DarkenForText{blueiceA}{20}
\DarkenForText{blueair}{20}

\DarkenForText{tealblue}{20}
\DarkenForText{deepbluecyan}{20}
\DarkenForText{azur}{20}
\DarkenForText{brightazure}{20}
\DarkenForText{lightazure}{20}
\DarkenForText{skylight}{20}

\DarkenForText{nightblue}{20}
\DarkenForText{slateblue}{20}
\DarkenForText{violetA}{20}
\DarkenForText{blueviolet}{20}
\DarkenForText{purpleA}{20}
\DarkenForText{lightpurple}{20}

\DarkenForText{lilac}{20}
\DarkenForText{indigoA}{20}
\DarkenForText{deepindigo}{20}
\DarkenForText{royalpurple}{20}

\DarkenForText{coolmint}{20}
\DarkenForText{deepsea}{20}

\DarkenForText{crimson}{20}
\DarkenForText{brickred}{20}
\DarkenForText{coral}{20}
\DarkenForText{salmon}{20}

\DarkenForText{magenta}{20}
\DarkenForText{fuchsia}{20}
\DarkenForText{violetB}{20}
\DarkenForText{indigoB}{20}

\DarkenForText{royalblueA}{20}
\DarkenForText{azure}{20}
\DarkenForText{skyblueB}{20}
\DarkenForText{tealB}{20}

\DarkenForText{marine}{20}

\providecommand\apxparameter{}

\apxproofmode{
\usepackage[\apxparameter]{apxproof}
\theoremstyle{plain}

\newtheoremrep{theorem}{Theorem}
\newtheoremrep{claim}[theorem]{Theorem}
\newtheoremrep{proposition}[theorem]{Proposition}
\newtheoremrep{lemma}[theorem]{Lemma}
\newtheoremrep{corollary}[theorem]{Corollary}
\newtheoremrep{resume}[theorem]{Summary}
}

\makeatletter
\apxproofmodesubsection{
\let\axp@section\subsection
\renewcommand\appendixprelim{\section{Proofs from main documents}}
}
\makeatother

\makeatletter
\newcommand\appendixapxproofsubsection{
\appendix
\let\apx@orig@appendix\appendix
\renewcommand{\appendix}{}
}
\makeatother

\usepackage{xcolor}
\definecolor{check}{RGB}{160,160,0}

\ifdefined\HEUREVERSION\else
\newcount\VERSIONheures
\newcount\VERSIONminutes
\newcommand\HEUREVERSION{%
  \VERSIONheures=\time
  \divide\VERSIONheures by 60
  \VERSIONminutes=\time
  \multiply\VERSIONheures by 60
  \advance\VERSIONminutes by -\VERSIONheures
  \divide\VERSIONheures by 60
  \the\VERSIONheures h\ifnum\VERSIONminutes<10 0\fi\the\VERSIONminutes}
\fi

\includeversion{NOTLFMI2020}
\includeversion{NOTCSE304}
\excludeversion{NOTDIFFUSIONFINALE}

\providecommand\ifnotmodeDIFFUSIONFINALE[1]{
\ifdefined\DIFFUSIONFINALE
\else
#1
\fi
}

\providecommand\ifmodeDIFFUSIONFINALE[1]{
\ifdefined\DIFFUSIONFINALE
\ifdefined\SAFEMODE
\else
#1
\fi
\fi
}

\ifnotmodeDIFFUSIONFINALE{\includeversion{NOTDIFFUSIONFINALE}}
\ifnotmodeDIFFUSIONFINALE{\newcommand\FINPROVISOIRELFMI[1]{}}

\providecommand\FINPROVISOIRELFMI{
  \section{Bibliographic notes}

  The current text is highly based (sometimes copied) from:
  
  \bibliographystyle{alpha}
  \bibliography{/Users/bournez/bibliographie/Bib-Files/bournez-avantbib2DOI,/Users/bournez/bibliographie/Bib-Files/perso}
  \end{document}
  }

\providecommand\ifnotPOLYCHAPTERMODE[1]{
\ifdefined\POLYCHAPTERMODE
\else
#1
\fi
}

\newcommand\MYCOURSEON[2]{ 
	\ifnotPOLYMODE{
		\ifdefined\MAINAUXFILE
			\externaldocument{\MAINAUXFILE}
		\fi
		}
	
	\ifdefined\PASLAPREMIEREFOIS
		\newpage
	\else
		\makeindex

\begin{document}
	\fi
	\gdef\PASLAPREMIEREFOIS{}

	\ifdefined\XPOLY
		\XPAGEDEGARDE{\TITRECOURS\ifPOLYCHAPTERMODE{\\[0.5cm] \Large \bf \LANGUE{Chapter:}{Chapitre:}  #2}}{\SUBTITRECOURS}
		\ifPOLYCHAPTERMODE{
		\renewcommand\thesection{\arabic{section}}
		\chapter*{#2}
		}
	\else
	
      \title{ \ifnotmodeDIFFUSIONFINALE{\LANGUE{Course}{Cours} -}
      \ifPOLYCHAPTERMODE{\LANGUE{Chapter:}{Chapitre:} }
      #2}
      \author{
        Olivier Bournez
        \\[0.5cm] 
      }
      \date{\LANGUE{Version of}{Version du} \today\ifnotmodeDIFFUSIONFINALE{\unskip, \HEUREVERSION}\unskip}

      \ifnotPOLYCHAPTERMODE{
      	\maketitle
		}
      \ifPOLYCHAPTERMODE{\ifdefined\DOPAGEDEGARDE
      			\maketitle
		\fi
		\renewcommand\thesection{\arabic{section}}
      		\chapter*{#2}}

      \fi

      \DOINDEXPRINT
      
      \ifmodeDIFFUSIONFINALE{
		\excludeversion{recherche}
	}

    }

%
%
%
%

\newcommand\MYCOURSE[1]{
  \MYCOURSEON{\csname numerocours#1\endcsname}{\csname nomcours#1\endcsname}
  
    \ifnotmodeDIFFUSIONFINALE{}

  }
  
  \newcommand\MYCOURSEnew[1]{
  \MYCOURSEON{???}{#1}
  
  \ifnotmodeDIFFUSIONFINALE{}

  }


\newcommand\PASPREMIEREPAGE[1]{}
\providecommand\mychapter[1]{
  \renewcommand\PASPREMIEREPAGE[1]{##1}  \hrule\hrule \section*{#1}
  \hrule\hrule \  \\[1cm]}
\providecommand\chapter[1]{\PASPREMIEREPAGE{\newpage} 
  \renewcommand\PASPREMIEREPAGE[1]{##1}  \hrule\hrule \section*{#1}
  \hrule\hrule \  \\[1cm]}


\providecommand\IFNOTSUB[1]{#1}
\IFNOTSUB{

\usepackage{versions}
\includeversion{pasweb}

}

%
%
%
%

\newcommand\COURS[3]{
  \expandafter\newcommand\csname numerocours#2\endcsname{#1}
  \expandafter\newcommand\csname nomcours#2\endcsname{#3}
}

\newcommand\NANCY[3]{
  \expandafter\newcommand\csname numerocours#2\endcsname{#1}
  \expandafter\newcommand\csname nomcours#2\endcsname{#3}
}

\newcommand\INFquatrecentdouze[3]{
  \expandafter\newcommand\csname numerocours#2\endcsname{#1}
  \expandafter\newcommand\csname nomcours#2\endcsname{#3}
}

\newcommand\INFcinqsixun[3]{
  \expandafter\newcommand\csname numerocours#2\endcsname{#1}
  \expandafter\newcommand\csname nomcours#2\endcsname{#3}
}

\newcommand\MPRI[3]{
  \expandafter\newcommand\csname numerocours#2\endcsname{#1}
  \expandafter\newcommand\csname nomcours#2\endcsname{#3}
}

\newcommand\COURSEINCLUDE[2][]{
  \renewcommand\IFNOTSUB[1]{}
  \renewcommand\MYCOURSE[1]{\chapter{\ifnotmodeDIFFUSIONFINALE{#1} \csname nomcours##1\endcsname}
    \ifnotmodeDIFFUSIONFINALE{\mychapter{Fichier: #2}}
  }
   \renewcommand\MYCOURSEnew[1]{\chapter{\ifnotmodeDIFFUSIONFINALE{#1} ##1 }
    \ifnotmodeDIFFUSIONFINALE{\mychapter{Fichier: #2}}
  }
  \input{#2}
}

\newcommand\STRESSCOURSEINCLUDE[1]{
  \COURSEINCLUDE[!STRESS!]{#1}
  }

  \newcommand\subCOURSEINCLUDE[1] {
    \NDLR{BEGIN:  J'inclus/ Je répète ICI #1 }
    \input{INCLUDE/#1}
    \NDLR{END:  J'inclus/ Je répète ICI #1 }
}

%
%
%
%


\newcommand\montruc[2][]{
  \ifnotmodeDIFFUSIONFINALE{
  \begin{maboitetruc}[#1]{\danger} #2
  \end{maboitetruc}
}
}

\newcommand\montrucdanger[3][]{
  \ifnotmodeDIFFUSIONFINALE{
                    \montruc[#1]{\danger #2: #3}
                    }
}

\newcommand\montrucdangermargin[3][]{
  \ifnotmodeDIFFUSIONFINALE{
  \montruc[#1]{\danger #2: #3}
\marginpar{\danger #2: #3}
}
 }

%
%
%
%

\makeatletter
\renewcommand\beginmarkversion{\montrucdanger[colback=yellow!10]{=== begin NOT}{}\@Vs@sffbox{\@currenvir$>$}}
 \renewcommand\endmarkversion{\@Vs@sffbox{$<$\@currenvir}\montrucdanger[colback=yellow!10]{end=== NOT}{}}
\makeatother

%
%
%
%

 \newcommand\NDLR[1]{\montrucdanger[colback=blue!20]{NDLR}{#1}}
 \newcommand\NLDR[1]{\NDLR{#1}}
 \newcommand\VOIRAUSSI[1]{\montrucdanger[colback=yellow!20]{VOIR AUSSI}{#1}}
 \newcommand\TODO[1]{\montrucdanger[colback=green!20]{TODO}{#1}}
 \newcommand\ATTENTION[1]{\montruc[colback=brown!20]{\danger ATTENTION#1}}
 \newcommand\COMPRENDPAS[1]{\montrucdanger[colback=yellow!20]{JE
     COMPRENDS PAS CA}{#1}}
  \newcommand\POSSIBLE[1]{\montrucdanger[colback=yellow!10]{
      Serait possible}{#1}}
 \newcommand\ENLEVE[1]{\montrucdanger[colback=gray!20]{
      ENLEVE}{#1}}

  \newcommand\DEPLACE[1]{
    \vspace{0.5cm}
    \hrule
    \hrule
      \vspace{0.5cm}
    \montrucdangermargin[colback=black!40]{DEPLACE/MISSING CONCEPT
    }{#1}
      \vspace{0.5cm}
    \hrule
    \hrule
      \vspace{0.5cm}
  }

\newcommand\BESOINDE[1]{\montrucdangermargin[colback=pink!20]{ATTTENTION
     BESOIN DE}{#1}}
\newcommand\MISSINGCONCEPT[1]{\montrucdangermargin[colback=pink!20]{MISSING CONCEPT
     }{#1}}
\newcommand\SAUTE[1]{\montrucdangermargin[colback=pink!20]{SAUTE
  }{#1}}
\newcommand\SHOULDBEHERE[1]{\BEGINTEMPENVCOLOR{blue!30}\montrucdanger[colback=pink!10]{
SHOULD
    BE HERE
  }{    
    \vspace{1cm}
    
    \centerline{$\vdots$}
    
   \centerline{ #1}
    
      \centerline{$\vdots$}
          
    \vspace{1cm}
    }\ENDTEMPENVCOLOR}
\newcommand\COULDBEHERE[1]{\montrucdangermargin[colback=pink!20]{COULD
    BE HERE
  }{#1}}

%
%
%
%

\newcommand\SOURCEURL[1]{
  \ifnotmodeDIFFUSIONFINALE{
    \marginpar{ \textcolor{magenta}{Source: \url{#1}} }
    }
}

\newcommand\SOURCETXT[1]{
  \ifnotmodeDIFFUSIONFINALE{
    \marginpar{ \textcolor{magenta}{Source: {#1}} }
    }
}

\newcommand\SOURCE[1]{
  \nocite{#1}
  \ifnotmodeDIFFUSIONFINALE{
    \marginpar{ \textcolor{magenta}{Source: \cite{#1}} }
    }
}


\newcommand\SOURCEJECH{\SOURCE{jech2003set}}
\newcommand\SOURCEKECHRIS{\SOURCE{kechris2012classical}}
\newcommand\SOURCENOTESSETTHEORY{\SOURCE{moschovakis2006notes}}
\newcommand\SOURCEKOZENCT{\SOURCE{LivreKozenTC}}
\newcommand\SOURCEMARKER{\SOURCE{marker2002descriptive}}
\newcommand\SOURCEMOSCHO{\SOURCE{moschovakis2009descriptive}}

%
%
%
%




\ifnotmodeDIFFUSIONFINALE{
\newenvironment{recherche}
{}{}
\tcolorboxenvironment{recherche}{colback=blue!5!white, colframe=red!75!black,fonttitle=\bfseries, colbacktitle=blue!85!black,enhanced,
attach boxed title to top center={yshift=-2mm},
  title={FAIRE RECHERCHE ICI}}
}







\ifnotmodeDIFFUSIONFINALE{
\newenvironment{commentaire}{\begin{maboitecommentaire}[colback=brown!30]{Commentaire} Commentaire :}{\end{maboitecommentaire}}
\newcommand\COMMENTAIRE[2][Commentaire ]{
\begin{maboitecommentaire}[colback=brown!30]{#1}
Commentaire : #2
\end{maboitecommentaire}
}
}




\ifnotmodeDIFFUSIONFINALE{
\newenvironment{pasbesoinpourlinstant}{NOT NEEDEED FOR
  THIS SECTION: }{ }
    \tcolorboxenvironment{pasbesoinpourlinstant}{colback=green!5!white, colframe=gray!75!black,fonttitle=\bfseries, colbacktitle=green!85!black,enhanced,
attach boxed title to top left={yshift=-2mm},
  title={PAS BESOIN POUR L'INSTANT}}
}


\ifnotmodeDIFFUSIONFINALE{
\newenvironment{bonus}{BONUS: }{
  }
  \tcolorboxenvironment{bonus}{colback=gray!5!white, colframe=gray!75!black,fonttitle=\bfseries, colbacktitle=gray!85!black,enhanced,
attach boxed title to top left={yshift=-2mm},
  title={BONUS}}
}


\ifnotmodeDIFFUSIONFINALE{
\newcommand\ATTENTIONREPETE[1]{
  \ATTENTION{Repete (attention modif): #1}
  \begin{remarkolivier}
Repete (attention modif): #1
\end{remarkolivier}
}
}

%
%
%
%

\newcommand{\x}{\times}
\newcommand{\s}{\sigma}
\newcommand\sep{.}
\newcommand\bzero{\zero}
\newcommand\bun{\un}
\renewcommand{\a}{\alpha}
\def\hd{{\sf hd}}
\def\tl{{\sf tl}}
\def\cons{{\sf cons}}
\def\Cons{{\sf Cons}}
\def\Pr{{\sf Pr}}
\def\Op{{\sf Op}}
\def\Rel{{\sf Rel}}
\def\Equal{{\sf Equal}}
\def\Eval{{\sf Eval}}
\def\Gate{{\sf Gate}}
\def\Selection{{\sf Selection}}
\def\Result{{\sf Result}}
\def\Circuit{{\sf Circuit}}
\def\next{{\sf next}}
\def\comp{{\sf comp}}
\def\finalcomp{{\sf finalcomp}}
\def\Pick{{\sf Pick}}
\def\Hd{{\sf Hd}}
\def\Tl{{\sf Tl}}
\def\RevHd{{\sf RevHd}}
\def\Node{{\sf Node}}
\def\Length{{\sf Length}}
\def\Tape{{\sf Tape}}
\def\Head{{\sf Head}}
\def\Time{{\sf Time}}
\def\unpad{{\sf unpad}}
\def\concat{{\sf concat}}
\def\rg{{\sf right}}
\def\lf{{\sf left}}
\def\st{{\sf state}}
\def\tope{{\sf top}}

%
%
%
%

\newcommand\buglst[1]{#1}



\ifnotSLIDE{
\LANGUE{
\spnewtheoremi{exemple}{\NDLR{ECRIT EXEMPLE AU LIEU DE
  EXAMPLE}}{\bfseries}{\rmfamily}{theorem} 
\spnewtheoremi{remarque}{\NDLR{ECRIT REMARQUE AU LIEU DE REMARK}}{\bfseries}{\rmfamily}{theorem} 
}
{
\spnewtheoremi{exemple}{Exemple}{\bfseries}{\rmfamily}{theorem} 
\spnewtheoremi{remarque}{Remarque}{\bfseries}{\rmfamily}{theorem} 
}
}
            
            \newcommand\PERSOSOURCEPOSSIBLE[1]{\PERSOSOURCE{#1}\PERSOCOMMENTAIRE{#1}}
\newcommand\PERSOPOSSIBLE[1]{\POSSIBLE{#1}}
\newcommand\PERSOETAIT[1]{\NDLR{ETAIT: #1}}

\providecommand\nl{}

\newcommand\PERSOSOURCE[1] { \SOURCETXT{#1}}
\newcommand\PERSOCOMMENTAIRE[1]{\NDLR{#1}}
\newcommand\PERSOCOMMENTAIREd[2]{\NDLR{#1 \textbf{#2}}}

\newenvironment{dessinpb}
{
\begin{center}\begin{tikzpicture}[baseline=(current bounding
      box.north)]
}
{\end{tikzpicture}
\end{center}}

\newcommand\PERSOMANQUE[1]{\NDLR{\textbf{-- manque --
      #1 --manque--}}}

\newcommand\PERSOENLEVE[1]{\ENLEVE{#1}}

%
%
%
%


\tikzstyle{nd} = [shape=circle,draw]

\newcommand\mmod{mod}

\newcommand\mmV{v}
\newcommand\mmVb{\overline{\mmV}}
\newcommand\mmVe{\mmVb}


\newcommand\exoref[1]{\ref{exo:#1-stt}}

\newenvironment{correction}[1]
{\par\medskip\noindent \label{exo:#1-cor} \hbox{\bf Exercice \ref{exo:#1-stt} 
  \   (page \pageref{exo:#1-stt}). }}
{\par\medskip}


\newcommand\DECISIONint[3]{
\begin{itemize}
\item[\textbf{Donnée}:] #2
\item[\textbf{Réponse}:] #3
\end{itemize}
}

\newcommand\INDECIDABLE[5]{
\DECISION{#1}{#2}{#3}

\begin{#4} \label{#5}
Le problème $#1$ 
est indécidable.
\end{#4}
}

\newcommand\INDECIDABLEenglish[5]{
\DECISIONenglish{#1}{#2}{#3}

\begin{#4} \label{#5}
The problem $#1$
is undecidable.
\end{#4}
}

\newcommand\INDECIDABLEq[5]{
\DECISION{#1}{#2}{#3}

\begin{#4} \label{#5}
Le problème $#1$
est ?
\end{#4}
}

\newcommand\INDECIDABLEqenglish[5]{
\DECISION{#1}{#2}{#3}

\begin{#4} \label{#5}
The problem $#1$
is ?
\end{#4}
}

\newcommand\DECISIONintenglish[3]{
\begin{itemize}
\item[\textbf{Input}:] #2
\item[\textbf{Answer}:] #3
\end{itemize}
}

\newcommand\DECISIONi[4]{
\item $#1$
\DECISIONint{#1}{#2}{#3}
}

\newcommand\DECISIONienglish[4]{
\item $#1$
\DECISIONintenglish{#1}{#2}{#3}
}

\newcommand\DECISION[3]{
\begin{definition}[$#1$]\ 

\DECISIONint{#1}{#2}{#3}
\end{definition}
}

\newcommand\DECISIONenglish[3]{
\begin{definition}[$#1$]\ 
%
\DECISIONintenglish{#1}{#2}{#3}
\end{definition}
}

\newcommand\NPC[4]{ 
\DECISION{#1}{#2}{#3}

\begin{#4}
Le problème $#1$
est $\NP$-complet. \LANGUE{\myindex{NP-completeness}}{\myindex{NP-completude@$\NP$-complétude}}
\end{#4}
}

\newcommand\NPCenglish[4]{ 
\DECISIONenglish{#1}{#2}{#3}
\begin{#4}
The problem $#1$
is $\NP$-complete. \LANGUE{\myindex{NP-completeness}}{\myindex{NP-completude@$\NP$-complétude}}
\end{#4}
}

\newcommand\NPCp[5]{
\DECISION{#1}{#2}{#3}

\begin{#4}[#5]
Le problème $#1$
est $\NP$-complet.
\end{#4}
}

\newcommand\NPCpenglish[5]{
\DECISIONenglish{#1}{#2}{#3}

\begin{#4}[#5]
The problem $#1$
is $\NP$-complete.
\end{#4}
}

\newcommand\BRUNOCOMMENTAIRE[1]{\NDLR{Bruno commentaire: #1}}

\ifnotSLIDE{
\spnewtheorem{notation}{Notation}{\bfseries}{\rmfamily}
}

 \newcommand\motnouv[1]{\emph{#1}\index{#1}}
\newcommand\INTmotnouv[1]{\emph{#1}}
\newcommand\myindex[1]{\index{#1}}

\newcommand\synonyme[2]{\kindex{#1|see{#2}}}
\newcommand\indexM[1]{\index{$#1$}}
\newcommand\idx[1]{#1\index{#1}}
\newcommand\idxm[1]{#1\index{$#1$}}
\newcommand\idxM[1]{$#1$\index{$#1$}}
\newcommand\idxmm[2]{#1\index{#2@$#1$}}
\newcommand\idxi[2]{#1\index{#2}}
 \newcommand\motnouvi[2]{\INTmotnouv{#1}\index{#2}} 
 \newcommand\motnouva[2]{\INTmotnouv{#1}\index{#2@#1}} 
\newcommand\motnouvm[1]{\INTmotnouv{#1}} 

\newenvironment{PERSOVERBATIM}{}{}
 \newcommand{\joNP}[3]{{ Problème \textbf{ { \motnouv{##1}}:}
            \begin{itemize}
            \item  \textbf{ Instance:} {##2} 
            \item   \textbf{ Question:} {##3}
            \end{itemize}
          }}

%
%
%
%

\newcommand\defin[1]{\textbf{#1}}


\newcommand\soustitre[1]{\textbf{#1}}


\providecommand\motnouv[1]{\emph{#1}}
\providecommand\motnouvi[2]{\emph{#1}}

\newcommand\papiernorm[1]{\myop{norm}}

\newcommand\olivier[1]{ \ATTENTION{Olivier: #1}}

\renewcommand\appendixprelim{\section{Proofs of the main results}}

\usepackage{tabularx}
\usepackage{enumitem}
\usepackage{microtype}
\usepackage{hyperref}

\hypersetup{
  colorlinks=true,
  linkcolor=blue!55!black,
  citecolor=blue!55!black,
  urlcolor=blue!55!black,
  pdftitle={One Polynomial Field for Every Euler Step and the Flow: Polynomial Dynamics for P and PSPACE},
  pdfauthor={Olivier Bournez}
}

\emergencystretch=2em
\setlist[itemize]{topsep=0.4em,itemsep=0.25em,parsep=0em}
\setlist[enumerate]{topsep=0.4em,itemsep=0.25em,parsep=0em}
\renewcommand{\arraystretch}{1.12}

\providecommand\Ptime{\mathrm P}
\providecommand\PSPACE{\mathrm{PSPACE}}
\providecommand\FP{\mathrm{FP}}
\providecommand\FPSPACE{\mathrm{FPSPACE}}
\providecommand\eps{\varepsilon}
\providecommand\restore{\varrho}
\providecommand\Len{\operatorname{Len}}
\providecommand\dist{\operatorname{dist}}
\providecommand\hardmap[1]{\mathcal H_{#1}}
\providecommand\ALP{\mathrm{ALP}}
\providecommand\GPVAL{\mathrm{GPVAL}}

\newif\ifinternalnotes
\internalnotesfalse
\newcommand{\workingnote}[1]{%
  \ifinternalnotes
  \par\smallskip\noindent
  \textbf{Internal proof check.}\ \emph{#1}\par\smallskip
  \fi}

\title{One Polynomial Field for Every Euler Step and the Flow:\\
Polynomial Dynamics for \(\Ptime\) and \(\PSPACE\)}
\author{Olivier Bournez\thanks{\affiliationofficielleolivier. \texttt{bournez@lix.polytechnique.fr}. ORCID: 0000-0002-9218-1130.}}
\date{}

\begin{document}

\begin{titlepage}
\maketitle
\thispagestyle{empty}
\begin{abstract}
We construct, for each deterministic Turing machine that halts on all inputs, one polynomial vector field with rational coefficients that performs the same logical computation at unit Euler step, at every smaller or variable step in \((0,1]\) with divergent sum, and along the flow.
This unifies discrete-time and continuous-time computation by polynomial dynamics, which earlier characterizations treated separately.
For a machine making \(T\) transitions in space \(H\), a logical scale \(h=2^{-s}\) with \(s=O(H)\) suffices, with physical time \(O(s(T+s))\), drift disturbances up to \(h^{2K}\) for a fixed exponent \(K\), and \(2Ks+3\) fractional bits at unit step.
On a fixed compact domain, polynomial time and robust eventual decision characterize \(\Ptime\) and \(\PSPACE\), respectively, with polynomial scale thresholds and no supplied stabilization horizon in the space case.
Without sign, floor or conditional primitives, margin amplification at bounded speed, overlapping polynomial gates and interval-product estimates give the common realization, from which we derive shared logical traces, fixed-field \(\PSPACE\)-complete prediction and a robust safety family with polynomial certificates common to all Euler steps and the flow.
\end{abstract}
\end{titlepage}

\setcounter{tocdepth}{1}
\pagenumbering{roman}
\tableofcontents
\clearpage
\pagenumbering{arabic}

\section{Introduction}\label{sec:introduction}

To a polynomial vector field \(F\), which we simply call a field, we can associate the ordinary differential equation \(\dot z=F(z)\), but also discrete-time dynamics, which can be regarded as its Euler schemes
\begin{equation}\label{eq:front-euler}
 z_{k+1}=z_k+\eta_k F(z_k),\qquad 0<\eta_k\le1.
\end{equation}
From this point of view, a field can be run in three ways: as a map, by Euler steps of size one; as a scheme, by Euler steps of any smaller or variable size; and as a differential equation, by its flow.
In general the three runs do not compute the same thing: the field \(F(z)=1-2z\) gives the unit-step map \(z\mapsto1-z\), which alternates between \(0\) and \(1\) from \(z_0=0\).
At step \(1/2\), every state reaches \(1/2\) in one update, while the flow converges to \(1/2\).
Our starting point is a surprising technical result: for each decider, that is, each deterministic Turing machine that halts on all inputs, we can construct one field and one polynomial initialization map with rational coefficients whose three runs perform \emph{the same computation}.

By the same computation, we mean that each run has its own observation times, given by an internal clock, at which its state approximately encodes the same successive configurations of the decider (compiled into a two-stack program), so that the discrete observations read the same word.
This holds at every logical scale \(h=2^{-s}\) for which \(s\) exceeds a threshold linear in the space of the decider.
Drift disturbances of sup norm at most \(h^{2K}\), for a fixed integer \(K\), are allowed on the coordinates that evolve, in each of the three runs. These are precision budgets, not noise floors: they shrink with the scale, while the input, the scale and the initial state are exact. The agreement between the runs is an agreement of their logical traces at the cycle sections, not of their states at common times.
The input is a raw integer, not an encoding of it: the field itself extracts its bits.

This result unifies discrete-time and continuous-time computation by polynomial dynamics, which earlier characterizations treated separately: the same field and the same initialization are correct for every admissible Euler mesh and for the flow, under the disturbances just described (Section~\ref{sec:related}).
We use it to characterize polynomial time and polynomial space without any sign, floor or conditional function, and to obtain logical traces and safety certificates common to all the runs.

Our fields have two frozen parameters, the input \(x\) and the scale \(h\), and we write \(F(x,h,z)\) in the dynamics given by~\eqref{eq:front-euler}.
The vector \(z\) contains the coordinates that move, called the active state.
The scale controls the logical resolution of the encoded configurations, in a fixed compact box.
The mesh is the sequence of steps \(\eta_k\), called admissible when its sum diverges, and physical time is \(t_k=\sum_{i<k}\eta_i\).
For the flow, a \emph{forcing} is a bounded measurable map \(\xi(t)\) added to the field on the active coordinates: the forced flow is the absolutely continuous solution of \(\dot z=F(x,h,z)+\xi(t)\) from the same initial state, and the exact flow is the case \(\xi=0\).
Branching and scheduling are implemented inside the field, which requires no external phase schedule. The observer uses thresholds for the final Boolean readout at \(1/2\) and for the phases of the clock at \(9/10\). Apart from the initialization, these observations are the only operations not performed by the polynomial field.

\myparagraph{Why agreement requires a construction}
In the example above, the unit-step alternation is not preserved under refinement.
This is a simple instance of the spurious periodic orbits studied in numerical dynamics~\cite{StuartHumphries1996}.
Convergence as \(\eta\to0\) alone does not give correctness at a fixed positive step. A direct Gr\"onwall error bound may require a step that shrinks with the horizon.
Slowing a field can turn a small step into a unit step, but it does not reduce the number of updates needed for the same computation. For example, a step of size \(2^{-s}\) requires \(2^s\) updates per unit of the original physical time.
The step bound must therefore be considered together with the cost of the execution. Our construction takes \(O(s)\) physical time per source transition for every admissible mesh, with step one allowed independently of the horizon, the input and the scale. The polynomial coefficients stay fixed and the active-coordinate speeds stay uniformly bounded. Correctness also requires ordered clock events and control of accumulated errors.

Computations have been compiled into chemical reaction networks~\cite{CMSB17vantardise,Vasi2020CRN} and realized in fluid dynamics~\cite{Cardona2021Turing}. In hybrid verification, relating discrete and continuous properties is a central question, with proof-theoretic correspondences established by Platzer~\cite{platzer2012complete}.
These settings motivate specifying which computations and safety properties survive a change in execution. Here we study this question for fields constructed from programs, with explicit time and precision bounds.

\myparagraph{A realization theorem with resources}
We first state our main technical result for a deterministic two-stack decider with finite control. Throughout, this discrete machine is the \emph{source} machine, and its exact configurations, transitions and running time are the source configurations, transitions and time, as opposed to the states and physical time of the polynomial realization. A one-tape Turing decider can be represented in this form with constant-factor time and space overhead, by storing the tape on the two sides of the head. An integer input is read in its canonical binary representation, with \(0\) represented by one bit. The compiler loads this word and makes the accepting and rejecting halts reset the working data (a \emph{resetting halt}).

The active state of the field contains data registers, which hold the encoded source configuration, together with auxiliary registers and an internal clock. This clock has four phases, and one cycle of the clock simulates one source transition. We observe the data at the beginning of each cycle, at times that we call the internal cycle sections, or simply sections. The sections of the text are always cited by their numbers, and we give the details in Section~\ref{sec:overview}.
Our proof lets the errors accumulate: it bounds their sum at each section by the local defects of the cycles, propagated by the products of the affine transitions over every interval of the run, and it chooses the logical scale so that this total stays inside the logical margins. This is what keeps the step at one, however long the computation is.

\begin{theoremrep}[A common polynomial realization]\label{thm:machine-resources}
Fix such a decider \(M\). For inputs of binary length \(n\), let \(T(n)\ge1\) bound its transitions and let \(H(n)\ge n\) bound each stack height, including the initial word and its fixed encoding overhead. There are rational polynomial maps \(F_M(h,z)\) and \(c_M(x,h)\), a fixed compact box \(\mathcal B\), an integer \(K\ge256\) and constants \(C,C',r_*>0\) with the following properties. The maps, box and constants are fixed by \(M\). One may take \(K=256\) for every \(M\), while the scale constant \(C\) may depend on \(M\).

For every nonnegative integer input \(x\in\N\) of binary length \(n\) and every integer
\begin{equation}\label{eq:source-resource-scale}
 s\ge C\bigl(1+n+H(n)+\log_2(T(n)+2)\bigr),\qquad h=2^{-s},
\end{equation}
start exactly at \(z_0=c_M(x,h)\). Every execution
\[
 z_{k+1}=z_k+\eta_k F_M(h,z_k)+\eta_k\xi_k,
 \quad 0<\eta_k\le1,\quad \sum_k\eta_k=\infty,
 \quad \|\xi_k\|_\infty\le h^{2K},
\]
stays in \(\mathcal B\). At every physical time \(t_k=\sum_{i<k}\eta_i\ge C's(T(n)+s)\), its output is within \(1/4\) of the answer of \(M\) on \(x\). Its active length through the first grid point at or beyond that horizon, that is, the sum of its increments in sup norm (Definition~\ref{def:length}), is \(O(s(T(n)+s))\), with a constant depending only on \(M\).

Before and after the resetting halt, the data at internal cycle sections are within \(r_*h\), in sup norm, of the successive encoded configurations of the compiled two-stack program. Consecutive cycle sections are separated by \(\Theta(1+s)\) physical time, with constants independent of the input, scale, mesh and disturbances. The same section and resource guarantees hold for the continuous system with any measurable forcing of sup norm at most \(h^{2K}\). Its solution stays in \(\mathcal B\) and has the same stable decision bound. Each execution has its own section times. In continuous time, after the initial section, cycle sections occur at every fourth successive phase crossing of the clock threshold \(9/10\). They are the limits of the corresponding constant-mesh Euler sections when the disturbances are cell averages of that forcing (Proposition~\ref{prop:logical-traces} and Lemma~\ref{lem:intrinsic-sections}).

Only active coordinates are perturbed. The input and scale are exact and frozen. The unit-step computation admits an implementation using \(O(s)\) fractional bits and \(O(s)\) workspace. The field receives neither \(T(n)\) nor \(H(n)\).
\end{theoremrep}
\begin{proofsketch}
Reading the input takes \(O(s+n)\) transitions. Valid affine transition products have norm at most \(G/h\), for a constant \(G\), so local cycle errors accumulate with a factor controlled by stack height. The scale bound makes their total smaller than the logical margin. Resetting halt gives an infinite correct tail, and compact Euler limits give the continuous conclusions.
\end{proofsketch}
\begin{proof}
We use the canonical raw reader of Lemma~\ref{lem-hard-normal-form}, followed by the fixed word program. Fixed encodings of the alphabet, and any reversal needed to load the word, change the time and the height by fixed factors. Hence the resulting source program uses at most \(a(T(n)+s+n+1)\) transitions and height at most \(a(H(n)+n+1)\), for a fixed integer \(a\). At zero, we use the canonical one-bit word, as in this lemma. We increase \(C\) so that~\eqref{eq:source-resource-scale} implies \(s\ge n+4\) and \(\lfloor(s-2)/2\rfloor\ge a(H(n)+n+1)\).

Fix for example \(K=256\), and let \(a_*=K/4-2>0\). For \(s\ge C\log_2(T(n)+2)\), we have \(T(n)+2\le2^{s/C}\). We also choose \(C\) so that \(1/C<a_*/2\). The left side of the stack budget~\eqref{eq:front-resource-condition} is then bounded by
\[
 C_1(1+s)(2^{s/C}+s+1)2^{-a_*s},
\]
where \(C_1\) depends only on the fixed program and on \(K\), and this tends to zero with \(s\). A final increase of \(C\), using the constant term in~\eqref{eq:source-resource-scale}, enforces the budget and every local smallness threshold, for all permitted scales. This also bounds the section error by \(r_*h\), with the fixed \(r_*\) of Theorem~\ref{thm:quantitative-transfer}. The source computation, including its resetting halting transition, therefore takes \(O(T(n)+s)\) cycles, each of time and length \(O(s)\), and a larger constant \(C'\) covers the horizon of the stable decision.

The prescribed initialization is the rational polynomial raw-input initialization of Lemma~\ref{lem-hard-normal-form}. The compiled field depends only on \(h\) and on the bounded active coordinates, and its degree, dimension and coefficients are fixed by the program. Theorem~\ref{thm:quantitative-transfer} gives the section estimates for fixed and variable meshes, including all permitted disturbances. After the resetting halt, the error is \(O(\Delta_\eps)\), and increasing \(C\) ensures that this error is also at most \(r_*h\) at all later sections. The argument on limiting sections in Proposition~\ref{prop:logical-traces} passes the quantitative section estimates to every measurably forced continuous execution of this resetting halting run, and Corollary~\ref{cor:forced-ode} gives the stable decision on its infinite tail. Both arguments only use the robust compact discrete trajectories that we just obtained: neither of them uses the present theorem. The speed, including the allowed forcing, is uniformly bounded on the fixed box. Since each step is at most one, the first grid point at or beyond the horizon overshoots it by at most one. This proves the bound on the active length, and the same estimate on the speed gives the corresponding continuous bound.

For unit steps, we initialize the rational vector exactly, we evaluate the fixed polynomial from the current rational or dyadic state, and we round the complete update to \(b=2Ks+3\) fractional bits. The resulting error is at most \(2^{-b}<h^{2K}\), so the rounded execution is an allowed trajectory, and compactness bounds its integer parts. The frozen input and scale, the exact initial vector and each evaluation of the fixed-degree polynomial use \(O(n+s)=O(s)\) bits. The bound \(T(n)+2\le2^{s/C}\) implies that a fixed exponential counter \(2^{c_1s}\) dominates the physical horizon, for a sufficiently large fixed integer \(c_1\). This counter uses \(O(s)\) bits, and it gives a stopping rule that does not require evaluating \(T(n)\). Notice that this is a statement on the resources of the execution: it is not an algorithm for discovering valid resource bounds of an arbitrary source program.
\end{proof}

For a fixed deterministic decider, the number of configurations at space \(H\) is \(2^{O(H)}\). An execution that halts cannot repeat a configuration. We may therefore take \(\log(T+2)=O(H)\), so \(s=O(H)\) suffices. Formula~\eqref{eq:source-resource-scale} keeps the separate time and space costs visible.

\myparagraph{Complexity as a consequence of the realization}
The converse directions use ordinary bit complexity. For polynomially many unit steps in a bounded box, polynomially many guard bits suffice. Under persistent perturbations of radius \(h^r\), for a fixed \(r>0\), rounding to \(O(s)\) bits gives a permitted execution on a finite state space. This yields a space bound even when no stabilization horizon is supplied.

\begin{corollary}[Unit-step and mesh-uniform complexity]\label{cor:front-complexity}
For rational polynomial systems with polynomial scale thresholds and a fixed compact active box, the following characterizations hold in the decision semantics of Section~\ref{sec:model-results}. At unit step, that is, when the polynomial map \(z\mapsto z+F(x,h,z)\) is iterated, polynomial physical time characterizes \(\Ptime\), and robust eventual decision, with disturbance tolerance \(h^r\) for a fixed \(r>0\), characterizes \(\PSPACE\).

Requiring correctness of the same field and initialization for every constant Euler step \(0<\eta\le1\) leaves both classes unchanged. Each language has a witness also valid for admissible variable meshes and in continuous time. For functions of polynomial output length, the unit-step and mesh-uniform finite-window semantics of Definition~\ref{def:function-window} characterize \(\FP\) with polynomial scale and time bounds.
\end{corollary}
\begin{proof}
The language assertions are Theorems~\ref{thm:main-P} and~\ref{thm:main-PSPACE}, with their continuous witnesses given by Theorem~\ref{thm:machine-resources}. Corollary~\ref{cor:functions-final} gives the function assertion. Correctness at every step implies correctness at step one by taking \(\eta=1\). The reverse implication at the level of classes uses compilation and may replace the original field.
\end{proof}

We do not claim refinement invariance for an arbitrary field that computes correctly at unit step: the reverse direction is an equality of language classes, which may replace the witness by compilation. The scale and resource bounds are semantic conditions, not a syntax that forces termination.

\myparagraph{Contributions}
\begin{itemize}
\item Theorem~\ref{thm:machine-resources} gives the common realization with explicit resources.
Its general interface, Theorem~\ref{thm:quantitative-transfer}, compiles separated affine programs (Definition~\ref{def:affine-program}) with a bound on transition products over every interval.
Corollary~\ref{cor:front-complexity} gives the characterizations of \(\Ptime\), \(\PSPACE\) and \(\FP\). The characterization of \(\PSPACE\) requires no supplied stabilization horizon (Theorem~\ref{thm:main-PSPACE}).
\item Proposition~\ref{prop:logical-traces} preserves logical observations under all admissible meshes and measurable continuous forcing.
Corollary~\ref{cor:segment-safety} gives a geometric condition for safety between sections, and Proposition~\ref{prop:implementation-contract} gives an error budget for approximate implementations.
\item A fixed field has a \(\PSPACE\)-complete robust prediction problem, with a horizon supplied in binary, whose hard instances agree under the temporal interpretations (Theorem~\ref{thm:prediction}).
Every safe member of a robust safety family has one rational polynomial certificate for all Euler steps and the flow. Since continuous-time safety does not imply Euler safety (Example~\ref{ex:flow-not-euler}), this is strictly more than a continuous certificate.
Uniformly polynomial certificate bit length would imply \(\mathrm{NP}=\PSPACE\) (Theorem~\ref{thm:common-certificates}).
\end{itemize}

\myparagraph{The two estimates behind the realization}
The compiler resolves separated affine tests by bounded-speed cubic amplifiers, the comparators, then uses polynomial weights to select an affine transition.
An autonomous clock controls auxiliary registers, called shadows, which receive the next configuration before it is copied back into the data.
The clock phases act as gates (they multiply the corresponding updates). These gates overlap, so the proof controls their integrated influence during each phase.
This gives a local cycle defect uniformly for every step at most one.

Over a long run, the defect is propagated through products of affine transition matrices over intervals.
For stacks, these products telescope with height, even though a single pop expands errors by four (stacks are encoded in base four).
Using the worst local gains would make the sufficient precision exponent grow with the number of transitions. This would lose the space bound for exponentially long computations.
The interval estimate instead gives an exponent controlled by stack height and the logarithm of the number of transitions.
The scale is chosen so that the sum of propagated defects stays inside every test margin.
The clock, comparator and copy estimates restore the full cycle invariant.
We prove these estimates directly for Euler updates, uniformly in the mesh, then obtain the forced continuous executions by compactness.
Section~\ref{sec:overview} gives the estimates and follows a stack parser through the construction.

\myparagraph{Comparison with prior realizations}
Gra\c{c}a, Campagnolo and Buescu give robust continuous simulations and polynomial ODE realizations of Turing machines~\cite{dsg05,GracaCampagnoloBuescu2008}.
In our polynomial-time characterization~\cite{JournalACM2017}, the flow simulates successive machine transitions through phases of computation and copying. Numerical evaluation then approximates that flow with a prescribed accuracy~\cite{PoulyGraca2016}.
Discrete ODE schemes give characterizations through recurrences: we developed this approach for polynomial time~\cite{MFCSJournal}, and Antonelli, Durand and Kontinen study small circuit classes~\cite{AntonelliDK26}.
To our knowledge, none of these results states one field and one initialization whose execution is correct for every Euler mesh in \((0,1]\) and for the forced flow.
The present construction gives the discrete updates themselves the same logical trace as the flow. The step bound remains one while the precision exponent depends on space and the logarithm of time. Table~\ref{tab:prior} compares the guarantees, including the earlier precision-robust discrete reachability results~\cite{JournalCSL}. Section~\ref{sec:related} compares the mechanisms and hypotheses in detail.

\begin{table}[t]
\centering\footnotesize
\begin{tabularx}{\linewidth}{@{}>{\raggedright\arraybackslash}p{3.0cm}>{\raggedright\arraybackslash}X>{\raggedright\arraybackslash}p{3.1cm}>{\raggedright\arraybackslash}X@{}}
\toprule
Prior result & Computational realization & Step bound in the stated guarantee & Relation to the present guarantee\\
\midrule
Siegelmann and Sontag~\cite{SS95}
& rational saturated-linear networks
& not applicable: one update map
& discrete update semantics; no flow\\
Koiran and Moore~\cite{KoiranMoore1999}
& analytic maps
& not applicable: one update map
& discrete update semantics; no flow\\
Gra\c{c}a, Campagnolo and Buescu~\cite{dsg05,GracaCampagnoloBuescu2008}
& analytic maps and flows; polynomial ODE realizations
& none stated
& robustness of the six-dimensional generating flow; unperturbed polynomial lifting\\
Our continuous characterizations of time~\cite{JournalACM2017} and space~\cite{BGDPRiccardo2022,Icalp2024}
& continuous time and space characterizations
& no Euler step bound stated
& exact-flow guarantee; resource characterizations\\
Our discrete ODEs and robustness results~\cite{BlancBournezMFCS2023Journal,JournalCSL}
& discrete schemes; precision-robust discrete reachability
& specified discrete recurrences
& time and precision bounds; no shared flow\\
Andrade, Frongillo and Piliouras~\cite{AndradeFrongilloPiliouras2023}
& replicator universality; bounded simulation by multiplicative weights
& from the horizon
& discrete transfer for bounded computations only\\
Cardona, Miranda and Peralta-Salas~\cite{Cardona2021Turing}
& polynomial flow on a sphere
& none stated
& exact-flow guarantee\\
This paper
& compact rational polynomial realizations with time and precision bounds
& one, above the logical-scale threshold
& one field, initialization and logical trace for every admissible Euler mesh and measurable continuous forcing\\
\bottomrule
\end{tabularx}
\caption{Executions covered by the cited results, as described in Section~\ref{sec:related}. ``None stated'' means that no Euler step bound is supplied. Separate numerical solvers~\cite{PoulyGraca2016} approximate a trajectory with a prescribed error, which is a different guarantee. Here \(\eta_k\le1\) is independent of the horizon, subject to the logical-scale threshold, and drift noise \(h^{2K}\) gives update error \(\eta_kh^{2K}\).}
\label{tab:prior}
\end{table}

\myparagraph{Scope and implementation cost}
Rounding requires \(2Ks+3\) fractional bits at unit step, and \(O(s+\log(1/\eta_k))\) at step \(\eta_k\): smaller steps may require more updates and more bits, despite the common physical-time bound.
Our fixed choice \(K\ge256\) is sufficient, since it makes \(3/K\) negligible against the cut exponents of Section~\ref{sec:proof-P}. With it a cycle lasts a constant times \(Ks\) physical time, so these are resource bounds, not a practical implementation claim.
The theorem concerns the compiled family, with its clock and phase separation, and we do not claim invariance for other numerical integrators.

\myparagraph{Organization}
In Section~\ref{sec:model-results}, we define the semantics and state the complexity and prediction results.
In Section~\ref{sec:overview}, we give the compiler interface and the proof ideas.
In Section~\ref{sec:related}, we compare with prior results, and in Section~\ref{sec:applications}, we develop the verification and implementation consequences.
In Sections~\ref{sec:proof-P} and~\ref{sec:proof-PSPACE}, we prove the results on time and space, and in Section~\ref{sec:common-certificates}, we prove the results on common safety certificates. The last sections treat continuous semantics, functions, local checks and open questions. Long proofs appear at the end of the paper, with a sketch after each corresponding statement, and Appendix~\ref{sec:supplement} collects two supplementary developments on certificates.

\section{Model and main results}\label{sec:model-results}

\subsection{Preliminaries: parameters and decision semantics}
For \(x\in\N^d\), with fixed \(d\), let \(\ell(x)\) be the maximum binary length of its coordinates, with \(\ell(0)=1\). We use the sup norm for vectors, and its induced norm for matrices. All the coefficients of fields and initializations are rational, and fixed independently of the input. The thresholds on scales and the polynomial bounds on time introduced below can be chosen as integer polynomials. We will always enlarge a scale threshold \(S\) (introduced in Definition~\ref{def:decision} below) so that \(S(n)\ge n+4\): this is harmless, since correctness is required for every larger scale.

\begin{definition}[Scale-controlled field]\label{def:scale-field}
A scale-controlled polynomial system consists of rational polynomial maps
$$
 F:\R^d\times\R\times\R^m\longrightarrow\R^m,
 \qquad c:\R^d\times\R\longrightarrow\R^m.
$$
At input \(x\), scale \(h=2^{-s}\) and temporal step \(0<\eta\le1\), its active orbit is
$$
 z_0=c(x,h),\qquad z_{k+1}=z_k+\eta F(x,h,z_k).
$$
The parameters \((x,h)\) are frozen, and the physical time of index \(k\) is \(k\eta\).
\end{definition}

\begin{definition}[Stable decision]\label{def:decision}
A system decides \(L\subseteq\N^d\) with threshold \(S\) and physical horizon \(Q\) if there are an output coordinate \(o\) and a fixed compact box \(\mathcal K\) such that, for all \(x\), all integers \(s\ge S(\ell(x))\) and every required mesh, the whole active orbit lies in \(\mathcal K\), and its output coordinate \(o_k\) at index \(k\) satisfies
$$
 |o_k-\mathbf1_L(x)|\le\frac14
 \quad\hbox{for all }k\ge\left\lceil Q(s)/\eta\right\rceil.
$$
The requirement is \emph{unit-step} when \(\eta=1\), and \emph{mesh-uniform} when it holds for every \(0<\eta\le1\), with the same field, initialization, box and horizon.
\end{definition}

\begin{definition}[Robust decision]\label{def:robust}
For a precision budget \(R\), the same requirement must hold for every orbit
$$
 \widetilde z_0=c(x,h),\qquad
 \widetilde z_{k+1}=\widetilde z_k+
 \eta F(x,h,\widetilde z_k)+\eta\xi_k,
 \qquad \|\xi_k\|_\infty\le2^{-R(s)}.
$$
The initial condition is exact, and the disturbances act only on the active coordinates (not on \(x\) or \(h\)). The box, the horizon and the output coordinate are uniform over all permitted sequences of disturbances.
\end{definition}

\begin{definition}[Eventual robust decision]\label{def:eventual-robust}
A system eventually decides \(L\) robustly with threshold \(S\) and precision exponent \(r>0\) if, for every input, every \(s\ge S(\ell(x))\) and every permitted orbit with drift disturbance at most \(h^r\), the whole active orbit stays in one fixed compact box, and its output eventually stays within \(1/4\) of \(\mathbf1_L(x)\). No stabilization horizon is given or assumed computable: the eventual index may depend on the orbit. Unit-step and mesh-uniform refer to the same choices of meshes as before, and the initialization remains exact.
\end{definition}

\begin{definition}[Active length]\label{def:length}
For an exact or perturbed finite orbit, define
$$
 \Len_\eta(z;N)=\sum_{k<N}\|z_{k+1}-z_k\|_\infty.
$$
For the perturbed scheme this equals
\(\eta\sum_{k<N}\|F(x,h,z_k)+\xi_k\|_\infty\).
\end{definition}

Frozen parameters have zero increments. They do not contribute to the length. This does not mean that their representation is free in a Turing simulation: the soundness proofs explicitly keep the input and the scale, with \(O(\ell(x)+s)\) bits. Notice also that a compact active state of constant dimension still carries \(O(s)\) bits when it must be resolved at a fixed power of \(h\).

The witnesses for languages constructed below have initializations that are bounded on allowed inputs, with output coordinate zero. By Lemma~\ref{lem-raw-translation}, we can therefore replace them by the raw state \((x,h,0,\ldots,0)\) through a polynomial conjugacy, which preserves the output coordinate and the active increments.

\subsection{Polynomial time and polynomial space}

\begin{theorem}[Polynomial time]\label{thm:main-P}
For every language \(L\subseteq\N^d\), the following are equivalent.
\begin{enumerate}
\item \(L\in\Ptime\).
\item \(L\) is decided by a unit-step scale-controlled rational polynomial field with polynomial scale threshold and polynomial physical time, while all active variables remain in a fixed compact box.
\item \(L\) is decided mesh-uniformly by such a field, with the same scale threshold, physical time, compact box and output coordinate for every \(\eta\in(0,1]\).
\end{enumerate}
The construction in Item~3 has active length polynomial in \(s\) on the prefix up to the decision horizon, uniformly in \(\eta\).
\end{theorem}
\begin{proofsketch}
A polynomial-time machine can be simulated with dyadic stacks, with a polynomial scale threshold and polynomially many source transitions, and we apply Theorem~\ref{thm:quantitative-transfer} with~\eqref{eq:front-resource-condition}. Conversely, the unit-step map can be followed with polynomially many guard bits for polynomially many iterations (no tolerance to noise is needed in this direction). Taking \(\eta=1\) also proves the soundness of the mesh-uniform formulation. The detailed proof is given in Section~\ref{sec:proof-P}.
\end{proofsketch}

\begin{theorem}[Polynomial space without a supplied horizon]\label{thm:main-PSPACE}
For every language \(L\subseteq\N^d\), the following are equivalent.
\begin{enumerate}
\item \(L\in\PSPACE\).
\item A unit-step scale-controlled rational polynomial system eventually decides \(L\) robustly, with polynomial scale threshold and precision exponent \(r>0\).
\item Such a system eventually decides \(L\) robustly and mesh-uniformly, with the same field, initialization, compact box and precision exponent for every \(0<\eta\le1\).
\end{enumerate}
The equivalence is unchanged if Items~2 and~3 additionally require a common physical stabilization horizon \(2^{cs}\), for a fixed \(c>0\). The constructed witnesses meet this stronger requirement, including for variable meshes and continuous forcing.
\end{theorem}
\begin{proofsketch}
The idea is to round the unit-step map to \(O(s)\) fractional bits: we get a permitted perturbed orbit on a finite set of \(2^{O(s)}\) states. Its eventual cycle can only contain correctly labeled states, since each of them repeats arbitrarily late. Following the orbit beyond the number of states therefore decides the input in polynomial space, without knowing its original stabilization time. Conversely, the stack compiler provides a robust witness with an exponential uniform horizon. We prove both assertions in Section~\ref{sec:proof-PSPACE}.
\end{proofsketch}

In the space theorem, robustness is a hypothesis on \emph{every} sequence of perturbations of the active state: it is not a floating-point experiment. Observe also that exact rational iteration, without rounding, is not the measure of space that we propose. We then get the following picture, where \(c\) is the exponent of the bound on source transitions for polynomial space, and \(c_0>c+1\) is a fixed integer (see the proof of Theorem~\ref{thm:main-PSPACE}):

\begin{center}
\begin{tabularx}{\textwidth}{@{}l>{\raggedright\arraybackslash}X>{\raggedright\arraybackslash}X@{}}
\toprule
 & Polynomial time & Polynomial space \\
\midrule
Source transitions & \(T(s)=\operatorname{poly}(s)\) & \(T(s)\le2^{cs}\) \\
Physical horizon & \(O((1+s)T(s))\) & not assumed; witnesses have \(2^{O(s)}\) \\
Soundness requirement & polynomially many exact steps in a bounded box & eventual stability under update error \(h^r\) \\
Precision in the construction & tolerance \(h^{2K}\) & tolerance \(h^{2K}\), with \(K/4>c+2\) \\
Euler step count & \(\lceil Q(s)/\eta\rceil\) & \(\lceil2^{c_0s}/\eta\rceil\) \\
\bottomrule
\end{tabularx}
\end{center}

At unit step, the physical horizon counts updates. A restriction \(\eta\le h^a=2^{-as}\), with fixed \(a>0\), would require at least \(2^{as}\) updates per unit of physical time. Unit steps avoid this exponential factor. The bound on physical time is common to the meshes, while their switching times may differ.
One may keep \(K=256\) by increasing the scale constant \(C\) in Theorem~\ref{thm:machine-resources}: its proof uses \(T(n)+2\le2^{s/C}\).

\myparagraph{A consequence for prediction}
From the space theorem, we also get a fixed dynamical prediction problem. The field is fixed once and for all: only the encoded initial state, the precision and the observation horizon vary. We show that robustness and compactness do not remove the computational difficulty, even on instances where the discrete and the continuous interpretations share the same decision.

\begin{theoremrep}[A fixed-field robust prediction problem]\label{thm:prediction}
There exist a fixed rational polynomial field \(F(h,Z)\), a fixed rational polynomial initialization \(c(x,h)\), an integer \(r>0\), a compact rational box and an output coordinate for which the following promise problem is \(\PSPACE\)-complete under polynomial-time many-one reductions.

An instance consists of a binary integer \(x\), a unary scale exponent \(s\ge\ell(x)+4\), and a positive binary horizon \(N\). It is promised that all unit-step executions from \(c(x,2^{-s})\), under active disturbances of norm at most \(2^{-rs}\), stay in the box and, at every index at least \(N\), all give an output within \(1/4\) of the same Boolean value. Decide that value.

The hard instances can be chosen so that their decision is also stable after physical time \(N\) for every admissible variable Euler mesh and for the continuous system with any measurable forcing of norm at most \(2^{-rs}\).
\end{theoremrep}
\begin{proofsketch}
For membership, we follow one permitted rounded orbit up to the binary horizon, in polynomial space. For hardness, we compile one fixed decider for a \(\PSPACE\)-complete language: its polynomial scale threshold and its exponential horizon have descriptions of polynomial size. All the temporal interpretations agree on these instances, by the same estimates on the compiler and on the forcing.
\end{proofsketch}
\begin{proof}
For membership, we keep the frozen input and scale exact, we initialize exactly, and we round complete unit updates to \(b=rs+3\) fractional bits. This is a permitted disturbance. Each evaluation of the polynomial has a polynomial cost in bits, and the state has \(O(s)\) bits inside the fixed box. A counter up to \(N\) has \(O(\log(N+1))\) bits. Reading at index \(N\) and comparing the output with \(1/2\) therefore uses space polynomial in the length of the instance. To get an algorithm bounded in space also on inputs that violate the promise, we stop and return an arbitrary answer if the initial state, or a rounded state, leaves the box. This check is made before the next evaluation, and evaluating once from a bounded state still uses polynomially many bits. Such a stop never happens on promised instances.

Fix a deterministic polynomial-space decider for true quantified Boolean formulas. We use any fixed effective binary encoding, with a leading sentinel bit to identify words with integers, and the decider rejects malformed encodings. Theorem~\ref{thm:main-PSPACE} constructs a single normalized field \(F(h,Z)\), its rational polynomial raw-input initialization, a polynomial integer threshold \(S\), an integer exponent \(r=2K\) and an integer \(c_0\) giving the physical horizon \(2^{c_0s}\). We choose a rational box containing the uniform compact bound. For a formula encoded by \(x\), we set
\[
 s=S(\ell(x)),\qquad N=2^{c_0s}.
\]
The unary description of \(s\), the denominator \(2^s\) and the binary description of \(N\) all have polynomial length, and the map from the formula to the instance is computable in polynomial time. All these instances satisfy the promise, and their answer is exactly the truth of the formula. This proves hardness, with the same fixed field, initialization, exponent and box for every instance. Corollaries~\ref{cor:variable-mesh} and~\ref{cor:forced-ode} give the last, stronger assertion on the hard instances. Notice that we need no claim on recognizing whether arbitrary instances satisfy the promise.
\end{proof}

\myparagraph{Where the hardness comes from}
The horizon \(N\) is given in binary: with \(N\) in unary, the same promise problem is in \(\Ptime\), since the unit-step map can be followed for \(N\) steps with polynomially many guard bits (Lemma~\ref{lem-truncated-simulation}).
Kawamura proves polynomial-space completeness for initial value problems with Lipschitz fields~\cite{Kawamura2009}. Our statement is about one fixed polynomial field under a promise of robustness, with agreement between the temporal interpretations on the hard instances.
The bound \(2^{-rs}\) on disturbances decreases with the instance, and the initial condition and the frozen parameters are exact (Section~\ref{sec:related} compares with models with a fixed noise).

\subsection{Continuous semantics and the scope of the bridge}

\begin{theorem}[Same field in continuous time]\label{thm:main-continuous}
Let \(F\) be the field constructed for Theorem~\ref{thm:main-P}. For every fixed input and permitted scale, the solution of
$$
 \dot z=F(x,h,z)
$$
decides the same language after polynomial physical time. The active arc length up to the decision is polynomial in \(s\).
\end{theorem}
\begin{proofsketch}
We apply the conclusion of Theorem~\ref{thm:quantitative-transfer} on the flow to the polynomial-time normal form. Compactness is uniform in the Euler mesh, and correctness holds on a common infinite tail. We detail the passage to the limit, and the estimate on the length that comes from the bounded speed, in Section~\ref{sec:continuous-bridge}.
\end{proofsketch}

\begin{corollary}[Polynomial-length analog recognition]\label{cor:ALP}
With the standard bounded encoding of binary words, the construction gives a weak analog computation of the language indicator. The polynomial-versus-generable and separator-to-recognizer results of~\cite{JournalACM2017} compile it into a strict polynomial-length analog recognizer, possibly enlarging the field.
\end{corollary}

The claim that the field is the same concerns the Euler executions and their continuous flow. The compilation into the standard recognizer may add coordinates and replace the initialization: we do not claim that the raw-input system is the recognizer of~\cite{JournalACM2017}. The estimates of compactness also cover variable meshes (Corollary~\ref{cor:variable-mesh}); for soundness, we use the unit mesh, and we do not assume that an arbitrary external mesh is computable.

\subsection{Shared traces and common certificates}\label{sec:front-traces-certificates}
All runs share more than the final decision. An observation with values in a finite alphabet, constant on a ball of radius proportional to \(h\) around each encoded configuration, reads the same finite word at the cycle sections of every run, including the exact and forced flows of a halting run (Proposition~\ref{prop:logical-traces}).
For safety, a common certificate is a polynomial \(V\) with \(V\le-1\) on the initial set and \(V\ge1\) on the unsafe set, whose sublevel set \(\{V\le0\}\) is invariant under every Euler update of step at most one, with any drift disturbance of sup norm at most \(\delta\); the flow then stays safe as well (Definition~\ref{def:common-certificate}).
A fixed field has a family of robust safety instances, \(\PSPACE\)-complete to decide, whose safe instances all have common certificates with rational coefficients. Unless \(\mathrm{NP}=\PSPACE\), no polynomial bounds their bit length, and unless \(\Ptime=\PSPACE\), no constant bounds their degree (Theorem~\ref{thm:common-certificates}). Section~\ref{sec:applications} develops these consequences.

\section{Proof architecture}\label{sec:overview}

\subsection{The compiler interface}

\begin{definition}[Separated affine program]\label{def:affine-program}
A separated affine program has finitely many rational affine tests \(u_j(Y,h)\) and branches
\(\mathcal H_\sigma(Y)=D_\sigma Y+b_\sigma\), indexed by their complete sign patterns. At scale \(h\), a set \(\mathcal C_h\subseteq[0,1]^m\) specifies its valid configurations. A valid prefix follows the selected branches and has all its configurations in \(\mathcal C_h\). The matrices and vectors are independent of \(h\). There are constants \(r_0,\kappa>0\) such that, for every valid configuration \(C\), all tests satisfy
\begin{equation}\label{eq:interface-margin}
 \|Y-C\|_\infty\le r_0h
 \ \Longrightarrow\
 \operatorname{sign}u_j(Y,h)=\operatorname{sign}u_j(C,h),
 \quad |u_j(Y,h)|\ge\kappa h.
\end{equation}
Tests are normalized to have modulus at most one on \([-2,3]^m\), for \(0<h\le1/16\). The condition also applies to unused tests. When decision semantics are used, a designated coordinate is Boolean in the halting configurations. A \emph{resetting halt} enters a valid configuration \(C_{\mathrm{halt}}\) by a constant branch, and the branch selected at \(C_{\mathrm{halt}}\) is also the constant map \(Y\mapsto C_{\mathrm{halt}}\).
\end{definition}

For a valid prefix \(C_0,\ldots,C_T\), an \emph{interval gain bound} is a number \(\Gamma\ge1\) such that
\begin{equation}\label{eq:front-interval-gain}
 \|D_{\sigma_{j-1}}\cdots D_{\sigma_i}\|_\infty\le\Gamma
 \qquad(0\le i<j\le T).
\end{equation}
Such a bound controls an error inserted at any intermediate transition (it does not need to bound products along invalid executions). Notice that the compiler below receives neither \(T\) nor \(\Gamma\). Let us first describe the compiled field.

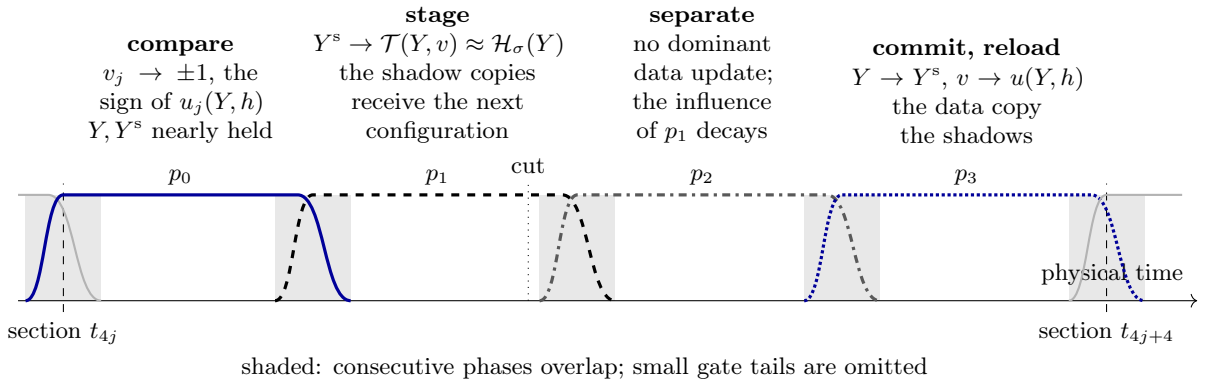
\begin{figure}[!b]
\centering
\begin{tikzpicture}[x=1cm,y=1cm,
 g0/.style={draw=blue!60!black,very thick},
 g1/.style={draw=black,very thick,dashed},
 g2/.style={draw=gray!70!black,very thick,dash dot},
 g3/.style={draw=blue!60!black,very thick,densely dotted},
 tail/.style={draw=gray!60,thick},
 lbl/.style={align=center,font=\footnotesize,text width=3.3cm,anchor=south}]
\foreach \a/\b in {0.1/1.1,3.4/4.4,6.9/7.9,10.4/11.4,13.9/14.9}
 \fill[gray!18] (\a,0) rectangle (\b,1.4);
\draw[->] (0,0) -- (15.6,0);
\node[font=\footnotesize,anchor=south east] at (15.55,0.08) {physical time};
\draw[dashed] (0.6,-0.15) -- (0.6,1.55);
\draw[dashed] (14.4,-0.15) -- (14.4,1.55);
\node[font=\footnotesize,anchor=north] at (0.6,-0.15) {section \(t_{4j}\)};
\node[font=\footnotesize,anchor=north] at (14.4,-0.15) {section \(t_{4j+4}\)};
\draw[dotted] (6.75,0) -- (6.75,1.55);
\node[font=\footnotesize,anchor=south] at (6.75,1.55) {cut};
\draw[tail] (0,1.4) -- (0.4,1.4) .. controls (0.75,1.4) and (0.75,0) .. (1.1,0);
\draw[g0] (0.1,0) .. controls (0.35,0) and (0.35,1.4) .. (0.6,1.4) -- (3.7,1.4) .. controls (4.05,1.4) and (4.05,0) .. (4.4,0);
\draw[g1] (3.4,0) .. controls (3.65,0) and (3.65,1.4) .. (3.9,1.4) -- (7.2,1.4) .. controls (7.55,1.4) and (7.55,0) .. (7.9,0);
\draw[g2] (6.9,0) .. controls (7.15,0) and (7.15,1.4) .. (7.4,1.4) -- (10.7,1.4) .. controls (11.05,1.4) and (11.05,0) .. (11.4,0);
\draw[g3] (10.4,0) .. controls (10.65,0) and (10.65,1.4) .. (10.9,1.4) -- (14.2,1.4) .. controls (14.55,1.4) and (14.55,0) .. (14.9,0);
\draw[tail] (13.9,0) .. controls (14.15,0) and (14.15,1.4) .. (14.4,1.4) -- (15.4,1.4);
\node[font=\footnotesize,anchor=south] at (2.15,1.4) {\(p_0\)};
\node[font=\footnotesize,anchor=south] at (5.55,1.4) {\(p_1\)};
\node[font=\footnotesize,anchor=south] at (9.05,1.4) {\(p_2\)};
\node[font=\footnotesize,anchor=south] at (12.55,1.4) {\(p_3\)};
\node[lbl] at (2.15,1.95) {\textbf{compare}\\\(v_j\to\pm1\), the sign of \(u_j(Y,h)\)\\\(Y,Y^{\mathrm s}\) nearly held};
\node[lbl] at (5.55,1.95) {\textbf{stage}\\\(Y^{\mathrm s}\to\mathcal T(Y,v)\approx\mathcal H_\sigma(Y)\)\\the shadow copies receive the next configuration};
\node[lbl] at (9.05,1.95) {\textbf{separate}\\no dominant data update;\\the influence of \(p_1\) decays};
\node[lbl] at (12.55,1.95) {\textbf{commit, reload}\\\(Y\to Y^{\mathrm s}\), \(v\to u(Y,h)\)\\the data copy the shadows};
\node[font=\footnotesize,anchor=north] at (7.5,-0.62) {shaded: consecutive phases overlap; small gate tails are omitted};
\end{tikzpicture}
\caption{One cycle of the autonomous clock, at unit step and without disturbance (schematic profiles, not numerical trajectories). Each phase lasts \(\Theta(\log(1/\eps))\) and gates the operation written above it: \(p_1\) weights the loading of the shadows, \(p_3\) the copy of the data and the tracking of the comparators, \(1-p_3\) the amplification of the comparators, and \(\mathcal T(Y,v)\) is the affine target selected by the comparators (Section~\ref{sec:polynomial-field}). The dotted line is the cut of Lemma~\ref{lem:cut-separation}.}
\label{fig:cycle}
\end{figure}

The compiled field, given explicitly in Section~\ref{sec:polynomial-field}, has an active state \(Z\) made of the data registers \(Y\), which hold the current configuration, of shadow registers \(Y^{\mathrm s}\) of the same dimension, which prepare the next configuration, of one comparator \(v_j\) for each test \(u_j\), and of a clock with four phases. One cycle of the clock simulates one transition of the program (Figure~\ref{fig:cycle}), and we write \(t_{4j}\) for the time at which cycle \(j\) starts: these are the internal cycle sections, at which we observe the data. A cycle is admissible when, at its section, the data are close to a source configuration, the shadows are close to the data, the clock is in its entry region, and each comparator has the sign of its test, with a magnitude \(\Omega(h)\) (this is the comparator-cone condition, see~\eqref{eq:cycle-admissible}).

\begin{theoremrep}[Quantitative compilation under temporal refinement]\label{thm:quantitative-transfer}
Fix a separated affine program \(M\) and an integer \(K\ge256\). There are a rational polynomial field \(F_{M,K}(h,Z)\), a rational polynomial initialization \(c_{M,K}(h,C_0)\), a fixed compact box and positive constants \(A,B,r_*,c_-,c_+,s_0\) with the following property.

Let \(h=2^{-s}\), \(s\ge s_0\), \(\eps=h^K\), \(0\le\nu\le\eps^2\), and set
\begin{equation}\label{eq:transfer-budget}
 \Delta_\eps=(1+\log(1/\eps))(\eps^{1/4}+\nu).
\end{equation}
For a valid prefix of \(T\ge1\) source transitions with interval gain bound \(\Gamma\), suppose
\begin{equation}\label{eq:transfer-condition}
 B\Gamma T\Delta_\eps\le r_*h.
\end{equation}
Starting exactly at the prescribed initialization, every trajectory
\[
 Z_{k+1}=Z_k+\eta F_{M,K}(h,Z_k)+\eta\xi_k,
 \qquad \|\xi_k\|_\infty\le\nu,\quad 0<\eta\le1,
\]
has internal cycle sections \(t_{4j}\), \(0\le j\le T\), with
\begin{align}
 \|Y(t_{4j})-C_j\|_\infty&\le B\Gamma j\Delta_\eps,
 \label{eq:transfer-error}\\
 c_-(1+s)\le t_{4j+4}-t_{4j}&\le c_+(1+s)\qquad(j<T).
 \label{eq:transfer-time}
\end{align}
At each section the shadow discrepancy \(\|Y^{\mathrm s}-Y\|_\infty\) is at most \(A\Delta_\eps\), and the clock and comparator-cone conditions are restored. The prefix stays in the fixed box and has active length \(O((1+s)T)\). The same conclusions hold for variable steps \(0<\eta_k\le1\) with \(\sum_k\eta_k=\infty\), using physical time \(t_k=\sum_{i<k}\eta_i\) and update noise \(\eta_k\xi_k\).

If the last transition resets to a fixed halting configuration, every subsequent cycle is admissible, and the data stay \(O(\Delta_\eps)\)-close to this configuration at every subsequent update (or micro-step). The exact polynomial flow gives the same stable decision, with the same bound on physical time. All constants are independent of the prefix, the input, the mesh and the sequence of disturbances.
\end{theoremrep}
\begin{proofsketch}
The autonomous clock separates the comparison, the loading of the shadows and the copy, and we control their overlaps by integrated bounds on the gates. The local cycle returns the full invariant of sections, with a defect \(B\Delta_\eps\). By variation of constants, these defects are propagated by interval products, which gives \(B\Gamma j\Delta_\eps\), and the budget keeps every section in the correct tube for the tests. The resetting halt removes any further accumulation, and we get the interpretation by the flow from the convergence of Euler schemes, by compactness.
\end{proofsketch}
\begin{proof}
We use the field of Section~\ref{sec:proof-P}, with the tests and the branches of the separated affine program. Its construction uses no operation specific to stacks: the selector is the polynomial interpolation of the finitely many affine branches. The arguments on the box and on the comparators use bounded normalized tests, and Lemma~\ref{lem-uniform-cycle} uses exactly the margin condition, bounded affine targets on a fixed neighborhood and a valid next configuration. This lemma therefore applies with the constants \(r_0,\kappa\) of Definition~\ref{def:affine-program}, and its own constants may depend on the fixed program and on \(K\).

Here \(r_*=r_0/(4D_*)\), where \(D_*=\max(1,\max_\sigma\|D_\sigma\|_\infty)\). Lemma~\ref{lem-P-accumulation}, in its general form with an interval gain, proves the error bound, and it gives the hypothesis on the ideal image at every cycle. The initialization is exact on the source data, with identical shadows, and with comparators initialized at their tests. The bounds on the clock give the two estimates on durations, because \(\log(1/\eps)=Ks\log2\). The field \(F_{M,K}(h,Z)\) has a uniformly bounded speed on the common compact set, so the actual perturbed increments give the stated length, as in Proposition~\ref{prop-P-length}.

The estimates on scalars and on locks hold one step at a time, and the bounds on weighted grids of Lemma~\ref{lem:weighted-grid} already allow variable steps: Corollary~\ref{cor:variable-mesh} gives the same induction on such a grid. For a resetting halt, Proposition~\ref{prop-contracting-halt} closes every later cycle without any accumulated past error. Consequently, the exact trajectories with constant mesh stay compact on the whole half-line, and Lemma~\ref{lem:euler-ode} gives the unique global flow. The closed bounds of the stable decision pass to its limit on each finite interval. Observe that we need no estimate of the distance between coarse trajectories and this flow.
\end{proof}

\myparagraph{The bounded-stack instance}
The source machines in Lemma~\ref{lem-hard-normal-form} represent a stack by
\[
 z_{\mathrm{stk}}=\sum_{i=1}^{H}d_i4^{-i},\qquad d_i\in\{1,2\},
 \qquad H\le\left\lfloor\frac{s-2}{2}\right\rfloor.
\]
In other words, a stack word \(d_1\cdots d_H\) is coded in base four. Push and pop are \((z_{\mathrm{stk}}+d)/4\) and \(4z_{\mathrm{stk}}-d\). The tests, the readers and the control of these machines satisfy Definition~\ref{def:affine-program}, and Lemma~\ref{lem-cumulative-gain} gives \(\Gamma=G/h\), for a constant \(G\). Hence the error of the compiler is \(GBj\Delta_\eps/h\), and its condition of validity is
\begin{equation}\label{eq:front-resource-condition}
 GB T\Delta_\eps/h^2
 \le 2GB(1+Ks\log2)T\,2^{-(K/4-2)s}\le r_*.
\end{equation}
This budget covers both polynomial-time and exponential-time simulations. The exponent \(K\) is fixed in the compiled field: it can be chosen once a bound \(T(s)\le2^{cs}\) is known, with \(K/4>c+2\).

The threshold on resources in Theorem~\ref{thm:machine-resources} makes this budget valid for the complete compiled computation, including the loading of the input and the resetting halt.

\myparagraph{A running example: a stack parser}\label{sec:running-example}
Consider balanced words over two types of parentheses. A program with finite control reads one symbol at a time: it pushes digit \(1\) for an opening round parenthesis and digit \(2\) for an opening square bracket, it pops on a closing symbol when the top digit matches, and it rejects otherwise. It accepts at the end of the input exactly when the stack is empty. Each symbol is encoded by two bits, after a sentinel bit \(1\), and the word is given as a raw integer; a second stack holds the unread word, and a buffer in the finite control assembles the pairs of bits. This is a fixed two-stack program with \(T(n)=O(n)\) transitions and height \(H(n)=O(n)\), for an input of binary length \(n\ge1\). For the word \(\texttt{([])}\), the work stack evolves as follows (the bounded number of control transitions per symbol is omitted):
\begin{center}
\begin{tabular}{clcc}
\toprule
Symbol & Logical action & Stack before & Stack after\\
\midrule
\(\texttt{(}\) & push \(1\) & \(0\) & \(1/4\)\\
\(\texttt{[}\) & push \(2\) & \(1/4\) & \(9/16\)\\
\(\texttt{]}\) & pop \(2\) & \(9/16\) & \(1/4\)\\
\(\texttt{)}\) & pop \(1\) & \(1/4\) & \(0\)\\
\bottomrule
\end{tabular}
\end{center}
The word \(\texttt{([)]}\) is rejected at its third symbol, which requests digit \(1\) while the top region is that of digit \(2\). These tests have constant margins, whereas the reader of the raw input has margins proportional to \(h\). At a scale \(s=O(n)\), Theorem~\ref{thm:machine-resources} gives \(O(n^2)\) physical time and \(O(n)\) fractional bits at unit step, and the same parser field works at every finer admissible mesh and under the permitted measurable continuous forcing, above the scale threshold of the theorem.

\subsection{Why the step can stay fixed: the two estimates}
We prove the logical execution directly, at every allowed step. Two estimates make this possible: a control of the integrated influence of the overlapping gates inside each cycle, which gives a local defect \(B\Delta_\eps\) uniformly in the mesh, and a bound on the products of transitions over every interval of the run, which propagates these defects without multiplying worst-case gains.

\myparagraph{Integrated influence of the gates}
The invariant of a cycle involves the data registers, their shadow copies \(Y^{\mathrm s}\), the signs of the comparators and the clock: each cycle loads the next configuration into the shadows, and then copies it into the data (Figure~\ref{fig:cycle}). Even a contractive update can copy an incorrect target if its gate opens too early. A bound on a gate at the end of a phase is not enough: an earlier leak may already have moved the target of another operation. In Lemma~\ref{lem:cut-separation}, we choose a cut inside each phase. Before this cut, the next gate has integrated mass \(O(\Delta_\eps)\), and after it, the preceding gate has integrated mass \(O(\Delta_\eps)\). The opposite gate has a small mass during the whole phase. Hence the shadows remain close to their previous value until the tests are reliable, and the data remain close to their old value until the new shadows are reliable. The estimate on double overlaps follows from this separation.

\myparagraph{The invariant is on the full state}
At a cycle section, the data must be close to a valid source configuration, the shadows must be close to the data, the clock must be in its entry region, and every comparator must have the correct sign and a magnitude \(\Omega(h)\). The local proof gives
$$
 \|Y^+-\mathcal H_\sigma(Y)\|\le C(e_0+\Delta_\eps),
 \qquad
 \|(Y^{\mathrm s})^+-Y^+\|\le C_{\mathrm{gap}}\Delta_\eps,
$$
where \(Y^+\) and \((Y^{\mathrm s})^+\) are the values at the next section, and \(e_0\) is the incoming discrepancy between shadows and data. We get the second constant from an estimate on bounded states, independently of the coefficient used to bound \(e_0\). Choosing the section constant first, and the defect constant second, closes the invariant. We do need two distinct constants: we do not assume a global contraction of the whole vector of data, and this would be false for a pop.

\myparagraph{Interval products, and the global argument}
For long computations, we use bounds on the products of transition matrices over \emph{intervals} of transitions. The normal form puts the tape on two bounded stacks. The tests on the top symbol have a constant gap, whereas the reader of the raw input has a gap proportional to \(h\). Once the pattern of the tests is fixed, the affine transition has separate linear blocks for each stack, each reader and the finite control. These blocks interact only through the choice of this pattern. A pop multiplies an error on a stack by four, and a push divides it by four, so that on a single stack, the coefficient of an interval of updates is
$$
 4^{\#\mathrm{pop}-\#\mathrm{push}}
 =4^{H_{\mathrm{start}}-H_{\mathrm{end}}}\le1/h.
$$
These products telescope with the height of the stack. The matrices of the reader also have bounded products, and the control blocks that reset contribute zero. This gives \(\Gamma=G/h\) in~\eqref{eq:front-interval-gain}, a property of valid operations on stacks, and not of arbitrary bounded affine updates. By variation of constants, the error at a section is a sum of local defects, propagated by these products. We get \(GBj\Delta_\eps/h\), instead of \(D_*^j\Delta_\eps\), where \(D_*=\max(1,\max_\sigma\|D_\sigma\|_\infty)\): multiplying a worst-case expansion at every transition would give a loss exponential in the number of transitions. A smaller tube for the induction gives the hypothesis on the image in the local cycle lemma, and it keeps every actual test on the intended branch. The same induction applies before the halt, for every mesh. At a resetting halt, the ideal affine image is constant: the incoming error on the data is discarded, and there is no further accumulation.

\myparagraph{Scalar invariance at unit step}
For a fixed disturbance \(\xi\), an Euler update can be written
\[
 Z+\eta(F(Z)+\xi)=(1-\eta)Z+\eta\bigl(Z+F(Z)+\xi\bigr).
\]
In other words, a convex set preserved by every allowed unit-step image is preserved at every smaller step.
This is the principle of convexity behind the forward-Euler condition in strong-stability-preserving methods~\cite{GottliebShuTadmor2001}.
Lemma~\ref{lem-invariant-boxes} checks the coupled scalar box, including the gates that perturbations make slightly negative or slightly larger than one.
Loading the data also needs the first-exit argument of the cycle, and invariance alone does not give the order or the duration of the clock events.

The normalized field and its derivative are uniformly bounded on a fixed compact neighborhood, since they are fixed polynomials in \((h,Z)\).
The construction resolves smaller margins by allowing longer cycles, and no coefficient grows with the requested precision.
We then prove the progress of the clock and the accuracy of copies from inequalities on one step.

\myparagraph{A threshold is computed, not supplied}
The scalar polynomial field \(\dot v=(v-v^3)/2\) amplifies an input of magnitude \(\Omega(h)\) toward the corresponding attractor (\(-1\) or \(+1\)), and its Euler update preserves \([-1,1]\) for every step at most one. Growth to a constant magnitude takes \(O(\log(1/h))\) physical time, and the subsequent contraction makes the error of the selector small enough for the current scale. During the copy phase, the comparator is coupled to a tracking term, so that it enters the next cycle with the \emph{new} sign, even when a source transition changes this sign.

\myparagraph{The clock provides its own schedule}
Four phase bits \(p_0,\ldots,p_3\) and four timers \(\tau_0,\ldots,\tau_3\) form an autonomous polynomial subsystem. Outside short switching intervals, one phase bit is close to one and the others are close to zero: this active bit plays the role of a token, passed from one phase to the next. A timer at rest is attracted to \(\eps=h^K\), and an active timer grows until it enables the next phase. One full cycle executes the test, the loading of the shadow copies, the separation and the copy back into the data (Figure~\ref{fig:cycle}). The lemmas on the clock (whose proofs are rather technical) show that the token does not disappear, that every phase has duration \(\Theta(\log(1/\eps))\), and that the same conclusions hold under the allowed disturbances. We insert no external time signal, no sine function and no exact piecewise gate into the field.

\myparagraph{Scope of the invariance theorem}
Theorem~\ref{thm:quantitative-transfer} is uniform over a class of source programs, but it concerns the outputs of our compiler, with its clock, its comparators, its shadows and its separation of phases: the estimates on locks are generic, the lemmas on the clock and on cycles are specific to this architecture, and exponentially many sections require the budget~\eqref{eq:transfer-condition}. Generic Euler error bounds do not give the result: on a compact set where the field and its derivative are bounded, an Euler interpolant has a drift defect at most \(C\eta\), and certifying it against a tolerance \(h^r\) by this estimate needs \(C\eta\le h^r\), a step that shrinks with the scale (a sufficient condition, not a lower bound for every discretization). Relaxation toward an invariant box does not give it either. For \(y'=4y(1-y)-y\) from \(y=1/2\), every Euler step in \([0,1]\) preserves \([0,1]\), where the field relaxes at rate one toward the polynomial target \(4y(1-y)\), but the unit-step orbit is \(1/2,1,0,0,\ldots\), whereas the flow \(y'=y(3-4y)\) increases to \(3/4\). An independent autonomous clock cannot change these trajectories of the data, whose discrete sections after the second update are zero while every continuous section is at least \(1/2\), even without disturbances. The compiled field has precisely the structure that is missing in this example. Notice also that the numerical states at sections are not identical: they approximate the same ideal configurations.

\myparagraph{The parser through the construction}
Each logical action of the parser is one cycle of Figure~\ref{fig:cycle}: the comparators resolve the test on the top digit, the shadows receive the pushed or popped stack, and the data copy them. The invariant of the source program is simple: after any nonrejecting prefix, the work stack contains exactly the unmatched openings, the most recent first, so that the stack is empty at the end exactly for the Dyck words with two types of parentheses. During copying, the data interpolate between the old and the new values of the stack, and the discrete observation of the parser is made at the internal sections: this distinction matters for the verification statements of Section~\ref{sec:applications}. The example gives a complete argument of correctness at the source level and a symbolic compiled trace; no numerical trajectory is used as evidence.

\section{Related work and comparison}\label{sec:related}\label{sec:not-corollary}

The closest works deal with three different questions: realizing computations with a dynamical model, preserving properties under approximation and bounding computations by the available precision. Each of these questions has been studied separately: our contribution is a quantitative construction that answers them simultaneously, for the Euler family that we consider. In what follows, we describe the guarantees stated in previous works (see also our survey of analog models~\cite{bournez2021survey}).

\myparagraph{Computational universality and the discretization obstruction}
Siegelmann and Sontag simulate multi-stack Turing machines in real time with finite recurrent networks that have rational weights and a saturated-linear activation~\cite[Theorem~1]{SS95}.
Koiran and Moore use analytic maps built from trigonometric terms, with a real-time simulation in dimension two or more~\cite{KoiranMoore1999}. Earlier, Moore's generalized shifts embed Turing machines in smooth dynamical systems~\cite{Moo90,Moo91}, and Koiran, Cosnard and Garzon simulate them by piecewise-linear maps in low dimension~\cite{KCG94}.
These classes of update maps are different from the globally polynomial maps that we require here. Their discrete simulations do not state a common guarantee for the same field under Euler refinement and in continuous time.

Andrade, Frongillo and Piliouras prove that replicator dynamics are Turing complete~\cite{AndradeFrongilloPiliouras2023}. In their full version (Section~5 and Theorem~2), they transfer bounded computations to multiplicative weights with a suitably chosen step, and controlling the error for unbounded computations remains a limitation of their argument. Our compiler gives an upper bound on the step that is independent of the horizon, but for a different field, and a longer horizon requires a finer logical scale. We do not prove the same property for multiplicative weights, nor for a prescribed game.

Branicky gives universality constructions for hybrid and smooth continuous systems~\cite{Branicky1995}.
Gra\c{c}a, Campagnolo and Buescu give robust simulations by analytic maps and flows~\cite{dsg05}, and realizations by polynomial ODEs~\cite{GracaCampagnoloBuescu2008}.
In the latter paper, Theorem~3 allows bounded perturbations of a generating field whose components are PIVP functions, and Theorem~4 realizes its unperturbed solution by a larger polynomial system, which gives Theorem~2.\footnote{These numbers refer to the \href{https://sqigmath.tecnico.ulisboa.pt/pub/GracaDS/06-GCB-Poly.pdf}{author manuscript}, whose Theorems~2--4 distinguish the polynomial realization, the robust generating flow and the lifting.} This lifting does not by itself assert robustness to independent perturbations of every auxiliary coordinate: such a guarantee would require additional estimates for the lifted system. Their guarantee, with a fixed noise, and ours, with a noise that depends on the scale and acts on all active coordinates, therefore have different domains.
Polynomial universality on compact spaces is also known: Cardona, Miranda and Peralta-Salas construct a Turing-complete polynomial field on a sphere and realize universal computation in incompressible Euler dynamics on a compact Riemannian manifold~\cite{Cardona2021Turing}.
This gives a fluid setting for computation. Preserving such a realization under numerical execution is a further question: our theorem concerns the constructed polynomial family, not a discretization of the fluid equations.
These results on flows do not state the simultaneous guarantee for the Euler family that we prove here.
In our construction, the estimates on the clock and the bounds on products of transitions control the compiled logical execution in the whole family.

Compilation into polynomial dynamics also appears in molecular programming. We proved the strong Turing completeness of continuous chemical reaction networks and compiled mixed analog-digital programs into them~\cite{CMSB17vantardise}. Vasi{\'c}, Soloveichik and Khurshid compile imperative programs into deterministic mass-action chemical reactions with CRN++~\cite{Vasi2020CRN}, and provide methods and tools for analyzing approximation errors.
Their clock species schedule operations whose reactions remain simultaneously active. Our gate estimates address the corresponding difficulty of bounding the total influence of an operation outside its intended phase, here uniformly over Euler meshes and the flow.
Chalk, Kornerup, Reeves and Soloveichik study composable CRNs whose computed function is independent of the reaction rate law~\cite{ChalkKornerupReevesSoloveichik2021}. This allows the individual reaction rates to vary. Our parameter \(\eta\) changes the time step applied to the whole fixed field, with a separate budget for drift errors. A chemical realization preserving this budget and the common traces would require a further construction.

\myparagraph{Numerical stability and invariant sets}
Stability under discretization is a very classical subject. Dahlquist studies absolute stability and error bounds over long times for multistep methods~\cite{Dahlquist1963}. Numerical schemes can also have spurious solutions, even when the method is consistent: Stuart and Humphries treat these phenomena, together with the preservation of qualitative dynamics~\cite[Chapters~5--7]{StuartHumphries1996}. For computations, one must also say which observations have to be preserved along a run.

For invariant sets, the hypotheses and the preserved objects are different from those of our compiler. Kloeden and Lorenz get nearby attracting sets from a uniformly asymptotically stable compact set, for consistent one-step discretizations with sufficiently small steps~\cite{KloedenLorenz1986}. Beyn proves the persistence of an invariant closed curve near a hyperbolic periodic orbit, for sufficiently small steps, with convergence at the order of the method and inherited stability~\cite{Beyn1987}. These results are about invariant sets of a given differential equation. In our case, we construct a field from a program, and we control its successive labeled configurations during the whole computation. We do not deduce this from the persistence of attractors, and we do not claim that the sufficient bounds on the step in these works necessarily tend to zero with our scale.

Strong stability-preserving methods transfer a stability property of forward Euler to suitable higher-order methods, by convex combinations of Euler stages, under a restriction on the step~\cite{GottliebShuTadmor2001}. Our estimates on the scalar box and on locks also use convexity. However, the compiler also requires growth away from zero, ordered events of the clock and a small integrated leakage between incompatible operations, and preserving a norm does not by itself preserve these temporal properties. Extending our theorem to Runge--Kutta methods would require such additional estimates. The dependence on the solver observed in learned models by Ott, Katiyar, Hennig and Tiemann~\cite{ott2020neural} motivates the question, but our construction neither trains a model nor certifies a prescribed neural architecture.

\myparagraph{Digitization and timed verification}
Henzinger, Manna and Pnueli give conditions under which verification over dense time reduces to verification with integer observations~\cite{HenzingerMannaPnueli1992}: digitization already provides a semantic link between discrete and continuous time. For timed automata, the reduction uses the closure of the implementation under digitization, and the closure of the specification under inverse digitization. Ouaknine and Worrell study these conditions, and they show that a robust timed semantics has quite different properties of decidability~\cite{OuaknineW03}.

In our setting, the clock is itself part of a polynomial differential equation. The perturbations act on the active state. We prove that the Euler family and the flow share the same word of observations at the internal sections. The times of the sections may differ between executions, so arbitrary timed specifications are not preserved. Unlike a digitization, the unit-step trace is not a rounding of the dense trace: the times of the sections of two executions may drift apart linearly in the number of cycles. For safety between sections, we need the additional geometric condition of Corollary~\ref{cor:segment-safety}. In other words, digitization and our theorem impose different hypotheses, and they preserve different observations.

\myparagraph{Time and space characterizations}
We proved in~\cite{JournalACM2017} that polynomial time is characterized by solutions of polynomial ODEs of polynomial length: the length controls the effect of time rescaling. The normalized field constructed here has a uniformly bounded speed, so its physical horizon bounds the length of the prefix. What is new is the correctness of the same field at step one and at every smaller step. The bridge to the standard recognizer uses the results of~\cite{JournalACM2017} on initializations and on the computation of functions, and it may enlarge the field.

Let us compare more precisely with~\cite{JournalACM2017}, both for the mechanism of simulation and for its proved semantics.
In~\cite{JournalACM2017}, we use a sample-and-hold mechanism, which depends on the time and on the requested accuracy. Numerical evaluation uses solvers for polynomial ODEs~\cite{PoulyGraca2016}.
We do not reset the whole error on the configuration after each step in that construction: in~\cite[Theorem~5.15]{JournalACM2017}, we bound the outgoing error by \(k\) times the incoming error, where \(k\) is the base of the encoding. We restore the finite control, while the error on the encoded tapes propagates through the step.
Our construction also stabilizes its clock and its comparators, and it contracts the errors of copies.
The error on the data is controlled through products of affine transitions over intervals, with a bound depending on the height of the stacks.
Together with the integrated estimates on gates, this gives correctness for all steps up to one, without comparing a coarse Euler trajectory to the exact flow.
In both constructions, we choose the precision from the horizon. In our proof of~\cite[Theorem~6.5, Section~9.1.2]{JournalACM2017}, the internal accuracy of sampling depends on the requested number of iterations, but it stays constant during the run. Time rescaling makes the duration of a cycle constant there, with length as the measure of cost. Here, \(s\) lengthens the cycles to \(\Theta(s)\), at bounded speed. Agreement at unit step also uses our estimates on the clock and on errors: it does not follow from the bounded speed alone. Of course, this compares the two constructions, and it is not an obstruction to other time rescalings.

For polynomial space, we obtained continuous characterizations in~\cite{BGDPRiccardo2022,Icalp2024}. In~\cite[Theorems~44--46]{JournalCSL}, we relate polynomial robustness to precision to \(\PSPACE\) for discrete-time reachability on compact domains. These results precede Theorem~\ref{thm:main-PSPACE} at unit step. Here, every perturbed execution must eventually keep the correct answer, but its stabilization time may depend on that execution. Soundness assumes no common horizon. Completeness gives a compact rational polynomial witness with a common horizon, correct for every Euler step in \((0,1]\) and for the forced flow.

We characterized polynomial time with linear length-ODE schemes in~\cite{MFCSJournal}, and we obtained analytic characterizations with discrete ODEs without sign, using \(\tanh\), composition and length-ODE schemes, in~\cite{BlancBournezMFCS2023Journal}. Avoiding a sign primitive has thus already been done. Here, the field is strictly polynomial, the scheduling is internal, and there is no external closure under composition. Our construction for primitive recursion in~\cite{MFCS2026bournez} provides polynomial iteration with a small parameter and raw integer inputs. On its diagonal \(\eta=h\), a physical duration \(T\) costs \(T/h\) updates, and this identification alone does not give the polynomial bound at unit step that we prove here.

\myparagraph{Noise, finite memory and space}
Maass and Orponen show that, in their model of noise, a fixed analog noise can reduce analog computation in discrete time to the power of finite automata~\cite{MO97}. Our completeness for polynomial space is consistent with this obstruction: the radius of the noise is \(h^r=2^{-rs}\), which decreases with the given scale, and not a positive floor of noise independent of the input. The initial condition and the frozen parameters are exact. We do not claim a physical universality with fixed precision.

Braverman, Schneider and Rojas formulate the space-bounded Church--Turing thesis, and they prove space bounds for the asymptotic behavior of classes of noisy systems~\cite{braverman2015space}. Their stochastic setting and their questions on invariant measures are different from our requirement that every adversarially perturbed orbit eventually give the same Boolean answer. However, the principle that finite memory bounds computation, even with unlimited time, is not new: it appears for perturbed Turing machines and hybrid systems~\cite{asarin01perturbed}, for hybrid automata under noise~\cite{Fraenzle1999}, and in our work on perturbed dynamical systems~\cite{DBLP:journals/jcss/BournezGH13}. Proposition~\ref{prop:eventual-PSPACE} gives a direct argument on a finite grid for our semantics of decision: we do not claim a new general principle relating precision and space. Its purpose is to remove from the characterization the assumption that a horizon is given.

\myparagraph{Approximate simulation and formal verification}
Platzer proves that differential dynamic logic is complete relative to its continuous fragment and relative to its discrete fragment~\cite{platzer2012complete}. These results give constructive reductions between proof problems for hybrid, continuous and discrete systems.
We construct a field and initialization whose Euler executions and flow share a word of logical observations, with explicit time and precision bounds. The correspondence thus concerns executions of the same field. For the robust family of Theorem~\ref{thm:common-certificates}, one polynomial certificate also proves safety for the whole Euler family and the flow.

Girard and Pappas develop approximate language inclusion, simulation and bisimulation for discrete and continuous systems, with applications to safety verification~\cite{girard2007approximation}. Pola, Girard and Tabuada construct approximately bisimilar symbolic models for nonlinear control systems, under hypotheses of incremental stability~\cite{PolaGirardTabuada2008}. Our direction is different: a discrete program is given, and we construct a field that realizes it. Proposition~\ref{prop:logical-traces} preserves a finite word of observations at the internal sections, under an explicit error budget. This is neither a theorem of approximate bisimulation for arbitrary plants, nor a procedure for the synthesis of controllers. Corollary~\ref{cor:segment-safety} adds the geometric hypothesis needed for safety between observations.

Gao, Avigad and Clarke get decidability by allowing a prescribed relaxation of bounded formulas over computable real functions~\cite{gao2012delta}: their \(\delta\)-decision problem permits an overlap between outcomes near a logical boundary. Our prediction problem instead promises one Boolean answer, shared by all trajectories under a specified perturbation of the state. Its \(\PSPACE\)-completeness is about a fixed polynomial field, with this promise and with temporal agreement: we do not claim that robust verification over the reals, or \(\PSPACE\) bounds for problems on ODEs, are new.

\myparagraph{Converse certificates and limits of certificate search}
Prajna and Rantzer formulate convex programs for temporal verification, and they prove a converse safety theorem~\cite{PrajnaRantzer2007}. In continuous time, Theorem~16 of Liu gives a robust \(C^1\) barrier at every strictly smaller level of disturbance (in its compact case, Assumption~15 requires the closure of the reachable set to be disjoint from the closure of the unsafe set). His Theorem~31 gives a converse in discrete time, for one fixed update map with additive disturbance~\cite{Liu2022Converse}. Reserves of robustness and smooth converse barriers are thus known ingredients. On a compact domain, with strict separation and a strict derivative on the boundary, one gets a rational polynomial by \(C^1\) approximation, as detailed in our proof: rationality alone is not our contribution.

Lemma~\ref{lem:common-converse} assumes a compact set invariant under \emph{all} maps \(z\mapsto z+\eta(f(z)+u)\), for \(0\le\eta\le1\) and \(\|u\|\le2\delta\). This is stronger than robust safety in continuous time, as shown by Example~\ref{ex:flow-not-euler}. We get one rational polynomial for all these steps, at disturbance level \(\delta\). Since the uncertainty on an update is \(\eta\delta\), which vanishes with the step, applying a converse for a fixed map separately for each \(\eta>0\) does not give the simultaneous assertion. What has to be proved in addition is that a decrease proportional to \(\eta\) survives smoothing and approximation. We prove this estimate directly, and we do not claim a first general converse theorem for common barriers or inclusions.

Ahmadi proves the strong \(\mathrm{NP}\)-hardness of the stability of cubic homogeneous systems, when the dimension varies (his Theorem~2.1). His Theorem~3.1 gives an \emph{unconditional} obstruction to a bounded degree of polynomial Lyapunov functions, already in dimension two and for fields of degree three, and his Remark~3.1 also allows rational coefficients in the field~\cite[full version, Sections II--III]{Ahmadi2012Cubic}. Our conditional obstruction on the degree does not strengthen this result. Ahmadi and Jungers also get an unbounded degree for common polynomial Lyapunov functions of stable pairs of planar linear maps, and they use quantifier elimination to turn bounds on certificates into decision procedures~\cite{AhmadiJungers2016}. Common certificates, and the principle of a reduction to complexity, are therefore not new by themselves.

Theorem~\ref{thm:common-certificates} deals with a different decision problem: compact robust safety for one fixed parametric field, with a nonzero initial box, a fixed unsafe halfspace and agreement between all Euler meshes and the flow. Every safe member has a certificate, but the bit length of its dense rational representation cannot be uniformly polynomial unless \(\mathrm{NP}=\PSPACE\), and a bounded degree over \(\R\) would imply \(\Ptime=\PSPACE\). Once the robust family and the converse are proved, the step to complexity uses standard real-algebraic bounds with a fixed number of variables~\cite{Renegar1992II}. These obstructions are conditional, and specific to this family: we prove neither an unconditional growth of the degree, nor exponential bounds on the size, nor bounds for a particular semidefinite relaxation.

\myparagraph{Implicit complexity and descriptive characterizations}
The bounded recursion of Cobham~\cite{Cob65}, the safe recursion of Bellantoni and Cook~\cite{BC92} and the ramified functionals of Leivant and Marion~\cite{LM97} give machine-independent characterizations, through restrictions on discrete computation.
We gave in~\cite{JournalACM2017} a characterization of polynomial time directly by rational polynomial ODEs, and in~\cite{BlancBournezMFCS2023Journal} analytic characterizations with discrete ODEs, without sign.
The absence of a sign primitive is thus not new by itself.
Here, polynomial dynamics implement comparisons and scheduling, and they preserve the computation under all the temporal interpretations.
Our characterizations keep explicit semantic conditions on resources: polynomial scale thresholds, polynomial time for \(\Ptime\), and robustness to precision for eventual decision in \(\PSPACE\).
Notice that the reader of raw integers extracts the bits sequentially, from the most significant one: we do not assume any primitive giving random access to the position of a bit.
The treatment of small circuit classes with discrete ODEs by Antonelli, Durand and Kontinen~\cite{AntonelliDK26} is about parallel resources, which do not exist in our sequential formulation in fixed dimension.
A syntax that enforces the bounds on resources (for example through ranks on the resets of timers) remains to be developed.

\section{Consequences for verification and implementation}\label{sec:applications}

Beyond the final answer, the common realization gives three guarantees: a shared word of logical observations, a condition for safety between observations and an error budget for approximate implementations. They complement the fixed-field prediction problem of Theorem~\ref{thm:prediction}. We will then see how a single polynomial certificate can prove safety for all the temporal interpretations, and what limits the size of such certificates.

\subsection{Traces, intermediate states and implementations}

\begin{propositionrep}[Logical traces under refinement and forcing]\label{prop:logical-traces}
In the setting of Theorem~\ref{thm:quantitative-transfer}, let \(\lambda_h\) be an observation with values in a finite alphabet, constant on each closed \(r_0h\)-neighborhood (in sup norm) of a valid configuration. Then the observations of \(Y\) at the cycle sections are exactly
\[
 \lambda_h(C_0),\lambda_h(C_1),\ldots,\lambda_h(C_T)
\]
for every allowed mesh and disturbance. For a resetting halting run and every measurable active forcing \(\xi\) with \(\|\xi(t)\|_\infty\le\nu\) almost everywhere, the continuous solution for that forcing has ordered cycle-section times with the same observations. Its data at these sections satisfy~\eqref{eq:transfer-error}, and their successive times satisfy~\eqref{eq:transfer-time}. After the initial section, they occur at every fourth successive phase crossing of the clock threshold \(9/10\). They are the limits of the corresponding Euler sections with constant mesh, when the forcing is averaged on cells (Lemma~\ref{lem:intrinsic-sections}). This includes the exact flow (take \(\xi=0\)). Every property of this finite word of observations is therefore preserved. Notice that we claim neither equality at common physical times, nor the preservation of properties of the unobserved intermediate states.
\end{propositionrep}
\begin{proofsketch}
The bound on the error at sections puts every observed state in the correct neighborhood. For a fixed measurable forcing, we use its averages on cells as the Euler disturbances, and we get the forced solution as a limit of interpolants, by compactness. We extract the finitely many section times together, and we pass their closed bounds to the limit: the positive lower bound on the duration of cycles keeps them ordered.
\end{proofsketch}
\begin{proof}
The budget and \(r_*\le r_0\) give \(\|Y(t_{4j})-C_j\|\le r_0h\), and the hypothesis on the observation immediately gives the displayed word, for fixed and variable meshes. For a finite control, one may take the control bits read at threshold \(1/2\): shrinking \(r_0\) if necessary makes this observation locally constant on valid neighborhoods. If the neighborhoods of configurations with different labels overlap, such an observation is not admissible: this separation is part of the hypothesis.

For the continuous assertion, we fix the input, the scale and the resetting halting run, and we fix a measurable forcing \(\xi\) with the stated bound. The input and the scale remain exact and frozen. We write \(Z_0=c_{M,K}(h,C_0)\) for the prescribed initial state. For constant meshes \(\eta\to0\), we set
\[
 \xi_k^{(\eta)}=\frac1\eta\int_{k\eta}^{(k+1)\eta}\xi(t)\,dt.
\]
Their sup norms are at most \(\nu\), so Theorem~\ref{thm:quantitative-transfer} applies to these Euler executions, and its conclusion on the resetting halt gives a common compact box \(\mathcal B\) on the whole half-line. Let \(Z_\eta\) be their piecewise-affine interpolants. Since \(\mathcal B\) is a box, hence convex, they stay in \(\mathcal B\). We choose bounds \(M\) and \(\Lambda\) for the norm and for the Lipschitz constant of \(F\) on \(\mathcal B\), with the frozen parameters fixed. Then \(Z_\eta\) is \((M+\nu)\)-Lipschitz.

On each complete cell of the grid, the averaged forcing has exactly the same integral as \(\xi\), and the difference on a final partial cell is at most \(2\eta\nu\). Freezing the field at the left endpoint of each cell adds at most \(\Lambda(M+\nu)\eta b\) on \([0,b]\). Hence
\[
 \sup_{0\le t\le b}\left\|Z_\eta(t)-Z_0
 -\int_0^t F(h,Z_\eta(u))\,du-\int_0^t\xi(u)\,du\right\|_\infty
 \le\Lambda(M+\nu)\eta b+2\eta\nu.
\]
By compactness and equicontinuity, we get uniform limits of subsequences on every bounded interval. A diagonal extraction and the displayed estimate give a global absolutely continuous solution of \(\dot Z=F(h,Z)+\xi\), which remains in \(\mathcal B\). The local Lipschitz continuity of \(F\) gives uniqueness for this fixed forcing: the difference of two solutions satisfies a Gr\"onwall inequality in which the forcing cancels. Every subsequence therefore has a further subsequence converging to the same \(Z\), and the whole family of interpolants converges uniformly to \(Z\) on bounded intervals. Observe that this argument does not need any continuity of \(\xi\).

We write \(t_{4j}^{(\eta)}\) for the Euler section times. The vectors \((t_0^{(\eta)},t_4^{(\eta)},\ldots,t_{4T}^{(\eta)})\) lie in the compact cube \([0,c_+(1+s)T]^{T+1}\), and we extract one subsequence on which they all converge, to \((\bar t_0,\ldots,\bar t_T)\). The closed bounds on durations give
\[
 \bar t_0=0,\qquad c_-(1+s)\le\bar t_{j+1}-\bar t_j\le c_+(1+s).
\]
Uniform convergence on this interval and the common bound on the speed imply the convergence of the states at these moving times: the displacement in time contributes at most \((M+\nu)|t_{4j}^{(\eta)}-\bar t_j|\). Passing the discrete bounds on the error at sections to the limit, we get
\[
 \|Y(\bar t_j)-C_j\|_\infty\le B\Gamma j\Delta_\eps\le r_0h
 \qquad(0\le j\le T).
\]
The hypothesis on the observation now identifies each label directly from its closed neighborhood: we need no continuity of the observation, which has discrete values, outside these neighborhoods. The forcing was arbitrary, so the conclusion holds for every permitted forcing, with the same quantitative constants. The section times depend on the forcing, but Lemma~\ref{lem:intrinsic-sections} identifies them with every fourth successive crossing of phases along \(Z\): they do not depend on the extracted subsequence. Equality of the finite words of observations gives the last assertion.
\end{proof}

\begin{lemmarep}[Intrinsic continuous sections]\label{lem:intrinsic-sections}
In the setting of Proposition~\ref{prop:logical-traces}, fix a measurable forcing \(\xi\) with \(\|\xi\|_\infty\le\nu\), and let \(Z\) be the continuous solution. Put \(\theta_0=0\) and, for each phase index \(j\ge0\) of the run, with indices modulo four,
\[
 \theta_{j+1}=\inf\{t\ge\theta_j:\ p_{j+1}(Z(t))\ge9/10\}.
\]
For each constant mesh \(\eta\), take as disturbance on a cell the average of this fixed \(\xi\) on the cell. Let \(t_0^{(\eta)}=0\), and let \(t_{j+1}^{(\eta)}\) be the first grid time after \(t_j^{(\eta)}\) at which \(p_{j+1}\ge9/10\) (these are the Euler phase entries). Then, for each fixed phase index \(j\), \(t_{j+1}^{(\eta)}\) converges to \(\theta_{j+1}\) as \(\eta\to0\), for the whole family of these meshes. With \(L=\log(1/\eps)\) and the constants of Lemma~\ref{lem-uniform-clock},
\[
 L-C_{\mathrm{dur}}\le\theta_{j+1}-\theta_j
 \le4L+C_{\mathrm{dur}}+D_{\mathrm{clk}},
\]
so these times have no finite accumulation point. The cycle-section times of Proposition~\ref{prop:logical-traces} (the limits \(\bar t_j\) of its proof) are \(\bar t_j=\theta_{4j}\). They are therefore defined from \(Z\) alone, and they do not depend on an extracted subsequence. The limit is taken at a fixed input and logical scale, for each fixed forcing.
\end{lemmarep}
\begin{proofsketch}
At the entry of a phase, the next bit is below \(1/10\), and before its first crossing on the grid, its affine interpolant is at most \(9/10+O(\eta)\). The bounds on the clock keep the erasure term small during the whole phase, so that the limiting bit has a derivative at least \(1/15\) whenever it lies in \([17/25,9/10]\). This rules out an earlier contact with the threshold, and it identifies every limit of a subsequence of crossing times.
\end{proofsketch}
\begin{proof}
We use the uniform convergence \(Z_\eta\to Z\) given by the argument of compactness and uniqueness in the proof of Proposition~\ref{prop:logical-traces} (this argument does not identify sections, and it does not use the present lemma). From Lemma~\ref{lem-uniform-clock}, the proof only uses the entry bound \(|p_{j+1}|\le\mu_0\) at each phase entry, the bounds~\eqref{eq:clock-duration} on durations, the bound in~\eqref{eq:clock-old-far} on the opposite bit during the whole phase, the definition of the phase entries as first crossings of \(9/10\) on the grid and the perturbed ranges of the coupled scalar box, all uniform in \(\eta\le1\) and \(\nu\le\eps^2\). Put \(V=M+\nu\), the common Lipschitz bound for the interpolants. We decrease the common smallness threshold so that \(\mu_0<1/10\), \(L-C_{\mathrm{dur}}>0\), \((C_{\mathrm{env}}\mu_0)^8\le10^{-3}\) and \(\nu\le10^{-4}\).

Fix \(j\), and assume by induction that \(t_j^{(\eta)}\to\theta_j\) (both are zero for \(j=0\)). The bounds on the durations of the clock place \(t_{j+1}^{(\eta)}-t_j^{(\eta)}\) in the fixed interval \([L-C_{\mathrm{dur}},4L+C_{\mathrm{dur}}+D_{\mathrm{clk}}]\). At the entry of the phase, \(p_{j+1}\le\mu_0\), and at every later grid point before \(t_{j+1}^{(\eta)}\), it is below \(9/10\). The last update has size at most \(V\eta\), so
\begin{equation}\label{eq:phase-crossing-overshoot}
 \begin{gathered}
 p_{j+1}(Z_\eta(t))\le9/10+V\eta
 \quad(t_j^{(\eta)}\le t\le t_{j+1}^{(\eta)}),\\
 9/10\le p_{j+1}(Z_\eta(t_{j+1}^{(\eta)}))\le9/10+V\eta.
 \end{gathered}
\end{equation}
The interpolant may cross the threshold inside the last cell. Bound~\eqref{eq:clock-old-far} holds at both endpoints of every cell, including the last cell of the phase, so it gives \(|p_{j+2}|\le C_{\mathrm{env}}\mu_0\) during the whole interpolated phase, by convexity of this bound on an interval.

Take any subsequence along which \(t_{j+1}^{(\eta)}\to t_*\). For \(t\in[\theta_j,t_*]\), let \(u_\eta\) be its projection onto \([t_j^{(\eta)},t_{j+1}^{(\eta)}]\). Then \(u_\eta\to t\), and
\[
 \|Z_\eta(u_\eta)-Z(t)\|_\infty
 \le V|u_\eta-t|+\sup_{0\le u\le b}\|Z_\eta(u)-Z(u)\|_\infty\longrightarrow0
\]
on any common bounded interval containing these phases. Passing the closed bounds to the limit, with \(q(t)=p_{j+1}(Z(t))\), we get
\[
 q(\theta_j)\le\mu_0<1/10,\qquad q(t)\le9/10\ (\theta_j\le t\le t_*),
 \qquad q(t_*)=9/10,
\]
and \(0\le E_{j+1}(Z(t))\le10^{-3}\) on this interval, where \(E_i=p_{i+1}^8\) is the clearing signal of phase \(i\). Hence the first crossing \(\theta_{j+1}\) exists, and it lies in \((\theta_j,t_*]\).

We now check the drift, with the perturbed scalar box. Since the interpolants lie in the closed box of Lemma~\ref{lem-invariant-boxes}, so does \(Z(t)\), for every \(t\). We write \(p=q(t)\), \(S=S_{j+1}(Z(t))\) and \(C=E_{j+1}(Z(t))\). The box gives
\[
 -2\nu\le S=p_j\tau_j^2\le1+9\nu.
\]
If \(17/25\le p\le9/10\), then \(a=p+S(1-p)-Cp\) satisfies
\[
 0<p-10^{-3}-2\nu\le a\le1+9\nu.
\]
On \([0,1]\), \(\restore(u)=3u^2-2u^3\) is increasing and \(\restore'\le3/2\), and for \(0\le d\le1/2\), \(\restore(1+d)=1-3d^2-2d^3\ge1-4d^2\). Considering separately the cases \(a\le1\) and \(a>1\), we get
\[
 \restore(a)\ge\restore(p)-\tfrac32(10^{-3}+2\nu)-324\nu^2.
\]
Almost everywhere in the band, the forcing coordinate included, this implies
\begin{equation}\label{eq:phase-crossing-drift}
 \dot q\ge p(2p-1)(1-p)-\tfrac3{2000}-4\nu-324\nu^2
 \ge\tfrac9{125}-\tfrac3{2000}-4\nu-324\nu^2>\tfrac1{15}.
\end{equation}
Indeed, the cubic is concave on \([17/25,9/10]\), so its minimum is the smaller of its two endpoint values, \(9/125\). At \(\nu=10^{-4}\), the last lower bound is \(1752419/25000000>1/15\).

Assume that \(\theta_{j+1}<t_*\). Then \(q(\theta_{j+1})=9/10\), by continuity and since \(q(\theta_j)<9/10\), and continuity gives a nonempty interval \([\theta_{j+1},\theta_{j+1}+d]\subset[\theta_j,t_*]\) on which \(q\ge17/25\). We already know that \(q\le9/10\) there. Absolute continuity and~\eqref{eq:phase-crossing-drift} then imply \(q(\theta_{j+1}+d)\ge9/10+d/15\), a contradiction. Hence \(t_*=\theta_{j+1}\). Since the possible phase-entry times lie in a compact set, we get the convergence of the whole family, by induction, for each fixed phase index. The inequalities on durations pass to the limit, and their positive lower bound excludes a finite accumulation point. Notice that we need no derivative at the crossing time itself. In particular, the extracted cycle-section times are \(\bar t_j=\theta_{4j}\).
\end{proof}

\begin{corollaryrep}[Safety between logical sections]\label{cor:segment-safety}
Put \(E=B\Gamma T\Delta_\eps\) and \(D_*=\max(1,\max_\sigma\|D_\sigma\|_\infty)\). If every ideal segment \([C_j,C_{j+1}]\) has distance at least \(d>D_*E+B\Delta_\eps\) from an unsafe data set \(U\), then both the data and shadows avoid \(U\) throughout the simulated prefix, for every permitted discrete execution. The exact flow of a halting run has the same guarantee between its limiting sections.
\end{corollaryrep}
\begin{proofsketch}
During a cycle, the data stay close to the segment joining the actual incoming data to their ideal affine image, and both endpoints of this segment are close to the corresponding source configurations. By convexity of a segment, the errors at the endpoints give a bound valid during the whole copy.
\end{proofsketch}
\begin{proof}
On cycle \(j\), we write \(Y_j=Y(t_{4j})\) and \(Y_*=D_{\sigma_j}Y_j+b_{\sigma_j}\). The section estimate gives \(\|Y_j-C_j\|\le E\) and \(\|Y_*-C_{j+1}\|\le D_*E\). Using the same convex coefficient on the two segments, we see that every point of \([Y_j,Y_*]\) is within \(D_*E\) of \([C_j,C_{j+1}]\). Lemma~\ref{lem-uniform-cycle} puts both the data and the shadows within \(B\Delta_\eps\) of the former segment, so their distance from \(U\) is at least \(d-D_*E-B\Delta_\eps>0\).

For the flow, we use the converging section times of Proposition~\ref{prop:logical-traces}. The distance to any nonempty set is a Lipschitz function, so the uniform convergence of the trajectories and the positive bound on the distance preserve the exclusion of \(U\) on the closed interval between limiting sections (the case of an empty unsafe set is immediate). Observe that the hypothesis on segments cannot be replaced by the safety of their endpoints alone: a safe set need not contain a segment joining two of its points.
\end{proof}

\begin{propositionrep}[A contract for approximate implementations]\label{prop:implementation-contract}
Fix a halting compiled computation at a permitted scale and write \(\nu=h^{2K}\). An implementation with exact prescribed initialization and updates
\[
 \widehat Z_{k+1}=\widehat Z_k+\eta_kF(h,\widehat Z_k)+e_k,
 \qquad \|e_k\|_\infty\le\eta_k\nu,
\]
preserves the compiler's guarantees whenever \(0<\eta_k\le1\) and \(\sum_k\eta_k=\infty\). In particular, one may replace \(F\) by any function \(\widehat F\) satisfying \(\|\widehat F-F\|_\infty\le\nu/2\) on the guaranteed compact box, and allow a further update error at most \(\eta_k\nu/2\).
\end{propositionrep}
\begin{proofsketch}
Dividing the complete error of an update by the step, we get exactly an admissible drift disturbance. An approximation of the field and an arithmetic error share this budget, and the guarantee of a compact box justifies the assumption on the approximation, by induction.
\end{proofsketch}
\begin{proof}
Set \(\xi_k=e_k/\eta_k\). Since \(\eta_k>0\), we have \(\|\xi_k\|\le\nu\), and we get an allowed trajectory in the compiler theorem. For the second assertion, as long as \(\widehat Z_k\) is in the box, the effective error is
\[
 e_k=\eta_k(\widehat F(h,\widehat Z_k)-F(h,\widehat Z_k))+a_k,
 \qquad \|a_k\|\le\eta_k\nu/2,
\]
and hence has norm at most \(\eta_k\nu\). The initial state belongs to the box. Any finite sequence of such effective disturbances can be extended by zeros into an allowed infinite sequence for the original field, and the guarantee of compactness for this sequence puts the next state in the box. By induction, we get both the validity of the assumption on the approximation and all the claimed guarantees. Notice that we need no assumption of continuity on the discrete approximation \(\widehat F\).
\end{proof}

\myparagraph{What precision means for an implementation}
If the update is evaluated exactly and then rounded,
\[
 b_k=\left\lceil 2Ks+\log_2(1/\eta_k)\right\rceil+2
\]
fractional bits suffice to make the complete error of rounding at most \(\eta_k h^{2K}/2\). When the arithmetic is itself approximate, its intermediate errors must also be included in the complete error of the update. At unit step, this is a precision \(O(s)\), but for arbitrarily small steps, the precision depends on \(\eta_k\). Adaptive steps that depend on the state are also covered, provided they remain at most one and their sum diverges (the cost of representing the mesh is a separate question). The theorem does not tolerate noise on the frozen input or on the scale, and, with its conservative choice \(K\ge256\), we certainly do not claim a practical execution in double precision.

\myparagraph{Uses and limits}
The statements on traces and on segments separate two tasks of verification: observing the correct logical execution, and excluding unsafe data at all intermediate times. The prediction theorem shows that compactness, polynomial dynamics and the stated robustness do not by themselves make prediction tractable: \(\PSPACE\)-hardness persists on instances where the discrete and the continuous decisions are the same. The implementation contract gives a sufficient test for replacing the exact evaluation of the field by an approximation. Finally, the size of the source program is a separate cost: the selector of Section~\ref{sec:polynomial-field} enumerates \(2^g\) sign patterns, where \(g\) is the number of tests. Our complexity theorems are for a fixed program: we claim neither a compiler of polynomial size as a function of \(g\), nor a numerical efficiency competitive with digital execution.

\subsection{Common safety certificates and their complexity}\label{sec:certificate-consequence}

Safety verification asks whether a set of initial states can reach an unsafe region. A polynomial certificate replaces this question on unbounded trajectories by algebraic conditions on one polynomial. Such certificates, and their convex relaxations, are classical tools in verification and control (see e.g.~\cite{PrajnaRantzer2007}). Here, the certificate must work simultaneously for a whole family of discrete updates and for its continuous interpretation.

\begin{definition}[Common polynomial safety certificate]\label{def:common-certificate}
Let \(f:\R^d\to\R^d\) be polynomial, \(\mathcal K\) a compact box, \(I\subset\operatorname{int}\mathcal K\) an initial set, \(U\subset\mathcal K\) an unsafe set, and \(\delta>0\). A polynomial \(V\) is a common certificate if
\begin{equation}\label{eq:certificate-separation}
 V\le-1\text{ on }I,\qquad
 V\ge1\text{ on }U\cup\partial\mathcal K,
\end{equation}
\begin{equation}\label{eq:certificate-invariance}
 z\in\mathcal K,\ V(z)\le0
 \quad\Longrightarrow\quad
 \begin{cases}
 z+\eta(f(z)+\xi)\in\mathcal K,\\
 V\bigl(z+\eta(f(z)+\xi)\bigr)\le0,
 \end{cases}
\end{equation}
where the implication holds for all \(0\le\eta\le1\) and \(\|\xi\|_\infty\le\delta\).
\end{definition}

In other words, the sublevel set of \(V\) is invariant under every allowed Euler update. It also contains the piecewise-linear interpolants, since a fraction of an update is itself an allowed smaller update. By Euler limits, we then get safety for every solution of \(\dot z=f(z)+\xi(t)\), with measurable \(\|\xi(t)\|_\infty\le\delta\). In Section~\ref{sec:common-certificates}, we prove this assertion, and a converse with a strict reserve of noise.

\begin{theoremrep}[Existence and complexity of common safety certificates]\label{thm:common-certificates}
There are a fixed active dimension \(m\), a fixed rational polynomial field \(F:\R\times\R^m\to\R^m\), a fixed rational polynomial initialization \(c(x,h)\), a fixed compact rational box \(\mathcal K\subset\R^m\), an active coordinate \(a\), a fixed integer \(K\) and an integer polynomial \(S\) with the following properties. For a binary integer \(x\), put
\[
 s=S(\ell(x)),\quad h=2^{-s},\quad
 \delta=\tfrac14h^{2K},\quad
 I_x=c(x,h)+[-\delta,\delta]^m,\quad
 U=\{Z\in\mathcal K:a(Z)\ge\tfrac12\}.
\]
All these instance data are computable in polynomial time from \(x\). Every execution from \(I_x\), with drift disturbance at most \(2\delta\), remains in a fixed compact subset of \(\operatorname{int}\mathcal K\), for every admissible Euler mesh and for continuous time. Exactly one of the following holds: either all these executions satisfy \(a\le1/4\) at all times, or all of them satisfy \(a\ge3/4\) after a common finite physical horizon. We call \emph{safe} the instances of the first alternative, and deciding this alternative is \(\PSPACE\)-complete.

Every safe instance has a common certificate \(V\in\Q[Z]\) for disturbance bound \(\delta\). A given certificate, encoded by its dense list of rational coefficients, is checkable in polynomial time in the instance and certificate lengths. Consequently, unless \(\mathrm{NP}=\PSPACE\), no polynomial in \(\ell(x)\) bounds the minimum certificate bit length on all safe instances. Moreover, unless \(\Ptime=\PSPACE\), no constant bounds their minimum degree, even if arbitrary real coefficients are allowed.
\end{theoremrep}
\begin{proofsketch}
We compile a fixed polynomial-space decider, and its accepting control bit plays the role of an alarm. A small initial box is admissible, and its extra error propagates through the same interval products. Rejection keeps the alarm below the unsafe threshold during every cycle. The closure of the reachable set under noise \(2\delta\) has a margin of safety, and Lemma~\ref{lem:common-converse} turns this reserve into a rational polynomial invariant for noise \(\delta\), uniformly down to step zero. With a fixed number of variables, real algebra decides the validity of a certificate and the existence of a certificate of fixed degree. Short certificates, or a uniform bound on the degree, would therefore collapse the stated complexity classes.
\end{proofsketch}
\begin{proof}
\myparagraph{The fixed family and perturbed initialization}
Fix a deterministic polynomial-space decider for true quantified Boolean formulas, with a leading sentinel in its binary encoding, and which rejects malformed inputs. We use the two-stack normal form of Lemma~\ref{lem-hard-normal-form}, including the resetting halt, and we let \(a\) be its accepting control bit. This bit is zero in every configuration of a rejecting run, and one in the accepting halt. It is a data coordinate of the compiled field, initially zero: we do not need to modify the polynomial clock or the selector.

We write \(T\le2^{cs}\) for the bound on the source running time, and we choose the fixed integer \(K\ge256\) with \(K/4>c+2\). We apply the cycle estimates with \(\nu=\eps^2=h^{2K}\), although the actual disturbances have norm at most \(2\delta=\nu/2\). It remains to check the additional freedom in the initial condition. A point of \(I_x\) has data error at most \(\delta\), shadow discrepancy at most \(2\delta\), phase bits within \(\delta\) of \((1,0,0,0)\) and timers within \(\delta\) of \(\eps\). For sufficiently large \(s\), these satisfy the scalar box, the conditions of the clock in phase zero and \(2\delta\le A\Delta_\eps\). Since \(\delta=o(h)\), the data lie in the \(r_0h\)-tube. The comparator initially differs by at most \(\delta\) from \(u_j(C_0,h)\), and the margin \(\kappa h\) and the tube condition give the same sign as \(u_j(Y,h)\), with modulus at least \(\kappa_0h\). Hence every point of the initial box is admissible.

The only change in the accumulation argument is the initial error on the data. If \(e_j=Y(t_{4j})-C_j\), expanding the recurrence gives
\[
 e_j=D_{\sigma_{j-1}}\cdots D_{\sigma_0}e_0
       +\sum_{i<j}D_{\sigma_{j-1}}\cdots D_{\sigma_{i+1}}\zeta_i,
 \qquad \|e_0\|\le\delta,\quad\|\zeta_i\|\le B\Delta_\eps,
\]
where the \(\zeta_i\) are the local defects of the cycles. Every interval product, including the empty product, has norm at most \(\Gamma=G/h\). We enlarge the polynomial scale threshold so that
\begin{equation}\label{eq:initial-box-budget}
 E:=\Gamma\delta+B\Gamma T\Delta_\eps
 \le2B\Gamma T\Delta_\eps\le r_*h.
\end{equation}
Here \(B,T\ge1\) and \(\delta\le\Delta_\eps\), and the last inequality follows from \(K/4>c+2\), keeping the factor \(1+s\) in \(\Delta_\eps\). Induction now proves all the hypotheses of cycles, exactly as in Lemma~\ref{lem-P-accumulation}. At the resetting halt, the incoming error is erased, so Proposition~\ref{prop-contracting-halt} applies indefinitely. These arguments are uniform in the initial point and in variable meshes. We choose a fixed rational box \(\mathcal K\) containing the resulting common compact bound in its interior, with a positive margin independent of \(x,s\).

\myparagraph{Safety at every physical time}
On a rejecting run, both endpoints of every ideal segment of data have alarm zero. Matching the endpoints of segments as in Corollary~\ref{cor:segment-safety}, we bound the actual alarm above by \(D_*E+B\Delta_\eps\) during the whole prefix, and after the halt, its modulus is at most \(2B\Delta_\eps\). We increase the scale threshold to make both bounds at most \(1/4\). The same bounds hold on Euler interpolants, by convexity of the halfspace. On an accepting run, the resetting tail has instead an alarm at least \(3/4\), uniformly after time \(2^{c_0s}\), for a fixed sufficiently large \(c_0\). The upper bound on the duration of cycles and \(T\le2^{cs}\) justify this common horizon, for all allowed initial points.

For continuous time, we fix an initial point in \(I_x\) and any measurable forcing bounded by \(2\delta\). Its averages over shrinking intervals of the mesh have the same bound. The proof of Corollary~\ref{cor:forced-ode}, by compactness and Euler limits, applies from this initial point, and it gives the same closed bounds on the alarm. The fixed box prevents any escape in finite time. Hence the dichotomy holds for all the temporal interpretations of the statement. The initial center, the radius and the frozen scale have polynomial bit length, since \(s=S(\ell(x))\). The safe binary inputs are precisely the complement of the true quantified Boolean formulas (with the chosen convention on malformed words). This language is \(\PSPACE\)-complete, because deterministic polynomial space is closed under complement. In particular, this is a total decision problem on the displayed generated family, and not a promise about arbitrary polynomial systems.

\myparagraph{Existence of certificates}
For a safe instance, let \(R\) be the closure of all the states that can be reached from \(I_x\) by finitely many updates with \(0<\eta\le1\) and disturbance at most \(2\delta\). Any such finite sequence extends to a divergent mesh, so all these states satisfy the preceding bounds. Hence \(R\) is compact, contained in \(\operatorname{int}\mathcal K\), and at distance at least \(1/4\) from \(U\). Concatenating one more update and passing to the closure, we get
\[
 r+\eta(F(h,r)+u)\in R
 \quad(r\in R,\ 0\le\eta\le1,\ \|u\|\le2\delta).
\]
Lemma~\ref{lem:common-converse} then gives a common rational polynomial certificate for the smaller bound \(\delta\) on disturbances.

\myparagraph{Certificate length and degree}
Proposition~\ref{prop:certificate-checking} gives the polynomial-time verifier. If one polynomial bounded the length of the shortest dense rational certificate on every safe input, a nondeterministic machine could guess this certificate and check it. Soundness rejects every unsafe input, and the hypothetical bound gives a witness for every safe input. The \(\PSPACE\)-complete safe language would then belong to \(\mathrm{NP}\), which gives \(\mathrm{NP}=\PSPACE\).

If instead a constant \(D\) bounded the degree of certificates on every safe input, even with real coefficients, the algorithm of Proposition~\ref{prop:certificate-checking} for the existence of certificates of fixed degree would decide this same language in polynomial time. On unsafe inputs, soundness forbids such a certificate, and on safe inputs, the hypothetical bound on the degree gives one. This would give \(\Ptime=\PSPACE\). Notice that neither argument assumes that the coefficients of a certificate of bounded degree have short rational representations.
\end{proof}

The theorem separates existence from size. Its positive assertion gives one certificate that can be reused for coarse discrete execution, for adaptive refinement and for continuous forcing. Its negative assertion is about the length of the description of such certificates in the worst case, even though the dimension of the state and the polynomial degree of the field are fixed, the initial box is nonzero, and the gap from the unsafe threshold is uniform. Corollary~\ref{cor:ordinary-barriers} gives the same obstructions for ordinary strict barriers in continuous time, and separately for unit-step invariants. In other words, the difficulty does not come from requiring one certificate for many meshes. Of course, the theorem does not rule out efficient certificates for structured applications, and it does not give a practical algorithm of synthesis.

The fixed object is the polynomial expression \(F(h,Z)\). The frozen parameter \(h\), the initial box and the radius of disturbances vary with the input (in particular, \(\delta=h^{2K}/4\) tends to zero), and each certificate is for one such instance and all its temporal steps. The theorem asserts neither a common certificate for all scales, nor hardness at a fixed positive radius of disturbances.

\myparagraph{An explicit certificate for a clocked implementation}
Appendix~\ref{sec:explicit-certificate} gives one explicit common certificate, for a clocked implementation of the evaluation of a discounted policy on two states, the contractive recurrence \(Y^{(j+1)}=\mathsf MY^{(j)}+b\). The implementation keeps the eight equations of the clock, but its gates have a positive floor, so it is not the unmodified output of the compiler. The certificate is a quadratic form in the data and the shadows, plus even powers of the normalized clock coordinates. Its degree is exponential in the scale exponent. It certifies a full initial box, disturbances on all twelve active coordinates, every admissible mesh and every forced flow, and it bounds the computed value under all of them. Its guarantee is different from the equality of finite traces at sections of Proposition~\ref{prop:logical-traces}: for this contractive program, the semantics of the value is proved directly, without the cycle lemma, and no equality of traces is claimed.

\subsection{A compact branching certificate for the unchanged compiler}\label{sec:branching-certificate}

Theorem~\ref{thm:common-certificates} rules out uniformly small certificates on the full state, under standard complexity assumptions. For a structured program, a certificate of constant size on the source suffices when it is asserted at the sections and combined with the contract of cycles, even for a program that never halts.
Inductive assertions allow one to verify a source program without enumerating its executions, as in abstract interpretation~\cite{cousot1977abstract}. Here, a strict invariant of the source can be reused for an autonomous implementation, under all meshes and under continuous forcing. The resulting certificate is \emph{compositional}: a small assertion at internal sections is combined with the proved contract of cycles. It is not a single polynomial satisfying Definition~\ref{def:common-certificate} at every micro-step. It is essential to keep these two formats distinct: Appendix~\ref{sec:supplement-branching} proves obstructions to low-degree certificates on the full state of the implementation.

\begin{propositionrep}[Strict section invariants lift without a time horizon]\label{prop:strict-section-lifting}
Fix a separated affine program with a compact valid set \(C\subseteq[0,1]^m\), independent of the scale, and a number \(\rho>0\) such that, if \(\sigma(Y)\) denotes the sign pattern of the tests at \(Y\),
\begin{equation}\label{eq:strict-source-invariant}
 \mathcal H_{\sigma(Y)}(Y)+[-\rho,\rho]^m\subseteq C
 \qquad(Y\in C).
\end{equation}
Use its unchanged compiled field with any fixed \(K\ge256\), and let \(B\) be the constant of Theorem~\ref{thm:quantitative-transfer} for this program. Put
\[
 \nu=\eps^2,\qquad
 d_s=B(1+\log(1/\eps))(\eps^{1/4}+\eps^2).
\]
For all sufficiently large scales satisfying \(d_s\le\min(\rho,r_0h)\), every execution starting at a cycle-entry state admissible above its own data \(Y_0\in C\) (in the sense of~\eqref{eq:cycle-admissible}) has infinitely many cycle sections, with \(Y_j\in C\) at every section. The conditions on scalars, on the clock, on shadows and on comparators are restored each time. The data and the shadows remain in
\(\operatorname{conv}(C)+[-d_s,d_s]^m\) at every physical time.
These statements hold for every divergent variable mesh and every drift disturbance of sup norm at most \(\nu\), and for every measurable continuous forcing with this bound. For the flow, sections are limits of internal Euler sections. Successive sections are separated by \(c_-(1+s)\) to \(c_+(1+s)\) physical time, with fixed positive constants.
\end{propositionrep}
\begin{proofsketch}
The idea is to use, at each section, the actual data as the new valid source configuration. The source image lies strictly inside the invariant, so the local defect of the cycle is absorbed by its margin, and the cycle restores every other part of the interface. This induction has no sum of errors indexed by the number of cycles. The bounds on segments give safety in continuous time, and Euler limits, by compactness, give the forced flow and its ordered sections.
\end{proofsketch}
\begin{proof}
We use the constants \(A,B,s_0\) of Lemma~\ref{lem-uniform-cycle} for this fixed program, and we increase the scale threshold to ensure its local smallness conditions and the displayed bound on \(d_s\). This is possible because \(\Delta_\eps/h\to0\), and neither \(\rho\) nor \(r_0\) depends on \(s\).

Suppose that the current section is admissible above its actual data \(Y_j\in C\). Set \(C^+=\mathcal H_{\sigma(Y_j)}(Y_j)\), which belongs to \(C\) by~\eqref{eq:strict-source-invariant}. The hypothesis of the cycle lemma on the ideal image holds with zero error, and its conclusions give
\[
 Y_{j+1}=C^++e_j,\qquad \|e_j\|_\infty\le d_s,
 \qquad \|Y^{\mathrm s}_{j+1}-Y_{j+1}\|_\infty\le A\Delta_\eps.
\]
The margin of the source puts \(Y_{j+1}\) in \(C\). The restored cones of the comparators concern the signs of \(u_l(Y_{j+1},h)\), as in~\eqref{eq:cycle-admissible}, and the conditions on the clock and on scalars are also restored. Hence the new section is admissible above \(Y_{j+1}\) itself. Observe that this induction needs no estimate relative to a single fixed ideal trajectory: in particular, the argument does not replace a sum of errors over an infinite time by a finite one.

The cycle lemma keeps the data and the shadows at distance at most \(d_s\) from \([Y_j,C^+]\), which lies in \(\operatorname{conv}(C)\). This closes a common compact bound on the whole active state, including at the next section, and the convexity of the displayed enlarged hull gives the same bound on Euler interpolants. The bounds on the durations of the clock, summed over four phases, give the stated lower and upper bounds on the duration of cycles. The weighted estimates of Corollary~\ref{cor:variable-mesh} apply at each cycle, so the induction works on every divergent variable mesh, with disturbances bounded by \(\nu\).

For the continuous statement, we fix any admissible initial state and any measurable forcing of sup norm at most \(\nu\). On a sequence of constant meshes tending to zero, we use its averages on intervals as the Euler disturbances. The compact bound that we just proved and the argument by Euler limits of Corollary~\ref{cor:forced-ode} give a unique global forced solution, with uniform convergence on every bounded interval of time. For each fixed \(j\), the times of the \(j\)-th section lie in a compact interval, by the bounds on durations, and a diagonal subsequence makes all the section times converge. Their limits have the same positive lower gaps and finite upper gaps, and they tend to infinity. At each such limit, the uniform convergence and the bounded speed of interpolants give \(Y_j\in C\), since \(C\) is closed. The other closed inequalities at sections pass to the limit as well. The condition on the signs of the comparators also passes: the margin of the tests is nonzero on \(C\) and \(|v_l|\ge\kappa_0h\), so the limit cannot have a zero product. Finally, the enlarged hull is closed, so its bound of safety holds for the whole flow. Notice that these limiting sections need not be first hitting times of a limiting threshold.
\end{proof}

\myparagraph{A controller that branches forever}
Consider the scalar feedback program
\begin{equation}\label{eq:branching-controller}
 \mathcal H_-(Y)=Y/2+1/2\quad(Y<1/2),\qquad
 \mathcal H_+(Y)=Y/2\quad(Y>1/2).
\end{equation}
It raises a low value and lowers a high one. With the normalized test \(u(Y)=(Y-1/2)/4\), its compiled equations for the data and the comparator are exactly
\begin{equation}\label{eq:unchanged-controller-field}
 \begin{split}
 v'&=(1-p_3)(v-v^3)/2+p_3\bigl((Y-1/2)/4-v\bigr),\\
 (Y^{\mathrm s})'&=p_1\bigl(Y/2+(1-v)/4-Y^{\mathrm s}\bigr),\\
 Y'&=p_3(Y^{\mathrm s}-Y).
 \end{split}
\end{equation}
The eight equations of the clock are unchanged, and there is no positive floor in the gates of the data. The active dimension is eleven.

The certificate of the source is the fixed quartic
\begin{equation}\label{eq:branching-quartic}
 P(Y)=\bigl(144(Y-1/2)^2-1\bigr)
       \bigl(16(Y-1/2)^2-1\bigr)
 =2304Y^4-4608Y^3+3296Y^2-992Y+105.
\end{equation}
Its nonpositive set is \(C=C_-\cup C_+\), where
\[
 C_-=[1/4,5/12],\qquad C_+=[7/12,3/4].
\]
The two branch checks are the rational interval inclusions
\begin{equation}\label{eq:controller-interval-certificate}
 \mathcal H_-(C_-)=[5/8,17/24]\subset C_+,\qquad
 \mathcal H_+(C_+)=[7/24,3/8]\subset C_-,
\end{equation}
each at distance \(1/24\) from the boundary of the indicated component. In other words, five integer coefficients and two checks of affine intervals give a strict invariant of the source, independently of the scale and of the length of the execution.

\begin{theoremrep}[A compact certificate for an indefinitely branching execution]\label{thm:branching-certificate}
Fix \(K\ge256\). Initialize the unchanged compiler~\eqref{eq:unchanged-controller-field} by
\[
 p=(1,0,0,0),\quad \tau_i=\eps,\quad
 Y=Y^{\mathrm s}=a,\quad v=(a-1/2)/4,
\]
where \(a\) is any rational number in
\([5/16,3/8]\cup[5/8,11/16]\).
Allow an initial error of sup norm at most \(\eps^2/16\) in all eleven active coordinates, and drift disturbances of sup norm at most \(\eps^2\).
For all sufficiently large \(s\), chosen independently of \(a\), the quartic certificate and the contract of cycles imply, for every allowed mesh and forced flow,
\begin{equation}\label{eq:controller-global-safety}
 Y(t),Y^{\mathrm s}(t)\in[11/48,37/48]\subset(1/5,4/5)
 \qquad(t\ge0).
\end{equation}
At successive internal sections, \(Y_j\) alternates between \(C_-\) and \(C_+\), and \(P(Y_j)\le-119/256\) for every \(j\ge1\).
If \(c_j\) alternates between \(1/3\) and \(2/3\), starting in the same component as \(a\), then
\begin{equation}\label{eq:controller-section-accuracy}
 |Y_j-c_j|\le2^{-j}|Y_0-c_0|+2d_s,\qquad d_s=B\Delta_\eps\le1/48.
\end{equation}
The certificate has degree four, and its list of coefficients has constant length. Its lifting uses the fixed proof of the compiler: we do not claim a small certificate on the full state in the sense of Definition~\ref{def:common-certificate}.
\end{theoremrep}
\begin{proofsketch}
The images of the branches have a margin \(1/24\), so a defect of the cycle at most \(1/48\) preserves the opposite component, and the whole path of the copy stays near \([1/4,3/4]\). The strict bound on the quartic follows by evaluating a quadratic in \((Y-1/2)^2\) at its endpoints. Each branch contracts by \(1/2\), which gives the error bound around the ideal two-cycle, without accumulation. The initial errors on the clock, the shadows and the comparators fit the interface of the compiler at sufficiently small scale.
\end{proofsketch}
\begin{proof}
The set \(C\) is a valid set of configurations for a separated affine program. Indeed, on \([-2,3]\), the test has modulus at most \(5/8\), and on \(C\), \(|Y-1/2|\ge1/12\). For \(h\le1/16\), a tube of radius \(r_0h\), with \(r_0=1/4\), therefore satisfies
\[
 |u(Y)|\ge\frac14\left(\frac1{12}-\frac1{64}\right)
       =\frac{13}{768}\ge\frac h4.
\]
The sign is constant on each such tube, so \(\kappa=1/4\) is valid, and the affine images of the branches in~\eqref{eq:controller-interval-certificate} prove~\eqref{eq:strict-source-invariant} with \(\rho=1/24\).

Let \(r_{\mathrm{init}}=\eps^2/16\). For sufficiently large \(s\), every allowed initial value of the data remains in the same component of \(C\) as \(a\), since the centers are at distance at least \(1/24\) from its boundary. The discrepancy of the shadows is at most \(2r_{\mathrm{init}}\le A\Delta_\eps\). Initially, \(|u(a)|\ge1/32\), and perturbing \(Y\) and \(v\) by \(r_{\mathrm{init}}\) preserves matching signs and gives \(|v|\ge1/32-r_{\mathrm{init}}\ge\kappa_0h\). The phase bits satisfy the scalar box and the entry inequalities of the clock in phase zero, and the timers lie in \([\eps-r_{\mathrm{init}},\eps+r_{\mathrm{init}}]\). Since \(r_{\mathrm{init}}=o(\eps)\), they lie in the required entry intervals of the timers for sufficiently small \(\eps\), with the constants of the clock chosen in Lemma~\ref{lem-uniform-clock}. Hence the full initial box is admissible above its actual data, and not only above the exact center.

We choose the scale threshold so that Proposition~\ref{prop:strict-section-lifting}, \(d_s\le1/48\) and the preceding inequalities on the initialization hold. All the constants depend only on the fixed program and on \(K\), and not on \(a\). The proposition gives infinitely many sections and compactness for all times. More precisely, a section in \(C_-\) has its successor in
\[
 [5/8,17/24]+[-1/48,1/48]=[29/48,35/48]\subset C_+,
\]
and a section in \(C_+\) has its successor in
\[
 [7/24,3/8]+[-1/48,1/48]=[13/48,19/48]\subset C_-.
\]
This proves the alternation at every section, for an unbounded number of cycles. Since \(\operatorname{conv}(C)=[1/4,3/4]\), the bound for all times on the data and the shadows is exactly~\eqref{eq:controller-global-safety}. Notice that the gap \((5/12,7/12)\) is not an unsafe region: the data pass through it during copying, and the quartic is only asserted at the sections.

At every section after the first,
\[
 25/2304\le (Y_j-1/2)^2\le121/2304.
\]
The expression \((144q-1)(16q-1)\) is convex in \(q\), so its maximum on this interval is the larger of its two values at the endpoints, namely \(-119/256\) and \(-805/768\). This proves the asserted strict margin of the quartic.

Finally, \(\mathcal H_-(1/3)=2/3\) and \(\mathcal H_+(2/3)=1/3\). At every section, the actual data and \(c_j\) select the same branch, whose slope is \(1/2\), and the defect of the cycle therefore gives
\[
 |Y_{j+1}-c_{j+1}|\le\tfrac12|Y_j-c_j|+d_s.
\]
Iterating, we get \(2^{-j}|Y_0-c_0|+2d_s(1-2^{-j})\), which implies~\eqref{eq:controller-section-accuracy}. All the conclusions pass to the sections of the forced flow, by the diagonal construction with Euler limits of Proposition~\ref{prop:strict-section-lifting}, and the gap of the test preserves the selected branch in the limit. Observe that we use neither an assumption of halting, nor a precision depending on a horizon.
\end{proof}

\myparagraph{Full-state certificates for this controller}
The quartic is an assertion on the source at the sections, transferred by the proved contract of cycles. It is not a certificate on the full state in the sense of Definition~\ref{def:common-certificate}. Such a certificate exists for this example, by Lemma~\ref{lem:common-converse} applied to the robust reachable set of Theorem~\ref{thm:branching-certificate}, with disturbance bound \(\eps^2/2\), but it is constrained: its sublevel set is nonconvex, so its degree is at least four, it must couple the clock to the data, and its normalized coefficients grow at least like \(1/\eps\), so that no single polynomial \(W(h,Z)\) serves all small scales (Propositions~\ref{prop:clock-nonconvexity} and~\ref{prop:certificate-coupling-scale} in Appendix~\ref{sec:supplement-branching}). Constructing an explicit low-degree certificate on the full state remains open for this unchanged controller.

\section{The polynomial compiler}\label{sec:proof-P}

\subsection{Soundness at unit step}

\begin{lemmarep}[Truncated simulation with an explicit precision budget]\label{lem-truncated-simulation}
Fix a rational polynomial field and a rational polynomial initialization. Suppose that the exact unit-step orbit on an input of length \(n\), at scale \(h=2^{-s}\), stays in a box of binary height \(H_{\mathrm{bin}}\) (its coordinates are bounded by \(2^H_{\mathrm{bin}}\)) for \(N\) steps. For every \(b_0\ge0\), its last state can be computed with error at most \(2^{-b_0}\), in time polynomial in \(n+s+H_{\mathrm{bin}}+N+b_0\). In particular, a polynomial threshold, a polynomial height and polynomially many steps imply membership in \(\Ptime\).
\end{lemmarep}
\begin{proofsketch}
We bound the Lipschitz constant of the unit-step map on a unit neighborhood of the exact orbit. Rounding after every step then costs a number of guard bits proportional to the number of steps times the logarithm of this constant. Notice that we need rational coefficients both for the field and for the initialization.
\end{proofsketch}
\begin{proof}
Write \(\Phi(z)=z+F(x,h,z)\), the frozen parameters being understood. Since the polynomial, its dimension and its degree are fixed, there is a constant \(C_F\) giving a Lipschitz bound
$$
 A\ge2,\qquad \log_2 A\le C_F(1+n+s+H_{\mathrm{bin}})
$$
on the unit neighborhood of the box, and evaluating the polynomial there involves integers of polynomial bit length. We compute the initial rational vector and each subsequent evaluation of the polynomial exactly, from the current rational or dyadic state, and we round each active coordinate to \(b\) fractional bits. The initial rational vector has \(O(n+s)\) bits before rounding. As long as the numerical orbit stays at distance at most one from the exact orbit, its error satisfies
$$
 e_0\le2^{-b},\qquad e_{k+1}\le A e_k+2^{-b},\qquad
 e_k\le 2^{-b}(k+1)A^k.
$$
Choose
$$
 b=b_0+\left\lceil (N+1)\log_2 A+\log_2(4(N+1))\right\rceil.
$$
The last bound is smaller than \(2^{-b_0}/2\) for every \(k\le N\), and, in particular, an induction on the first exit justifies the neighborhood in which we used the Lipschitz estimate. A polynomial of fixed degree on dyadics with \(b\) fractional bits can be evaluated exactly with intermediate integers of \(O(b+n+s+H_{\mathrm{bin}})\) bits, up to constants depending on the polynomial, and the denominator coming from its rational coefficients has a fixed bit length. Each evaluation and each rounding therefore take polynomial time, and there are \(N\) of them. For a decision, we choose \(b_0=4\), and we compare the numerical output with \(1/2\): the assumed exact error of \(1/4\), together with the numerical error of at most \(1/16\), leaves a strict separation. Taking \(s=S(n)\) and \(N=\lceil Q(s)\rceil\) proves soundness.
\end{proof}

\subsection{A dyadic-stack normal form for the source machine}\label{sec:hard-normal-form}

A stack word \(d_1\cdots d_H\), with \(d_i\in\{1,2\}\), is represented by
$$
 z_{\mathrm{stk}}=\sum_{i=1}^{H}d_i4^{-i},\qquad
 H\le H_s:=\left\lfloor\frac{s-2}{2}\right\rfloor.
$$
The empty stack has code zero. The operations
$$
 \operatorname{push}_d(z_{\mathrm{stk}})=(z_{\mathrm{stk}}+d)/4,
 \qquad \operatorname{pop}_d(z_{\mathrm{stk}})=4z_{\mathrm{stk}}-d
$$
are only used on a valid stack, and, for a pop, with its actual top digit. The empty stack, the stacks with top digit one and the stacks with top digit two occupy respectively
$$
 \{0\},\qquad[1/4,5/12),\qquad[1/2,2/3).
$$
The thresholds \(1/8\) and \(11/24\) therefore have constant margins.

\myparagraph{Why these digits}
Two constraints fix the set of digits. Zero is the code of the empty stack, so the digits of stacks are nonzero: a push of digit \(0\) on the empty stack would return the empty code. The cylinders of the first digit must be separated by open gaps of constant width, so that each comparator is only used at a fixed distance from its threshold. For digits \(d<d'\) and largest digit \(d_{\max}\) in base \(k\), the gap between the cylinders of \(d\) and \(d'\) is \((d'-d)/k-d_{\max}/(k(k-1))\). Base four is the smallest base with two nonzero digits with a positive gap, and it has exactly two such pairs, \(\{1,2\}\) with gap \(1/12\) and \(\{1,3\}\) with gap \(1/4\). We take \(\{1,2\}\): this is the encoding of words we used in~\cite{JournalACM2017}, whose alphabet \(\{0,\ldots,k-2\}\) leaves the largest digit of the base unused, so that the standard encoding \(\psi_4\) of a binary word is literally a stack code, and Section~\ref{sec:word-inputs} loads a word without conversion. The digits \(\{1,3\}\) of the Cantor-like encoding of Siegelmann and Sontag~\cite{SS95} would triple the second margin, but margins only enter the constants.

\myparagraph{A reader with margins also after the last bit}
For one raw integer \(x\), we initialize \(y=xh\) and \(\omega=1/4\). While \(\omega\ge h\), we read \(b=[y\ge\omega]\), and we set
$$
 y\leftarrow y-b\omega,\qquad \omega\leftarrow\omega/2.
$$
At the end, the reader is frozen at \((y,\omega)=(0,h/2)\). The reader is sequential (one cycle per bit, from the most significant bit down), and the scale marks the end of the word: it gives no random access to the position of a bit. Before a common normalization, we use the following affine forms:
\begin{center}
\begin{tabular}{lll}
\toprule
Test & Affine form & Guaranteed margin\\
\midrule
stack nonempty & \(z_{\mathrm{stk}}-1/8\) & \(1/8\)\\
top digit two & \(z_{\mathrm{stk}}-11/24\) & \(1/24\)\\
reader bit & \(y-\omega+h/4\) & \(h/4\)\\
reader still active & \(\omega-3h/4\) & \(h/4\)\\
control bit & \(c-1/2\) & \(1/2\)\\
\bottomrule
\end{tabular}
\end{center}
Observe that the shift of the reader bit is \(h/4\), and not \(h/2\). While reading, \(y-\omega\) is an integer multiple of \(h\), so the new form gives exactly the same bit. At the terminal state of the reader, it is \(-h/4\), and at a resetting halting state \((y,\omega)=(0,0)\), it is \(h/4\). Hence an unused test of the reader never becomes zero. The form for an active reader has values \(-h/4\) and \(-3h/4\) in these two states. In other words, every comparator, including those that are not consulted by the current instruction, keeps a margin.

We divide all the displayed affine forms by eight to define the actual tests \(u_j\). On the fixed enlarged box \([-2,3]^m\) of the data, with \(0<h\le1/16\), every normalized test has modulus at most one (for example, the form of the reader bit has modulus less than \(3/4\) there). We may take
\begin{equation}\label{eq:hard-margin-constants}
 r_0=\frac1{64},\qquad \kappa=\frac1{64}
\end{equation}
in the implication
\begin{equation}\label{eq:hard-margin}
 \|Y-C\|_\infty\le r_0h
 \ \Longrightarrow\
 \operatorname{sign}u_j(Y,h)=\operatorname{sign}u_j(C,h),
 \quad |u_j(Y,h)|\ge\kappa h.
\end{equation}
Indeed, the largest norm of a coefficient of an active variable is \(1/4\), whereas the smallest normalized margin of the reader is \(h/32\). The margins of the stacks and of the control satisfy the same implication for \(h\le1/16\).

\begin{lemmarep}[Time and space normal forms for the source machine]\label{lem-hard-normal-form}
A language \(L\subseteq\N^d\), for fixed \(d\), is in \(\Ptime\) if and only if it is decided by a two-stack machine with finite control of the preceding form, for every \(s\ge S(\ell(x))\), in at most \(T(s)\) source transitions, with polynomial \(S,T\) and stack heights at most \(H_s\). If \(L\in\PSPACE\), the same normal form has a polynomial \(S\) and at most \(2^{cs}\) transitions, for a constant \(c\). All the active data registers lie in \([0,1]\), and the accepting and rejecting configurations may be fixed resetting configurations.
\end{lemmarep}
\begin{proofsketch}
We read each raw integer with decreasing dyadic weights, and we ignore leading zeroes. We simulate a Turing tape by the stacks on the left and on the right of the head, with binary blocks of fixed length for a finite tape alphabet. A sufficiently large polynomial scale bounds all the heights of stacks. Counting the source configurations, which have \(O(s)\) bits, gives the exponential bound on time for a halting computation bounded in space.
\end{proofsketch}
\begin{proof}
We represent the part of the tape on each side of its head by a stack, and the finite control stores the current symbol of the tape and any block of fixed length currently being transferred. Each Turing step makes a bounded number of pushes, pops and changes of control. A fixed finite tape alphabet is encoded by words of fixed length over \(\{1,2\}\), so this only changes the time and the height of stacks by fixed factors. Empty stacks represent the unused tape beyond the stored part.

For one integer of binary length at most \(n\), \(s\ge n+1\) ensures \(xh<1/2\). The reader visits the weights \(1/4,1/8,\ldots,h\), in \(s-1\) steps, and leading zeroes are ignored thanks to a flag in the finite control. If every bit is zero, the control writes the canonical one-bit representation of zero. Otherwise, it stores exactly the canonical binary word, and a transfer with finite control can reverse the stored word when necessary. For fixed \(d\), we use \(d\) pairs of readers, all initialized polynomially as \((x_i h,1/4)\), and we read them successively, inserting fixed delimiters between the canonical words. This takes \(O(ds+n)\) source transitions, and it stores \(O(dn)\) digits, instead of \(O(ds)\) leading zeroes.

Let \(H(n)\) bound the maximal number of digits on stacks in this preprocessing and in the chosen Turing simulation. We choose an integer polynomial \(S(n)\ge\max(n+1,2H(n)+4,4)\). For every larger \(s\), we get the same input word, all heights are at most \(H_s\), and the only extra preprocessing consists of ignored leading zeroes. Consequently, a polynomial-time Turing simulation takes a number of source transitions polynomial in \(s\).

For a polynomial-space simulation, \(H(n)\) is still polynomial. At a fixed scale, each stack has at most \(2^{H_s+1}\) possible words. The values of the readers belong to the dyadic grid \((h/2)\Z\cap[0,1]\), there is a fixed number of readers, and there are finitely many states of the control. Hence the total number of source configurations is at most \(2^{cs}\), after increasing \(c\) to absorb fixed constants and small values of \(s\). A deterministic halting run cannot repeat a configuration before halting, so the same bound applies to its duration.

On a transition entering a halting state, we reset every work stack and every register of the readers to zero, and we set the control to the corresponding canonical accepting or rejecting state. We apply the same assignment on later halting transitions, and the output is the accepting control bit. This makes the two halting configurations fixed, and it does not change the answer. Every test remains separated from zero, as checked above.

Conversely, a source configuration has \(O(s)\) bits, and its affine transitions and comparisons can be evaluated exactly in polynomial time and in space linear in \(s\). A polynomial bound on the source time therefore gives \(\Ptime\) at \(s=S(n)\). This also gives the upper bound on space for the source model.
\end{proof}

For every complete sign pattern \(\sigma\), including control tests, one source transition has the affine form
$$
 \mathcal H_\sigma(Y)=D_\sigma Y+b_\sigma.
$$
We choose the identity for impossible patterns. A valid transition has separate linear blocks for each stack, for each pair of readers and for the finite control. In particular, the digit pushed on a stack is determined by \(\sigma\), and not by an analog multiplication between two coordinates of stacks.

\begin{lemmarep}[Cumulative gain]\label{lem-cumulative-gain}
Along every valid source computation and for every subinterval of transitions,
$$
 \|D_{\sigma_{k_2-1}}\cdots D_{\sigma_{k_1}}\|_\infty\le G/h
$$
for a fixed constant \(G\). The same bound holds after resetting halt.
\end{lemmarep}
\begin{proofsketch}
On a stack, the logarithm of the linear gain is exactly the decrease of its height. The matrices of the readers have a bounded off-diagonal term (a geometric series). The matrices of resets are zero on the relevant blocks, and they cannot increase an interval product.
\end{proofsketch}
\begin{proof}
On a stack block, write \(H_{\mathrm{stk}}(k)\) for the height of the stack after \(k\) transitions. An interval containing no reset has product
$$
 (1/4)^{\#\mathrm{push}}4^{\#\mathrm{pop}}
 =4^{H_{\mathrm{stk}}(k_1)-H_{\mathrm{stk}}(k_2)}\le4^{H_s}\le1/(4h).
$$
If it contains a reset, its product is zero, and instructions that do not touch the stack contribute the identity.

On a reader block, the nontrivial matrices are
$$
 M_b=\begin{pmatrix}1&-b\\0&1/2\end{pmatrix},\qquad b\in\{0,1\}.
$$
After \(r\) reading operations, interspersed with identities, their product is
$$
 \begin{pmatrix}
 1&-\sum_{j=0}^{r-1} b_j2^{-j}\\0&2^{-r}
 \end{pmatrix}
$$
for the bits in chronological order, and its maximal row sum is at most three. Resets again give zero. The control bits are assigned constants on every valid transition, so their linear part is zero. The complete product is block diagonal, and taking the maximum of the preceding estimates gives, for example, \(G=3\). Notice that this bound holds for products over every interval, and not only from the initial time: it applies to an error inserted at any intermediate cycle.
\end{proof}

\subsection{The polynomial field}\label{sec:polynomial-field}

Fix a separated affine program (the source stack machine above is one such program), and let \(u_1,\ldots,u_g\) be its normalized affine tests. For a sign pattern \(\sigma\in\{-1,1\}^g\), we define

$$
\beta_\sigma(v)
=
\prod_{j=1}^g\frac{1+\sigma_jv_j}{2}
$$

and

$$
\mathcal T(Y,v)
=
\sum_{\sigma\in\{-1,1\}^g}
\beta_\sigma(v)\bigl(D_\sigma Y+b_\sigma\bigr).
$$

The pattern \(\sigma\) is only an index of summation: the field below is one fixed polynomial, the same for every input, every scale and every executed branch. The comparators select the branch through the weights \(\beta_\sigma(v)\), which are the weights of multilinear interpolation on \([-1,1]^g\). They sum to one, and at a vertex \(v=\sigma_0\), they reduce to the indicator of \(\sigma=\sigma_0\).

Let

$$
\restore(a)=3a^2-2a^3,
\qquad
\eps=h^K,
$$

where \(K\) is the fixed integer exponent (\(K\ge256\), see Lemma~\ref{lem-uniform-cycle}). The active state contains four phase bits \(p_i\), four timers \(\tau_i\), the comparators \(v_j\), the data registers \(Y\) and their shadow registers \(Y^{\mathrm s}\). The indices of the clock are taken modulo four. We set

\begin{align*}
p_i'&=\restore(a_i)-p_i,
&
a_i&=p_i+S_i(1-p_i)-E_ip_i,\\
S_i&=p_{i-1}\tau_{i-1}^2,
&
E_i&=p_{i+1}^8,\\
\tau_i'&=p_i\tau_i(1-\tau_i)+(1-p_i)(\eps-\tau_i),\\
v_j'&=(1-p_3)\frac12(v_j-v_j^3)+p_3\bigl(u_j(Y,h)-v_j\bigr),\\
(Y^{\mathrm s})'&=p_1\bigl(\mathcal T(Y,v)-Y^{\mathrm s}\bigr),\\
Y'&=p_3\bigl(Y^{\mathrm s}-Y\bigr).
\end{align*}

The Euler update of mesh \(\eta\) is \(Z^+=Z+\eta F(Z)\), and the field is a polynomial with rational coefficients. More explicitly, for \(m\) data registers and \(g\) tests, the normalized core has \(2m+g+8\) active coordinates, and total degree at most
\[
 \max\{27,K+1,g+2\}
\]
in \((h,Z)\). Indeed, \(a_i\) has degree at most nine and \(\restore\) has degree three, the timer contains \((1-p_i)h^K\), and each selected affine target has degree at most \(g+1\), before multiplication by the loading gate. The other components have degree at most four. This count is before the optional translation to a raw state below. The formula of the selector can contain \(2^g\) branches, so the bound on the dimension is not a polynomial bound on the size of its expanded description.

The intended initial section is

$$
p=(1,0,0,0),
\qquad
\tau_i=\eps,
\qquad
Y=C_0(x,h),
\qquad
Y^{\mathrm s}=Y,
\qquad
v_j=u_j(Y,h),
$$

where \(C_0(x,h)\) is the initial source configuration (it contains in particular \(y=xh\)). This section is polynomial in \((x,h)\).

\begin{lemma}[Raw-state translation]\label{lem-raw-translation}
Let \(c(x,h)\) be a rational polynomial initialization and let \((x,h)\) remain frozen. Put
$$
 \widehat F(x,h,w)=F(x,h,w+c(x,h)).
$$
Then \(w=z-c(x,h)\) conjugates the orbit of \(F\) starting at \(c(x,h)\) to the orbit of \(\widehat F\) starting at zero, for every mesh and with the same additive disturbances. It preserves the active increments and the length. If \(c\) is bounded on the allowed inputs, it also preserves compactness, up to a fixed enlargement. The output coordinate itself is preserved when \(c_o(x,h)=0\), as in our witnesses for languages.
\end{lemma}
\begin{proof}
The vector \(c(x,h)\) is constant along an orbit. Subtracting it from the initial state and from both sides of each update, we get exactly the displayed field, with the same disturbance, and the differences of successive states are unchanged. In general, the original output is \(w_o+c_o(x,h)\): we need the last hypothesis to identify it with the coordinate \(w_o\) itself.
\end{proof}

\subsection{A common calculus for exact and perturbed schemes}

We prove the analytic estimates simultaneously for
\begin{equation}\label{eq:common-perturbation}
 Z^+=Z+\eta F(Z)+\eta\xi,\qquad
 0<\eta\le1,\qquad \|\xi\|_\infty\le\nu\le\eps^2,
 \qquad \eps=h^K.
\end{equation}
Only the active coordinates are perturbed, and the exact scheme is the case \(\nu=0\). In the whole proof, a constant may depend on the fixed source machine, but not on the input, on the number of the phase, on \(s\) or on \(\eta\). All the requirements of smallness are imposed on one threshold \(\eps_0>0\), after these constants have been chosen.

An interval \([a,b)\) of the grid contains precisely the updates starting in it. We write
$$
 \int_a^b f(t)\,d_\eta t
 :=\eta\sum_{a\le k\eta<b}f(k\eta).
$$
For variable steps, this notation means \(\sum_{a\le t_k<b}\eta_k f(t_k)\), with \(t_{k+1}-t_k=\eta_k\). A hitting time is always a grid time, and passing to the first grid time after a prescribed physical time adds at most one unit. Lemma~\ref{lem:weighted-grid} controls the weighted sums and products used below, without a lower bound on the steps. Corollary~\ref{cor:variable-mesh} assembles the extension to variable meshes.

For a nonnegative sequence, the recurrence
$$
 a^+\le(1-\lambda\eta)a+\eta b,
 \qquad 0<\lambda\le1,
$$
gives \(a(t)\le e^{-\lambda t}a(0)+b/\lambda\). Conversely,
\(a^+\ge(1+\lambda\eta)a\) implies
\(a(t)\ge(1+\lambda)^t a(0)\), because
\(\log(1+\lambda\eta)\ge\eta\log(1+\lambda)\).

\begin{lemmarep}[Weighted-grid estimates]\label{lem:weighted-grid}
Let \(t_{k+1}-t_k=\eta_k\in(0,1]\), and let \(m<n\). For \(b>0\),
\begin{align}
 \sum_{k=m}^{n-1}\eta_k e^{-b(t_k-t_m)}&\le e^b/b,
 \label{eq:grid-decay}\\
 \sum_{k=m}^{n-1}\eta_k e^{b(t_k-t_m)}&\le
       \bigl(e^{b(t_n-t_m)}-1\bigr)/b.
 \label{eq:grid-growth}
\end{align}
If \(0<\lambda<a\), then
\begin{equation}\label{eq:grid-convolution}
 \sum_{k=m}^{n-1}\eta_k e^{-\lambda(t_n-t_{k+1})}
             e^{-a(t_k-t_m)}
 \le \frac{e^a}{a-\lambda}e^{-\lambda(t_n-t_m)}.
\end{equation}
Products of factors \(1-\lambda\eta_k\), for \(0<\lambda\le1\), are at most
\(e^{-\lambda(t_n-t_m)}\). Products of \(1+\lambda\eta_k\) are at least
\((1+\lambda)^{t_n-t_m}\).
\end{lemmarep}
\begin{proofsketch}
We compare each value at a left endpoint with the integral on its own cell of the grid. Since the width of cells is at most one, we get a uniform exponential factor for decreasing functions. The convolution reduces to a decreasing exponential with rate \(a-\lambda\).
\end{proofsketch}
\begin{proof}
For \(u\in[t_k,t_{k+1}]\),
\(e^{-b(t_k-t_m)}\le e^b e^{-b(u-t_m)}\), whereas
\(e^{b(t_k-t_m)}\le e^{b(u-t_m)}\). Integrating on this cell and summing, we get
\eqref{eq:grid-decay} and \eqref{eq:grid-growth}. The summand in
\eqref{eq:grid-convolution}, without \(\eta_k\), is
$$
 e^{-\lambda(t_n-t_m)}e^{\lambda\eta_k}
                    e^{-(a-\lambda)(t_k-t_m)}.
$$
Use \(e^{\lambda\eta_k}\le e^\lambda\) and \eqref{eq:grid-decay} with rate
\(a-\lambda\). Finally \(1-z\le e^{-z}\), and concavity of the logarithm gives
\(\log(1+\lambda\eta_k)\ge\eta_k\log(1+\lambda)\), and multiplying proves the bounds on products.
Observe that we use no lower bound on an individual step.
\end{proof}

\begin{lemmarep}[Coupled scalar box]\label{lem-invariant-boxes}\label{lem:perturbed-box}
Assume \(|u_j(Y,h)|\le1\). For sufficiently small \(\eps\), the scheme \eqref{eq:common-perturbation} preserves
\begin{equation}\label{eq:scalar-box}
 -\nu\le p_i\le1+\nu,\qquad
 \eps/2\le\tau_i\le1+3\nu,\qquad
 |v_j|\le1+4\nu.
\end{equation}
For the exact scheme, the sharper intervals \([0,1]\), \([\eps,1]\) and \([-1,1]\) are invariant.
\end{lemmarep}
\begin{proofsketch}
An Euler step of size at most one is a convex combination of the current state and of its full unit-step update, disturbance included. It is therefore sufficient to check the unit-step image of the proposed box. This also deals with the corners where a perturbed gate is slightly negative, or slightly larger than one.
\end{proofsketch}
\begin{proof}
Put \(r(a)=3a^2-2a^3\), \(A(t)=2t-t^2\) and \(g(v)=(3v-v^3)/2\). For a fixed admissible disturbance, the update is the convex combination of the current state and of \(Z+F(Z)+\xi\), and we check the latter.

For phase bits, on \eqref{eq:scalar-box} and for \(\nu\) sufficiently small,
$$
 -2\nu\le S_i=p_{i-1}\tau_{i-1}^2\le1+9\nu,
 \qquad 0\le E_i=p_{i+1}^8\le1+9\nu.
$$
Clipping \(p,S,C\) to \([0,1]\) changes \(a=p(1-S-C)+S\) by at most \(24\nu\), since its three partial derivatives are uniformly bounded by \(1.1\) on this neighborhood, and the clipped value is in \([0,1]\). Hence \(a\in[-24\nu,1+24\nu]\subseteq[-1/2,3/2]\). On the latter interval, \(0\le r(a)\le1\), so \(p^+=r(a)+\xi_p\in[-\nu,1+\nu]\).

At unit step, a timer is updated by
$$
 \tau^+=p A(\tau)+(1-p)\eps+\xi_\tau.
$$
On \([\eps/2,1+3\nu]\), the concave quadratic \(A\) lies in
\([\eps-\eps^2/4,1]\) (its value at the other endpoint is \(1-9\nu^2\), which is larger than the displayed lower bound). Extending \(p\) from \([0,1]\) to \([-\nu,1+\nu]\) changes a convex combination of \(A(\tau)\) and \(\eps\) by at most \(\nu\). Consequently,
$$
 \eps-\eps^2/4-2\nu\le\tau^+\le1+2\nu.
$$
For \(\eps\) small and \(\nu\le\eps^2\), this interval is contained in \([\eps/2,1+3\nu]\).

At unit step, a comparator is updated by
$$
 v^+=(1-p_3)g(v)+p_3u+\xi_v.
$$
For small \(\nu\), the cubic \(g\) maps \([-1-4\nu,1+4\nu]\) into \([-1,1]\): this follows from its critical points \(\pm1\), from
\(g(1+4\nu)=1-24\nu^2-32\nu^3\), and from oddness. Since \(|u|\le1\), extrapolating the convex combination by at most \(\nu\) at either endpoint adds at most \(2\nu\), hence \(|v^+|\le1+3\nu\). The three estimates use the same box, so they prove the coupled invariance. At \(\nu=0\), the same calculations preserve \([\eps,1]\) for the timers. Notice that there is no exceptional division when \(\eta=p_3=1\): the update of the comparator is simply \(v^+=u\).
\end{proof}

\begin{lemmarep}[Scalar bounds with uniform slack]\label{lem-timer}\label{lem-phase-bit}
The following estimates hold on \eqref{eq:scalar-box}, for sufficiently small \(\eps\). We write \(p\) for a phase bit, \(\tau\) for its timer, \(S\) and \(C\) for its activation and clearing terms \(S_i\) and \(E_i\), \(S_+=\max(S,0)\), \(p_+=\max(p,0)\) and \(e_+=\max(1-p,0)\) (these are only notations for the analysis). There is an absolute constant \(C_0\) such that:

\myparagraph{Phase bits}
When \(p\le1/10\) and \(S\le(1+\nu)/16\), the region \(p\le1/10\) is invariant and
$$
 p_+^+\le(1-3\eta/8)p_++7\eta(S_+)^2+C_0\eta\nu.
$$
When \(p\ge9/10\) and \(C\le1/20\), the region \(p\ge9/10\) is invariant and
$$
 e_+^+\le(1-3\eta/8)e_++7\eta C^2+C_0\eta\nu.
$$
If \(C\ge43/100\), \(S\le1/100\), and \(p\in[-\nu,1+\nu]\), then
$$
 p_+^+\le(1-3\eta/8)p_++2\eta S_++C_0\eta\nu.
$$
The barrier \(p\ge24/25\) is invariant when \(C\le(17/25)^8\). If \(S\ge18/25\), \(C\le10^{-3}\), and \(p<17/25\), then
\begin{equation}\label{eq:clock-switch-affine}
 p^+\ge(1-\eta)p+(79/100)\eta.
\end{equation}
If \(17/25\le p\le9/10\) and \(C\le10^{-3}\), then
\begin{equation}\label{eq:clock-switch-drift}
 p^+\ge p+(69/1000)\eta.
\end{equation}
Hence the bit reaches \(17/25\) in bounded physical time under the first hypotheses, and then \(9/10\) in bounded additional time under the second ones.

\myparagraph{Timers}
At rest, \(p\le1/10\) implies
$$
 \tau^+\le(1-4\eta/5)\tau+\eta(\eps+2\nu).
$$
While \(p\ge9/10\) and \(\eps/2\le\tau\le1/2\),
$$
 \tau^+\ge(1+17\eta/50)\tau.
$$
There is a constant \(C_1\) such that always
$$
 (\tau+\eps+C_1\nu)^+
 \le[1+\eta(1+C_1\nu)](\tau+\eps+C_1\nu).
$$
If \(p\ge24/25\) and \(\tau\ge1/2\), then
$$
 (1-\tau)_+^+
 \le(1-12\eta/25)(1-\tau)_++\eta/25+C_0\eta\nu.
$$
\end{lemmarep}
\begin{proofsketch}
All the bounds follow from the cubic polynomial of restoration and from the quadratic update of timers. We keep a strict slack in the numerical constants before adding noise. The invariance of thresholds is checked at the threshold itself: we do not infer it from an imprecise stationary bound.
\end{proofsketch}
\begin{proof}
Let us first consider \(\nu=0\). For a low phase bit,
$$
 r(a)\le3(p+S)^2\le6p^2+6S^2\le(3/5)p+6S^2.
$$
The corresponding estimate for a high bit follows by replacing \(p\) with \(1-p\), and using \(1-a\le(1-p)+C\). For erasure, monotonicity and \(r'\le3/2\) on \([0,1]\) give
$$
 r(a)\le r(57p/100)+(3/2)S\le(121/200)p+(3/2)S.
$$
The second inequality holds because
\(r(57p/100)/p\) is increasing on \((0,1]\), and its maximum is
\(302157/500000<121/200\). Hence the available rate of erasure is greater than \(3/8\).

The strict threshold checks are
$$
 r((9/10)(19/20))>9/10,\qquad
 r(1/10+(9/10)/16)<1/10,
$$
and
$$
 r((24/25)(1-(17/25)^8))>24/25.
$$
For the low interval, the largest possible \(a\) occurs at \(p=1/10,S=1/16,C=0\), and for a high interval, the smallest possible \(a\) occurs at the lower endpoint, with \(S=0\) and the largest permitted \(C\). The displayed strict inequalities therefore bound the images of the complete intervals, and not only the drift at one point of the boundary. Euler steps are convex combinations with these unit-step images. On \([17/25,9/10]\),
$$
 r(p)-p=p(1-p)(2p-1)\ge0.072.
$$
Replacing \(p\) in this expression by \(p(1-C)\), with \(C\le10^{-3}\), loses at most \(3/2000\).

For perturbations, we project the arguments of \(r\) to their unit intervals. The derivatives of the fixed polynomial are bounded on \eqref{eq:scalar-box}, so the drift on one step changes by at most a constant times \(\nu\), and the actual update by at most this constant times \(\eta\nu\). Taking positive parts preserves the upper inequalities, because \(1-3\eta/8\ge0\), and the strict inequalities on thresholds that we just displayed persist for small \(\nu\). Increasing \(C_0\), and replacing the coefficients \(6\) and \(3/2\) and the stronger rates of contraction by the displayed weaker constants, we get the bounds of contraction.

For completeness, the bounds for switching also allow arguments of restoration just above one. Take \(\nu\le10^{-4}\). When \(p<17/25\), \(S\ge18/25\) and \(C\le10^{-3}\), the scalar box gives \(719/1000\le a\le1+10\nu\). On this interval, the minimum of \(r\) is at an endpoint, and both values at the endpoints exceed \(4/5\). Hence \(r(a)+\xi_p\ge79/100\), which proves~\eqref{eq:clock-switch-affine}. When \(17/25\le p\le9/10\), the bounds \(S\ge-2\nu\) and \(C\le10^{-3}\) give
\[
 p-10^{-3}-2\nu\le a\le1+9\nu.
\]
Using \(r'\le3/2\) on \([0,1]\) and \(r(1+d)\ge1-4d^2\) for \(0\le d\le1/2\), we see that the complete noisy drift is at least
\[
 \frac9{125}-\frac3{2000}-4\nu-324\nu^2
 \ge\frac{1752419}{25000000}>\frac{69}{1000},
\]
which proves~\eqref{eq:clock-switch-drift} without assuming that \(S\) is nonnegative.

For timers, write
$$
 b=(2p-1)\tau-p\tau^2+(1-p)\eps.
$$
For \(0\le p\le1/10\), the drift is at most \(-(4/5)\tau+\eps\). For \(-\nu\le p<0\), we write instead \(b=-\tau+\eps+p(A(\tau)-\eps)\), where \(A(\tau)=2\tau-\tau^2\in[0,1]\) on the scalar box, hence \(b\le-\tau+\eps+\nu\). In both cases, adding \(\xi_\tau\) proves the resting inequality, with \(2\nu\).

For \(9/10\le p\le1\), \(\tau\le1/2\) gives \(b\ge(7/20)\tau\). Extrapolating \(p\) by at most \(\nu\) and adding noise loses at most \(2\nu\). Since \(\tau\ge\eps/2\) and \(\nu\le\eps^2\), the rate \(17/50\) follows, for example, whenever \(\eps\le1/400\). The same unit-step estimate also excludes a downward crossing of a threshold of the timer that has already been reached: for \(9/10\le p\le1+\nu\),
$$
 \tau\in[1/4,1+3\nu]\ \Longrightarrow\ \tau^+\ge1/4,
 \qquad
 \tau\in[1/2,1+3\nu]\ \Longrightarrow\ \tau^+\ge1/2,
$$
for sufficiently small \(\eps\). Indeed, \(A\) is concave, its values at the lower endpoints are \(7/16\) and \(3/4\), and \(A(1+3\nu)=1-9\nu^2\): we multiply the corresponding minimum by \(9/10\), and we absorb \(2\nu\). This justifies that the argument of switching can be continued after either hitting time.

For the global bound, \(b+\xi_\tau\le(1+4\nu)\tau+\eps+2\nu\). Indeed, if \(p\ge0\), we drop the term \(-p\tau^2\), and if \(p<0\), we bound \(-p\tau^2\le2\nu\tau\). The remaining coefficients are bounded by \(p\le1+\nu\), so one may take \(C_1=4\) in the inequality of global growth. Notice that these are bounds for the complete noisy update, and not only for its noiseless drift. Finally, for \(c=(1-\tau)_+\) and \(p\ge24/25\), the logistic term gives \(-pc(1-c)\le-(12/25)c\) when \(0\le c\le1/2\), and the resting term and the disturbance contribute at most \(1/25+O(\nu)\). Values \(\tau>1\) are covered by the same estimate on positive parts.
\end{proof}

\begin{lemmarep}[Locks with slightly perturbed gates]\label{lem-lock}
Suppose \(X,C\) stay in a fixed box and
$$
 X^+=X+\eta G(C-X)+\eta\zeta,\qquad
 -\nu\le G\le1+\nu,\qquad \|\zeta\|\le\nu.
$$
Let \(\bar G=\min(1,\max(0,G))\) (this clipped gate is only used in the proof). Replacing \(G\) by \(\bar G\) changes the bound on the disturbance to \(C_B\nu\), for a constant \(C_B\) depending only on the box. For a fixed point \(C_*\) and a target satisfying \(\|C-C_*\|\le d\),
\begin{equation}\label{eq:lock-kernel}
 \|X_b-C_*\|
 \le \prod_{a\le k\eta<b}(1-\eta\bar G_k)\|X_a-C_*\|
      +d\left(1-\prod_{a\le k\eta<b}(1-\eta\bar G_k)\right)
      +C_B\nu(b-a).
\end{equation}
If \(\bar G\ge\gamma>0\) on the whole interval, the last term can instead be bounded by \(C_B\nu/\gamma\). Without a fixed target, the total variation is at most
$$
 C_B\int_a^b |G|\,d_\eta t+C_B\nu(b-a).
$$
\end{lemmarep}
\begin{proofsketch}
Clipping the gate is an estimate, and not a modification of the polynomial dynamics. Variation of constants for the resulting convex lock gives a telescoping kernel. In particular, a small mass of the gate controls a closed lock, even when the target changes arbitrarily.
\end{proofsketch}
\begin{proof}
Write
\(r=\zeta+(G-\bar G)(C-X)\), so that \(\|r\|\le C_B\nu\). We subtract \(C_*\) and we iterate. The weights of the inhomogeneous terms of the target sum to
\(1-\prod(1-\eta\bar G_k)\), as one sees by telescoping the products. The terms of disturbance have a total weight at most the physical length of the interval, or at most \(1/\gamma\) when all gates are at least \(\gamma\). Summing the original increments proves the bound on the variation. The same kernel proves the following fact, used below: if the targets stay at distance at most \(d\) from a fixed convex segment, and if the initial value is at distance at most \(d_0\), then the lock stays at distance at most \(\max(d,d_0)+C_B\nu(b-a)\) from this segment.
\end{proof}

\subsection{The autonomous clock, with perturbations included}

Put \(L=\log(1/\eps)\), \(\mu_0=\eps^{1/4}\) and \(\lambda_{\mathrm{clk}}=3/8\) (here \(\mu_0\) has nothing to do with a requested precision on the output). The indices of phases are cyclic modulo four. We set \(t_0=0\), and, once \(t_j\) is defined, we let \(t_{j+1}\) be the first subsequent grid time with \(p_{j+1}\ge9/10\). We let \(t_j^b\) be the first grid time after \(t_j\) with \(\tau_j\ge1/4\).

\begin{lemmarep}[Uniform clock]\label{lem-uniform-clock}\label{lem:robust-clock}
There are absolute constants \(A_0,B_0,C_{\mathrm{dur}},C_{\mathrm{env}},D_{\mathrm{clk}}>0\) and \(\eps_0>0\) such that the following holds for \(0<\eps\le\eps_0\), \(0\le\nu\le\eps^2\) and every mesh \(0<\eta\le1\). On the scalar box, the entry conditions
\begin{equation}\label{eq:clock-entry}
 \begin{gathered}
 p_j\ge9/10,\quad |p_{j+1}|,|p_{j+2}|\le\mu_0,\quad
 \eps/2\le\tau_j\le A_0\eps,\\
 \tau_{j+1}\le2\eps,\qquad \tau_{j+2}\le B_0\eps^{4/5}
 \end{gathered}
\end{equation}
are mapped to the same conditions with \(j\) replaced by \(j+1\) (no additional condition is imposed on the preceding bit or on its timer). Moreover,
\begin{equation}\label{eq:clock-duration}
 L-C_{\mathrm{dur}}\le t_j^b-t_j\le4L+C_{\mathrm{dur}},\qquad
 0\le t_{j+1}-t_j^b\le D_{\mathrm{clk}}.
\end{equation}
The active bit is at least \(9/10\) outside a terminal interval of length at most \(D_{\mathrm{clk}}\), and the next timer remains \(O(\eps)\) during the whole phase. The following bounds hold at every grid point of the closed interval \([t_j,t_{j+1}]\) of the phase, its last point included:
\begin{equation}\label{eq:clock-old-far}
 \begin{split}
 |p_{j-1}(t)|&\le C_{\mathrm{env}} e^{-\lambda_{\mathrm{clk}}(t-t_j)}+C_{\mathrm{env}}\eps^{37/20}+C_{\mathrm{env}}\nu,\\
 |p_{j+2}(t)|&\le C_{\mathrm{env}}\mu_0.
 \end{split}
\end{equation}
Before \(t_j^b\),
\begin{equation}\label{eq:clock-next}
 |p_{j+1}(t)|\le C_{\mathrm{env}}(\mu_0+\tau_j(t)^4+\nu),
 \qquad
 \tau_j(t)\le C_{\mathrm{env}}\eps e^{(1+4\nu)(t-t_j)}.
\end{equation}
The initial clock \(p=(1,0,0,0)\), \(\tau_i=\eps\) satisfies the entry conditions.
\end{lemmarep}
\begin{proofsketch}
During the long waiting part, a small next bit keeps the active bit on, and the active bit makes its timer grow and erases the preceding bit. When the timer saturates, the next bit crosses its self-maintaining threshold before it can extinguish the current bit. The two resting timers give the extra small factors needed to close the induction, also under noise.
\end{proofsketch}
\begin{proof}
All the estimates refer to the common perturbed scheme. We fix \(D_{\mathrm{clk}}=32\) and \(D_1=13\): the argument of switching below gives a terminal duration at most \(24<D_{\mathrm{clk}}\), including all the overshoots of the grid. We put \(C_D=4e^{2(D_{\mathrm{clk}}+1)}\), and we choose \(A_0>C_D\). We then choose \(C_{\mathrm{dur}}\) from the bounds on waiting times, then
\[
 B_0>2e^{(4/5)(C_{\mathrm{dur}}+D_1)}+2,
\]
then \(C_{\mathrm{env}}\), and finally one sufficiently small \(\eps_0\). All the restrictions on \(\nu\) below follow from \(\nu\le\eps^2\), by decreasing this last threshold. With this order, \(D_{\mathrm{clk}}\) and \(C_{\mathrm{dur}}\) do not depend on \(B_0\).

\myparagraph{Waiting region}
We start from \eqref{eq:clock-entry}, and we stop at the first failure of
$$
 p_j\ge9/10,\quad p_{j+1}\le1/10,\quad
 |p_{j+2}|\le\mu_0,\quad \tau_{j+1}\le2\eps,
$$
or at the first crossing of \(\tau_j=1/4\). Before this crossing,
\(S_{j+1}\le(1+\nu)/16\), and the estimate for a low bit preserves \(p_{j+1}\le1/10\). Hence \(C_j=p_{j+1}^8\le10^{-8}\), and the estimate for a high bit preserves \(p_j\ge9/10\). The next timer is at rest, with stationary bound
\((5/4)(\eps+2\nu)<2\eps\), and its contribution to the source of the opposite bit is at most \(O(\eps^2)\). The recurrence for a low bit, applied to this opposite bit, has stationary bound \(O(\eps^4+\nu)=o(\mu_0)\), and its negative part is at most \(\nu\). Hence \(|p_{j+2}|\le\mu_0\) is also preserved. These estimates rule out every proposed exit before the crossing of the timer.

The same bounds hold at \(t_j^b\), since the update reaching it is evaluated before the crossing of the timer. The active timer increases at rate at least \(17/50\) while it is below \(1/2\), and its growth over an elapsed time \(u\) is at least \((67/50)^u\). Since \(\log(67/50)>1/4\), it reaches \(1/4\) by time \(4L+1\), the last grid step included. The upper estimate of global growth, with \(C_1=4\), gives
\[
 \frac14\le(A_0+2)\eps\exp\bigl((1+4\nu)(t_j^b-t_j)\bigr).
\]
For \(4\nu L\le1\), this implies
\(t_j^b-t_j\ge L-\log(4(A_0+2))-1\).
We choose \(C_{\mathrm{dur}}\ge\log(4(A_0+2))+2\), and we decrease \(\eps_0\) so that \(L-C_{\mathrm{dur}}>D_1+1\). The recurrence for a low bit, with the increasing timer as a bound on the past forcing, gives~\eqref{eq:clock-next}. Notice that these choices do not use \(B_0\).

The opposite timer remains below \(B_0\eps^{4/5}\), by its resting recurrence. Consequently, the source of the preceding bit satisfies
$$
 |S_{j-1}|=|p_{j+2}|\tau_{j+2}^2
 \le B_0^2\eps^{1/4+8/5}=B_0^2\eps^{37/20}.
$$
We require \(B_0^2\eps^{37/20}\le1/100\). The active bit gives \(C_{j-1}\ge(9/10)^8>43/100\). More precisely, the inequality of erasure and the lower bound of the scalar box give
\begin{equation}\label{eq:clock-explicit-old}
 |p_{j-1}(t)|\le2e^{-(3/8)(t-t_j)}
        +C_{\mathrm{bit}}(B_0^2\eps^{37/20}+\nu),
\end{equation}
where \(C_{\mathrm{bit}}\) is an absolute constant of the scalar estimates, independent of \(B_0\). After decreasing \(\eps_0\), the second term is at most \(1/100\). Let \(d_j\) be the first grid time at least \(t_j+12\). Then
\[
 d_j-t_j\le13=D_1,\qquad
 |p_{j-1}(t)|\le2e^{-9/2}+1/100<1/10
 \quad(d_j\le t\le t_j^b).
\]
The extra unit accounts for the rounding to the grid: the reset of the timer must start at \(d_j\), and not at the time \(t_j+12\), which may not be on the grid. The fixed bound \(D_1\) is independent of \(B_0\). The inequality of erasure proves~\eqref{eq:clock-old-far} up to \(t_j^b\), with the sharper residual exponent \(37/20\) needed for the next entry.

\myparagraph{A fixed window before proving switching}
We first bound the opposite bit on the whole window of length \(D_{\mathrm{clk}}+1\) starting at \(t_j^b\), without assuming that switching happens there. The global estimate on timers and \(\tau_{j+1}(t_j^b)\le2\eps\) give
\[
 \tau_{j+1}(t)\le4\eps e^{2(D_{\mathrm{clk}}+1)}=C_D\eps
 \quad(0\le t-t_j^b\le D_{\mathrm{clk}}+1).
\]
Here \(\eps\le1/4\) and \(\nu\le\eps^2\) suffice for the displayed coarse bound. Consequently, \((S_{j+2})_+\le2C_D^2\eps^2\), independently of the active bit. At a proposed first exit of \(|p_{j+2}|\le\mu_0\), the preceding update satisfies the inequality for a low bit. We choose the common threshold so that
\[
 2C_D^2\eps^2\le1/16,\qquad
 28C_D^4\eps^4+C_0\nu\le(3/8)\mu_0,\qquad \nu\le\mu_0.
\]
This update cannot leave \([-\nu,\mu_0]\). Hence \(|p_{j+2}|\le\mu_0\) at every grid point of the window, including any candidate for an exit, and
\begin{equation}\label{eq:clock-terminal-small}
 E_{j+1}=p_{j+2}^8\le\mu_0^8=\eps^2\le10^{-3}.
\end{equation}
The opposite timer stays at most \(B_0\eps^{4/5}\), by its resting inequality, so \(|S_{j-1}|\le B_0^2\eps^{37/20}\) still holds. The preceding bit was already below \(1/10\) at \(t_j^b\), and the barrier for a low bit keeps it there during this whole window. Notice that these bounds come before the estimate on the time of switching: they do not assume it.

\myparagraph{Switching without loss of the token}
In the waiting region, the inequality for a high bit has a stationary deficit at most
\((7\cdot10^{-16}+C_0\nu)/(3/8)\), and we make this at most \(1/1000\). From the first grid time at least \(t_j+3\), the deficit is at most \((1/10)e^{-9/8}+1/1000<1/25\). Hence \(p_j(t_j^b)\ge24/25\).

Let \(t^c\) be the first grid time after \(t_j^b\) with \(p_{j+1}\ge17/25\). Up to this crossing, included, the stronger high barrier preserves \(p_j\ge24/25\), since every update before the crossing is evaluated with \(|p_{j+1}|\le17/25\). If \(t^c\) has not yet happened, the following stages bound its waiting time.
\begin{enumerate}
\item From \(\tau_j\ge1/4\), the growth by factors \(1+(17/50)\eta\) reaches \(1/2\) within four units of physical time, since \((67/50)^3>2\) and rounding to the grid costs at most one. The timer cannot cross either threshold downwards while the active bit stays high.
\item Starting at this crossing, the deficit \(c=(1-\tau_j)_+\) satisfies
\[
 c^+\le(1-12\eta/25)c+(41/1000)\eta,
\]
after requiring \(C_0\nu\le1/1000\). Its stationary bound is \(41/480<1/10\), and at the first grid time at least ten units later,
\(c\le\tfrac12 e^{-24/5}+41/480<1/10\).
This stage costs at most eleven units, and the same recurrence preserves \(c\le1/10\) until \(t^c\).
\item Now \(S_{j+1}\ge(24/25)(9/10)^2>18/25\), and~\eqref{eq:clock-terminal-small} bounds the term of erasure. Until \(t^c\), inequality~\eqref{eq:clock-switch-affine} gives, after an elapsed time \(u\),
\[
 p_{j+1}\ge79/100-(79/100+\nu)e^{-u}.
\]
At \(u\ge3\), this exceeds \(17/25\). Including the overshoot of the grid, this stage costs at most four units.
\end{enumerate}
If the crossing happens during an earlier stage, the later stages are not needed. Hence \(t^c-t_j^b\le4+11+4=19\). If \(p_{j+1}(t^c)<9/10\), inequality~\eqref{eq:clock-switch-drift} makes it increase at rate at least \(69/1000\) until its first crossing of \(9/10\), and it cannot drop below \(17/25\) during this interval. Since \(4(69/1000)>9/10-17/25\), this last stage costs at most five units, its last grid step included. Therefore
\[
 t_{j+1}-t_j^b\le4+11+4+5=24<D_{\mathrm{clk}}.
\]
All the estimates only used the fixed window, which was already justified: the window does not depend on the unknown time of switching. This also proves that \(\tau_{j+1}(t_{j+1})\le C_D\eps<A_0\eps\).

For the preceding bit after \(t_j^b\), we use the inequality for a low bit instead of the one for erasure. When \((S_{j-1})_+\le1/100\), its source term satisfies \(7(S_{j-1})_+^2\le2(S_{j-1})_+\), so the recurrence used for~\eqref{eq:clock-explicit-old} holds on both sides of \(t_j^b\), and its envelope continues up to \(t_{j+1}\), with the same rate of decay. The preceding bit remains below \(1/10\) from \(d_j\) to \(t_{j+1}\), which also justifies the use of the estimate of its resting timer on this whole interval.

\myparagraph{Recovery of the entry conditions}
At \(t_{j+1}\), the new active bit is at least \(9/10\), its timer is between \(\eps/2\) and \(A_0\eps\), and the old opposite bit is at most \(\mu_0\). The old preceding bit satisfies
$$
 |p_{j-1}(t_{j+1})|
 \le C_{\mathrm{env}} e^{-\lambda_{\mathrm{clk}}(L-C_{\mathrm{dur}})}+C\eps^{37/20}+C\nu\le\mu_0,
$$
for small \(\eps\), because \(\lambda_{\mathrm{clk}}=3/8>1/4\), and the old opposite timer satisfies
$$
 \tau_{j+2}(t_{j+1})
 \le B_0\eps^{4/5}e^{-(4/5)(L-C_{\mathrm{dur}})}
       +(5/4)(\eps+2\nu)\le2\eps.
$$
The old preceding bit is below \(1/10\) from the grid time \(d_j\), with \(d_j-t_j\le D_1=13\). Starting the bound of its resting timer at \(d_j\), from the upper bound of the scalar box, we get
$$
 \tau_{j-1}(t_{j+1})
 \le(1+3\nu)e^{-(4/5)(L-C_{\mathrm{dur}}-D_1)}
       +(5/4)(\eps+2\nu)\le B_0\eps^{4/5},
$$
after choosing \(B_0\) sufficiently large. These are exactly the conditions \eqref{eq:clock-entry}, with indices shifted by one. Every remaining condition of smallness is a finite inequality of the form \(C\eps^b\le\eps^{b'}\) with \(b>b'\), or \(\nu L\le c\), and it is satisfied by one common positive \(\eps_0\). This completes the induction, for arbitrarily many phases.

To be clear, the closure uses three different mechanisms. The reset of the old bit absorbs
\(C\eps^{3/8}+C\eps^{37/20}+C\nu\) into \(\eps^{1/4}\), because all the exponents are strictly larger than \(1/4\). The reset of the opposite timer absorbs
\(C B_0\eps^{8/5}\) into the spare part of the bound \(2\eps\), once its stationary value \((5/4)\eps+O(\nu)\) is subtracted. In contrast, the reset of the preceding timer contains a term with exponent exactly \(4/5\), whose coefficient cannot be made smaller by shrinking \(\eps\). This is why \(B_0\) is chosen only after \(C_{\mathrm{dur}},D_1\), for example larger than
\(2e^{(4/5)(C_{\mathrm{dur}}+D_1)}+2\). Its appearance in the source of the old bit is then harmless, because this source has exponent \(37/20\). In other words, we distinguish the choices of coefficients from the choices of small parameters, and this closes the order of the quantifiers.
\end{proof}

\subsection{A cut that separates successive operations}

Set \(\vartheta=7/8\), and let \(t_j^{\mathrm{cut}}\) be the first grid time at least \(t_j+\vartheta L\) (the cut of phase \(j\)). For small \(\eps\), the lower estimate on durations gives \(t_j^{\mathrm{cut}}<t_j^b\). Put
\begin{equation}\label{eq:small-clock-mass}
 \Delta_\eps=(1+L)(\eps^{1/4}+\nu).
\end{equation}
In particular, for every fixed \(\alpha<1/4\), \(\Delta_\eps=o(\eps^\alpha)\) uniformly for \(0\le\nu\le\eps^2\).

\begin{lemmarep}[Cut separation]\label{lem:cut-separation}
On every phase \(I_j=[t_j,t_{j+1})\) of the exact or perturbed clock,
\begin{align}
 \int_{t_j}^{t_j^{\mathrm{cut}}}|p_{j+1}|\,d_\eta t&\le C \Delta_\eps,
 &\int_{t_j^{\mathrm{cut}}}^{t_{j+1}}|p_{j-1}|\,d_\eta t&\le C \Delta_\eps,
 \label{eq:cut-masses}\\
 \int_{t_j}^{t_{j+1}}|p_{j+2}|\,d_\eta t&\le C \Delta_\eps,
 &\int_{t_j}^{t_{j+1}}|p_{j-1}|\,d_\eta t&\le C.
 \label{eq:opposite-old-masses}
\end{align}
The active gate is at least \(9/10\) on an interval of length at least \(L-C\), starting at the entry of the phase.
\end{lemmarep}
\begin{proofsketch}
We use the envelope of the growing timer before the cut, and the envelope of the decaying preceding bit after it. The choice \(\vartheta=7/8\) satisfies both \(4(1-\vartheta)>1/4\) and \((3/8)\vartheta>1/4\), and these two strict inequalities separate the operations with a positive power of \(\eps\).
\end{proofsketch}
\begin{proof}
Write \(u=t-t_j\). Since \(\nu L=o(1)\), the envelope of the timer in \eqref{eq:clock-next} gives
\(\tau_j(t)\le C'\eps e^u\) before \(t_j^{\mathrm{cut}}\). Therefore
$$
 \int_{t_j}^{t_j^{\mathrm{cut}}}|p_{j+1}|\,d_\eta t
 \le C L(\eps^{1/4}+\nu)+C\eps^4 e^{4(\vartheta L+1)}
 \le C \Delta_\eps+C\eps^{1/2}.
$$
The weighted sums are bounded with Lemma~\ref{lem:weighted-grid}, whose constants are uniform, even for nonuniform meshes bounded by one. On the late interval, \eqref{eq:clock-old-far} gives
$$
 \int_{t_j^{\mathrm{cut}}}^{t_{j+1}}|p_{j-1}|\,d_\eta t
 \le C\eps^{\lambda_{\mathrm{clk}}\vartheta}+CL(\eps^{37/20}+\nu),
 \qquad \lambda_{\mathrm{clk}}\vartheta=21/64>1/4.
$$
Both bounds are at most a constant times \(\Delta_\eps\). The estimate on the opposite gate follows from its pointwise bound and from the duration \(O(L)\) of the phase. Integrating the envelope of the old gate from zero gives a constant plus \(CL(\eps^{37/20}+\nu)\), which is uniformly bounded. The last assertion follows from the lower bound on durations and from the bounded terminal window of the clock lemma.
\end{proof}

\begin{corollary}[Double overlap, including perturbations]\label{lem-double-overlap}\label{lem:perturbed-overlap}
For \(P_{j+1}(t)=\int_{t_j}^{t}|p_{j+1}(u)|\,d_\eta u\),
$$
 \int_{t_j}^{t_{j+1}}|p_{j-1}(t)|
       \min(1,P_{j+1}(t))\,d_\eta t\le C \Delta_\eps.
$$
In particular this is \(O(\eps^\alpha+L\nu)\) for every fixed \(\alpha<1/4\).
\end{corollary}
\begin{proof}
Before the cut, \(P_{j+1}\le C\Delta_\eps\), and the old gate has a bounded total mass. After the cut, we bound the minimum by one, and we use the small late mass of the old gate. These are exactly the two estimates of Lemma~\ref{lem:cut-separation}.
\end{proof}

\subsection{Comparators: purification, retention and preparation}

The next lemma makes explicit the three different uses of a comparator. The symbol \(u\) denotes the value of its test, which depends on time: it is not a new variable of the state.

\begin{lemmarep}[Comparator estimates]\label{lem-sign-amplifier}\label{lem:comparator-control}
Consider
$$
 v^+=v+\eta\left[(1-p)\frac{v-v^3}{2}+p(u-v)\right]+\eta\xi_v,
 \quad |\xi_v|\le\nu,
$$
on the scalar box, with \(|u|\le1\).

\myparagraph{Cone preservation}
For any fixed positive margin constant of the test, suppose that \(u\ge m>0\) on an interval. If \(0<r\le m/4\), \(r\le1/4\) and \(\nu\) is smaller than a fixed constant times \(r\), then the region \(v\ge r\) is invariant. The reflected assertion holds for a negative test.

\myparagraph{Purification during a comparison phase}
Suppose that \(u\ge m\ge c h\), \(v(0)\ge c'h\), and that the tracking gate satisfies the envelope of the old gate of Lemma~\ref{lem-uniform-clock}. For every fixed \(K\ge256\), at the time \(\vartheta L+O(1)\) of the cut,
$$
 |v-1|\le C\eps^{1/4}.
$$
After a bounded initial delay, it takes at most a time \(9\log(1/h)+C\) to reach \(v\ge1/2\).

\myparagraph{Retention under small tracking mass}
If \(|v(a)-\sigma|\le d\), \(\sigma\in\{-1,1\}\), and
\(d+\int_a^b|p|\,d_\eta t+(b-a)\nu\) is sufficiently small, then
$$
 \sup_{a\le t\le b}|v(t)-\sigma|
 \le C\left(d+\int_a^b|p|\,d_\eta t+(b-a)\nu\right).
$$
Here we need no assumption on the sign of \(u\).

\myparagraph{Preparation under an open tracking gate}
If \(p\ge9/10\) and \(u\ge m\ge ch\), then, starting from any value of the comparator in its scalar box, a time at most \(2\log(1/h)+C\) suffices to get \(v\ge m/2\). As long as \(u\ge m\) holds, the resulting smaller cone (for example \(v\ge m/4\)) is preserved, even if the tracking gate changes arbitrarily in its permitted interval.
\end{lemmarep}
\begin{proofsketch}
The unit-step amplifier \(g(v)=(3v-v^3)/2\) repels zero, and it attracts the appropriate sign. A positive target gives an invariant lower barrier, and near a sign, the amplifier cannot increase the error until the error is macroscopic. For the preparation of a new sign, an open tracking gate overcomes even a comparator saturated with the opposite sign. Each argument uses physical time, and it survives an additive error \(O(\eta\nu)\).
\end{proofsketch}
\begin{proof}
We reflect \((u,v)\) if necessary. The constants of this proof may depend on the fixed margin constants. Excursions of the gate outside \([0,1]\) change the drift on the bounded scalar box by at most \(C\nu\), and we account for them together with \(\xi_v\).

\myparagraph{Invariant cone}
For \(0<r\le1/4\), \(g(r)\ge(47/32)r\). On \([r,1+4\nu]\), the minimum of \(g\) is \(g(r)\) for small \(\nu\), since the value near the other endpoint is close to one. The unit-step target of the comparator, with the gate clipped to \([0,1]\), is a convex combination of \(g(v)\) and of \(u\ge4r\), so it is at least \(r+(15/32)r\). Including the disturbance, and then taking the convex combination for step \(\eta\), we get
$$
 v^+\ge r+\eta\bigl((15/32)r-C\nu\bigr)\ge r.
$$
This proves the barrier, without any accumulation of a small loss over many steps.

\myparagraph{Growth and saturation}
The envelope of the old gate is below \(1/11\) after a bounded delay, and until then, the preceding barrier preserves a cone \(\Omega(h)\). For \(0<v\le1/2\),
$$
 v^+-v\ge\eta v\left(\frac38-\frac{11}{8}\bar p\right)-C\eta\nu,
 \qquad \bar p=\min(1,\max(0,p)).
$$
When \(\bar p\le1/11\) and \(v\ge C'\nu\), this is at least \(\eta v/8\), and the latter threshold is automatic in the preserved cone, because \(\nu\le h^{2K}=o(h)\). Hence the time to reach \(1/2\) is at most
\(\log(1/(2c'h))/\log(9/8)+C\le9\log(1/h)+C\).
The barrier at \(1/2\) is also invariant while \(\bar p\le1/11\), since the positive drift of the amplifier there strictly dominates the worst drift of tracking.

Put \(e=(1-v)_+\) once \(v\ge1/2\). After weakening the rate slightly, the cubic identity gives
$$
 e^+\le(1-\eta/3)e+C\eta(|p|+\nu).
$$
The part with \(v>1\) is bounded by the scalar box. If \(t_*\le9\log(1/h)+C\) is the hitting time, the convolution with the envelope of the old gate gives
$$
 |1-v(t)|\le C e^{-(t-t_*)/3}
       +C e^{-(t-t_j)/3}+C(\eps^{37/20}+\nu).
$$
At the cut, its first exponent in \(\eps\) is at least
$$
 \frac{\vartheta}{3}-\frac3K
 \ge\frac7{24}-\frac3{256}=\frac{215}{768}>\frac14.
$$
All the constants and the bounded delays are absorbed by decreasing \(\eps_0\). This proves purification.

\myparagraph{Retention}
Inside the half-neighborhood \(|v-\sigma|\le1/2\), the pure amplifier points toward \(\sigma\), and its Euler map cannot increase this distance. The tracking term and the disturbance increase it by at most \(C\eta(|p|+\nu)\), and the same estimate holds for the part \(O(\nu)\) outside \([-1,1]\), by the bound on the derivative of the polynomial. We sum the increments, and we consider a first exit from the half-neighborhood: the assumption of smallness in the statement prevents this exit.

\myparagraph{Tracking a new sign}
For \(v\le0\), \(p\ge9/10\), and \(u\ge m\), the drift obeys
$$
 v^+\ge(1-17\eta/20)v+(9/10)\eta m-C\eta\nu.
$$
The homogeneous factor is nonnegative, and the iterates are bounded below by a negative term decaying exponentially, plus a positive constant multiple of \(m\). Since \(\nu=o(m)\), zero is crossed within a time \(2\log(1/h)+C\). When \(0\le v\le m/2\), the drift of tracking is at least \((9/20)m\), and the amplifier is nonnegative, apart from the correction \(O(\nu)\) already accounted for. Hence \(v\ge m/2\) is reached after a bounded additional time. The argument of cones, with the lower barrier \(m/4\), proves the last assertion.
\end{proof}

\subsection{The uniform cycle lemma}

Fix a separated affine program as in Definition~\ref{def:affine-program}. Its tests are normalized on a fixed neighborhood of the convex hull of the valid states, with margin constants \(r_0,\kappa>0\), and we put \(\kappa_0=\kappa/16\). For the instance with stacks, the constants are the ones given by the two-stack normal form.

Let \(A>0\) be the section constant chosen in the cycle lemma below. For the level \(\nu\) of disturbances, a state at the entry of a cycle is \emph{admissible above \(C\)} if the scalar box and the entry conditions of the clock for phase zero hold, and
\begin{equation}\label{eq:cycle-admissible}
 \|Y-C\|_\infty\le r_0h,\qquad
 \|Y^{\mathrm s}-Y\|_\infty\le A\Delta_\eps,\qquad
 v_j u_j(Y,h)>0,\quad |v_j|\ge\kappa_0h.
\end{equation}
The initialization has identical data and shadows. The constant \(A\) only bounds the gap between shadows and data, and the separate constant \(B\) below bounds the defect on the data: we require neither equality nor contraction between these two constants.

\begin{lemmarep}[Mesh-uniform cycle, including perturbations]\label{lem-uniform-cycle}\label{lem:robust-cycle}
For every fixed integer \(K\ge256\), there are constants \(A,B,s_0>0\) such that, for \(s\ge s_0\), \(\eps=h^K\), \(0\le\nu\le\eps^2\), and every \(\eta\in(0,1]\), the following holds.

Let the cycle start at an admissible state above \(C\), whose source sign pattern is \(\sigma\). Set \(Y_0=Y(t_{4c})\) (the cycle has index \(c\)) and \(Y_*=D_\sigma Y_0+b_\sigma\), and assume
$$
 \|Y_*-C^+\|_\infty\le r_0h/2
$$
for the next source configuration \(C^+\). At the next section,
\begin{align}
 \|Y(t_{4c+4})-Y_*\|_\infty&\le B\Delta_\eps,
 \label{eq:cycle-defect-new}\\
 \|Y^{\mathrm s}(t_{4c+4})-Y(t_{4c+4})\|_\infty&\le A\Delta_\eps.
 \label{eq:cycle-shadow-new}
\end{align}
The conditions on scalars, on the clock and on the cones of comparators are restored. During the whole cycle, both the data and the shadows stay within \(B\Delta_\eps\) of the segment \([Y_0,Y_*]\), and all the active coordinates stay in a fixed compact box. A cycle has physical duration \(O(L)\), uniformly in the mesh.
\end{lemmarep}
\begin{proofsketch}
We use the lemma of cut separation twice. In phase zero, the loading gate has a negligible mass until the comparators are purified, and afterwards, the copy gate has a negligible mass. In phase two, the roles are reversed, which protects the completed value of the shadows before the data start changing. Phase three contracts the difference between shadows and data, independently of its incoming error, and it prepares the next cones of comparators. Hence the outgoing gap of the shadows is controlled independently of its incoming value: we choose its section constant first, and then the constant for the defect on the data.
\end{proofsketch}
\begin{proof}
We relabel the four phases as \(I_0,I_1,I_2,I_3\), their entries as \(t_0,\ldots,t_4\), and their cuts as \(t_0^{\mathrm{cut}},\ldots,t_3^{\mathrm{cut}}\). We write \(W=Y^{\mathrm s}\), \(e_0=\|W(t_0)-Y_0\|\) and \(d_\eps=e_0+\Delta_\eps\). We first prove the estimates under the provisional condition \(d_\eps\le c_{\mathrm{loc}}h\), for a sufficiently small positive constant \(c_{\mathrm{loc}}\) depending only on the fixed program and on \(K\). In the intermediate estimates, the constants depend neither on \(c_{\mathrm{loc}}\) nor on the coefficient used to bound \(e_0\). In particular, we get a bound \(C_{\mathrm{loc}}d_\eps\) on the data and on segments, and a bound \(C_{\mathrm{gap}}\Delta_\eps\) on the gap of the shadows. These constants may depend on the fixed program, on its normalized tests, on the compact neighborhood and on the clock, but they are independent of \(A\). At the end, we choose
\begin{equation}\label{eq:cycle-constant-order}
 A=2C_{\mathrm{gap}},\qquad
 B\ge\max\{1,C_{\mathrm{loc}}(A+1)\},\qquad
 (A+1)\Delta_\eps\le c_{\mathrm{loc}}h.
\end{equation}
The last inequality is enforced by choosing \(s_0\) last, since \(\Delta_\eps/h\to0\) uniformly for \(\nu\le\eps^2\). Hence the provisional condition is a consequence of admissibility, and not an additional invariant assumed without justification.

We work until the first exit of \((Y,W)\) from the neighborhood of radius \(1/2\) of \([Y_0,Y_*]^2\). For large \(s\), both endpoints belong to \([-1/4,5/4]^m\), so this neighborhood lies inside the fixed box \(\mathcal Q=[-2,3]^m\) of the data. On \(\mathcal Q\times[-2,2]^g\), let \(M_{\mathcal T}\) bound \(\|\mathcal T(Y,v)\|\), and let \(L_u=\max_j\|\partial_Yu_j\|_1\): both are fixed constants depending on the program. The target is uniformly Lipschitz there, the tests have modulus at most one, and the scalar box applies. By Lemma~\ref{lem-lock}, clipped gates give ordinary convex locks, with an effective additive disturbance at most \(C\nu\). Similarly, clipping \(v\) to \([-1,1]^g\) changes \(\mathcal T\) by at most \(C\nu\). These are only analytical estimates: the actual field is unchanged. Every bound below is also valid at the first update that may exit, since it uses the preceding state, which is in the neighborhood. These bounds will place \(Y,W\) within \(Cd_\eps\) of the segment, strictly inside this neighborhood, when \(c_{\mathrm{loc}}\) and then \(h\) are small enough.

\myparagraph{Phase zero, before the cut}
The mass of the loading gate on \([t_0,t_0^{\mathrm{cut}})\) is at most \(C\Delta_\eps\), so the variation of the shadows is at most \(C\Delta_\eps\), including the term \(CL\nu\) of disturbances. The data form a lock toward \(W\), and initially \(\|W-Y_0\|=e_0\). Formula \eqref{eq:lock-kernel} therefore gives
\begin{equation}\label{eq:phase0-freeze}
 \sup_{t_0\le t\le t_0^{\mathrm{cut}}}\|Y(t)-Y_0\|\le C d_\eps.
\end{equation}
Notice that this does not require the old copy gate to be small pointwise: its target is already close to the data. The entry condition gives \(\sigma_j u_j(Y_0,h)\ge\kappa h\), and by affineness and~\eqref{eq:phase0-freeze}, we get, on the whole interval,
\[
 \sigma_j u_j(Y(t),h)\ge\kappa h-L_u C d_\eps
 \ge\kappa h/2,
\]
after choosing \(c_{\mathrm{loc}}\) small enough. This argument also covers an incoming \(Y_0\) on the boundary of the admissible tube: we do not assume that the moving data remain in this tube before proving their margins for the tests. Consequently, the estimates on cones and on purification of Lemma~\ref{lem:comparator-control} give
$$
 \|v(t_0^{\mathrm{cut}})-\sigma\|_\infty\le C\eps^{1/4}.
$$
The bound on cones used here is uniform up to the cut, including the bounded interval before the old gate closes.

\myparagraph{Phase zero after the cut, and phase one}
The remaining mass of the copy gate \(p_3\) in phase zero is at most \(C\Delta_\eps\). In phase one, it is the opposite gate, again with mass at most \(C\Delta_\eps\). The data therefore move by at most \(C\Delta_\eps\) after \eqref{eq:phase0-freeze}, and the retention of the comparators on the same interval gives
$$
 \sup_{t_0^{\mathrm{cut}}\le t\le t_2}\|v(t)-\sigma\|\le C \Delta_\eps.
$$
Since the selectors sum to one at \(v=\sigma\), and select precisely the affine branch \(\sigma\), the Lipschitz continuity of polynomials on the fixed box gives
\begin{equation}\label{eq:correct-loading-target}
 \|\mathcal T(Y(t),v(t))-Y_*\|\le C d_\eps,
 \qquad t_0^{\mathrm{cut}}\le t\le t_2.
\end{equation}
On phase one, the loading gate is at least \(9/10\) during a physical time at least \(L-C\), and the kernel of the open lock gives
\begin{equation}\label{eq:shadow-loaded}
 \|W(t_2)-Y_*\|\le C e^{-(9/10)(L-C)}+Cd_\eps\le C d_\eps.
\end{equation}
The bounded terminal interval cannot invalidate this estimate, because its target still satisfies \eqref{eq:correct-loading-target}. Before the contraction is complete, this target and the initial shadow lie near the two endpoints of the segment, and the version of the lock lemma for convex segments gives the claimed bound on segments during these intervals.

\myparagraph{Phase two, before the cut}
Here, the copy gate \(p_3\) is the next gate, and its mass before \(t_2^{\mathrm{cut}}\) is at most \(C\Delta_\eps\). Hence the data still satisfy \(\|Y-Y_0\|\le Cd_\eps\), and the retention of comparators still gives \(\|v-\sigma\|\le C\Delta_\eps\), so the target in \eqref{eq:correct-loading-target} remains accurate. Starting from \eqref{eq:shadow-loaded}, a lock with such targets keeps \(W\) within \(Cd_\eps\) of \(Y_*\), whatever the total mass of the old loading gate.

\myparagraph{Phase two, after the cut}
The remaining mass of the loading gate is now at most \(C\Delta_\eps\). Its target may change arbitrarily when the data and the comparators move, but the estimate on the variation with a bounded target shows that \(W\) changes by only \(C\Delta_\eps\). In particular,
\begin{equation}\label{eq:handover-shadow}
 \sup_{t_2\le t\le t_3}\|W(t)-Y_*\|\le Cd_\eps.
\end{equation}
The copy lock moves \(Y\) toward this shadow, which is already accurate, and \(Y\) stays near \([Y_0,Y_*]\) (we do not claim that it remains frozen during phase two).

\myparagraph{Phase three, copy accuracy}
The loading gate is the opposite gate on this phase, with mass \(O(\Delta_\eps)\), so \eqref{eq:handover-shadow} remains valid up to \(t_4\). The copy gate is at least \(9/10\) during a time at least \(L-C\), and the kernel of its lock gives
$$
 \|Y(t_4)-Y_*\|
 \le C e^{-(9/10)(L-C)}+Cd_\eps
 \le C(e_0+\Delta_\eps)\le B\Delta_\eps.
$$
The last constant \(B\) is fixed after the machine, \(K\) and the section constant \(A\).

\myparagraph{A stronger bound on the shadow difference}
Let \(d=W-Y\) on phase three. We have exactly
$$
 d^+=(1-\eta p_3)d+\eta p_1(\mathcal T(Y,v)-W)
                    +\eta(\xi_W-\xi_Y).
$$
After clipping the gate in this estimate, the forcing has a total norm at most \(C\Delta_\eps\), and the initial \(d\) is bounded by a fixed constant. The homogeneous product over the long interval where the copy is open is at most \(C e^{-(9/10)L}\), and over the bounded terminal interval, its absolute value is at most \(e^{C\nu D_{\mathrm{clk}}}\). Consequently,
$$
 \|d(t_4)\|\le C e^{-(9/10)L}+C\Delta_\eps\le C_{\mathrm{gap}}\Delta_\eps.
$$
The coarse bound on the initial \(d(t_3)\) follows from \(Y,W\in\mathcal Q\), and the bound on the forcing uses \(M_{\mathcal T}\) and the masses of the clock. Hence \(C_{\mathrm{gap}}\) depends on the fixed program and on the compact box, but not on \(e_0\) or \(A\): it can indeed be chosen before \(A\). The choice of the constants in~\eqref{eq:cycle-constant-order} proves \eqref{eq:cycle-shadow-new}, with slack, after increasing \(s_0\). The estimate on the data may multiply the incoming gap by \(\|D_\sigma\|\), and its larger constant is allowed, because it is \(B\), and not \(A\). This is the mechanism that closes a single scale of errors, without a circular choice of constants.

\myparagraph{Preparation of the new comparator cones}
On phase three, after a time at most \(3\log(1/h)+C\), the estimate on the copy gives \(\|Y-Y_*\|\le C h^{27/10}+C d_\eps\le r_0h/4\), by decreasing \(c_{\mathrm{loc}}\) and then increasing the scale threshold. The assumption on \(Y_*\) and the margin of the source tests then imply that every new test has a fixed sign, and a modulus at least \(m=\kappa h/2\), and this remains true up to \(t_4\). While the copy gate is open, the preparation of the comparators takes a further time \(2\log(1/h)+C\), and it gives a modulus at least \(m/2\). The combined time is \(5L/K+C\), strictly shorter than \(L-C\) for the chosen \(K\) and sufficiently large \(s\). The barrier of cones with \(r=\kappa_0h\le m/4\) then applies up to the terminal transition of the clock, with the same fixed \(\kappa_0\). Observe that no unspecified multiplicative loss on the constant of cones is iterated.

The clock lemma restores its entry conditions. The bounds on segments in all four phases close the argument of first exit from the compact neighborhood, and their proof only used \(Cd_\eps<1\), \(Cd_\eps<ch\) and the scalar estimates already proved. The duration of the clock gives a physical time \(O(L)\). Hence all the conclusions, including the ones needed for the next cycle, hold uniformly in the sequence of perturbations and in the mesh.
\end{proof}

\subsection{Accumulation and the safety tube}

Let
$$
 D_* = \max(1,\max_\sigma\|D_\sigma\|_\infty),
 \qquad r_* = r_0/(4D_*).
$$
In particular, an error at most \(r_*h\) at a section puts its ideal affine image within \(r_0h/4\) of the next source configuration, strictly inside the hypothesis of the cycle lemma.

\begin{lemmarep}[Accumulation with an interval-product bound]\label{lem-P-accumulation}
Consider a valid prefix of \(T\) transitions of a separated affine program, with interval gain bound \(\Gamma\ge1\). If
\begin{equation}\label{eq:general-error-budget}
 B\Gamma T\Delta_\eps\le r_*h,
\end{equation}
then each cycle of this prefix is admissible, for the exact or the perturbed scheme, and
\[
 \|Y(t_{4k})-C_k\|_\infty\le B\Gamma k\Delta_\eps\le r_*h.
\]
For the stack normal form, one can use \(\Gamma=G/h\), so the condition becomes
\begin{equation}\label{eq:global-error-budget}
 GB T\Delta_\eps/h^2\le r_*.
\end{equation}
For a polynomial \(T=T(s)\), this holds for all sufficiently large \(s\) whenever \(K/4>2\), in addition to the local requirement \(K\ge256\).
\end{lemmarep}
\begin{proofsketch}
We maintain the smaller tube \(r_*h\), which gives the hypothesis of the local cycle lemma on the ideal image, and we expand the error as a sum of local defects propagated by interval products. The bound uses the gain of these products: we never multiply bounds on the norms of individual matrices.
\end{proofsketch}
\begin{proof}
The initial source coordinates are exact. Suppose that the conclusions hold up to section \(k\). By~\eqref{eq:interface-margin}, the tests have there the source sign pattern, and the ideal affine image is within \(D_*r_*h\le r_0h/4\) of \(C_{k+1}\). The local cycle lemma applies, and it gives
\[
 Y_{k+1}=D_{\sigma_k}Y_k+b_{\sigma_k}+\delta_k,
 \qquad \|\delta_k\|_\infty\le B\Delta_\eps,
\]
with all the conditions on the clock, scalars, shadows and cones restored. Subtracting the exact recurrence, we get
\[
 Y_{k+1}-C_{k+1}
 =\sum_{i=0}^k D_{\sigma_k}\cdots D_{\sigma_{i+1}}\delta_i.
\]
The empty product is the identity, with norm at most \(\Gamma\), and every other product is bounded by the assumed interval gain. Hence the error is at most \(B\Gamma(k+1)\Delta_\eps\), inside the smaller tube by~\eqref{eq:general-error-budget}. This closes the induction. Notice that the agreement of the actual and ideal branches is obtained at each step: it is not assumed for the whole approximate execution.

For stacks, Lemma~\ref{lem-cumulative-gain} gives \(\Gamma=G/h\). Since \(\nu\le\eps^2\le\eps^{1/4}\),
\begin{equation}\label{eq:intrinsic-budget}
 \Delta_\eps\le2(1+Ks\log2)h^{K/4}.
\end{equation}
For a polynomial \(T(s)\), the relative budget is bounded by a constant times \(T(s)(1+s)2^{-(K/4-2)s}\), which tends to zero. Observe that we keep the logarithmic factor everywhere.
\end{proof}

\begin{propositionrep}[Resetting halt and an infinite correct tail]\label{prop-contracting-halt}
Suppose that every halting branch is the constant map \(Y\mapsto C_{\mathrm{halt}}\). For sufficiently large \(s\), once a halting cycle is completed, all the subsequent sections are admissible above \(C_{\mathrm{halt}}\), with error at most \(B\Delta_\eps\), and at every subsequent micro-step, the data are within \(C\Delta_\eps\) of the same source configuration. This holds for all perturbations of the permitted size.
\end{propositionrep}
\begin{proofsketch}
The ideal image no longer carries the incoming error on the data, because its linear part is zero. The cycle lemma can therefore be applied indefinitely, without a growing sum of past defects. Its estimate on segments controls the whole trajectory between sections, which we need for a stable readout.
\end{proofsketch}
\begin{proof}
At a halting branch, \(Y_*=C_{\mathrm{halt}}\) exactly. The cycle lemma returns a section with an error on the data at most \(B\Delta_\eps\), and it restores the other conditions of admissibility. Since \(\Delta_\eps=o(h)\), this error on the data is at most \(r_*h\) for large \(s\). We apply the same argument again at the next cycle, and no previous error on the data appears in its ideal image.

Between two such sections, the segment in the cycle lemma is
\([Y_k,C_{\mathrm{halt}}]\). Its endpoints are within \(B\Delta_\eps\) of \(C_{\mathrm{halt}}\), and the trajectory is within \(B\Delta_\eps\) of it, so the distance during the whole cycle is at most \(C'\Delta_\eps\). The output is the accepting control bit, so it stays within \(C'\Delta_\eps\) of its exact Boolean value at every micro-step, and not only at the sections of cycles. The same argument covers arbitrarily many disturbed cycles.
\end{proof}

\subsection{Completion of the polynomial-time proof}

\begin{proof}[Proof of Theorem~\ref{thm:main-P}]
The implication \(2\Rightarrow1\) is Lemma~\ref{lem-truncated-simulation}, and \(3\Rightarrow2\) follows by taking \(\eta=1\). For \(1\Rightarrow3\), we choose the two-stack normal form with a polynomial \(T(s)\), and we fix an integer \(K\ge256\), with the constants of the unified cycle lemma. We increase the polynomial threshold so that \(s\ge4\), the readers of the input are valid, all the scalar estimates and the estimates on cycles apply, and \eqref{eq:global-error-budget} holds.

The initialization is an admissible section: the source data are exact, the shadows are equal to the data, the timers are equal to \(h^K\), and the comparators are equal to the values of their normalized affine tests. Accumulation therefore simulates the complete source computation, with the correct branches. The duration of one cycle is at most \(C_K(1+s)\), for a fixed constant \(C_K\), and a fixed integer polynomial
$$
 Q(s)=C(1+s)(T(s)+1)
$$
can be chosen larger than the time at which the halting cycle is completed, uniformly in \(\eta\). The proposition on the resetting halt then gives a compact active trajectory and an infinite correct tail. We choose the threshold still larger, so that \(C'\Delta_\eps\le1/4\). Then, at every grid index \(k\ge\lceil Q(s)/\eta\rceil\), the output decides the language.

All the coefficients of the field and of the initialization are rational. In the constructed initialization, the output coordinate is zero, so the translation to a raw state preserves the output coordinate itself. The scale threshold is a polynomial in the length of the input, and not a function of its contents. Notice that no computation is hidden in a parity of the time of observation, or in the digits of a real coefficient. This completes the characterization.
\end{proof}

\begin{propositionrep}[Active length up to a specified horizon]\label{prop-P-length}
One cycle of the constructed dynamics has active length \(O(1+s)\), uniformly in the mesh, and in the polynomial-time case, the prefix up to the decision horizon \(Q(s)\) has a polynomial active length. These bounds also hold for the perturbed trajectories, with the length defined by their actual increments.
\end{propositionrep}
\begin{proofsketch}
After the polynomial normalization of the input, the core field only depends on the bounded active state and on the frozen scale, so its speed is uniformly bounded. We multiply this bound by the physical duration, and not by the number of micro-steps.
\end{proofsketch}
\begin{proof}
In the normalized coordinates, the core is \(F(h,Z)\), which does not depend on the unbounded raw integer. The permitted scales and the active states belong to a fixed compact set, so \(\|F(h,Z)\|_\infty\le M\) there, for a fixed \(M\). A disturbed step has norm at most \(\eta(M+\nu)\). Consequently, a cycle of duration at most \(C_K(1+s)\) has length at most \((M+1)C_K(1+s)\), and the prefix ending at the first grid time after \(Q(s)\) has length at most
$$
 (M+1)(Q(s)+1).
$$
This is polynomial in \(s\) when \(Q\) is polynomial. The translation to a raw initial state does not change the increments, so it preserves the length. Of course, we do not bound the length of the infinite tail: the autonomous clock keeps cycling after the decision, and this tail need not have a finite length.
\end{proof}

\begin{corollary}[Nonuniform temporal meshes]\label{cor:variable-mesh}
The constructed field also computes under any sequence of temporal steps \(0<\eta_k\le1\) with \(\sum_k\eta_k=\infty\), with the same bounds in physical time \(t_k=\sum_{i<k}\eta_i\). The permitted disturbances have norm at most \(\eta_k h^{2K}\) at step \(k\).
\end{corollary}
\begin{proof}
Every estimate on boxes, locks and scalars is an inequality on one step, valid for every step at most one. For the nonnegative factors of contraction used in the proofs, the products are at most \(e^{-\lambda\sum\eta_k}\). For the factors of growth, the concavity of \(\log\) and \(1+u\le e^u\) give the two different bounds
\[
 (1+\lambda)^{\sum\eta_k}
 \le\prod_k(1+\lambda\eta_k)
 \le e^{\lambda\sum\eta_k},\qquad \lambda\ge0.
\]
The lower bound gives the upper bound on the waiting times for growth, and the upper bound gives the lower bound on these waiting times. All the overshoots of hitting times are at most one. The proofs on the clock, on cut separation and on cycles therefore hold with the weighted sums over this nonuniform grid, and the divergence of the physical time ensures that every required phase is reached after finitely many indices. Soundness for the associated mesh-uniform class still uses the unit mesh: we need no assumption on the computation of an individual arbitrary mesh.
\end{proof}

\section{Robust polynomial-space simulation}\label{sec:proof-PSPACE}

The estimates on the box, the clock, the separation by cuts, the comparators and the cycles were proved above for every \(0\le\nu\le\eps^2\), so the argument for space does not require a second, schematic, repetition of these proofs. We now fix \(\nu=\eps^2=h^{2K}\), and we verify the bounds on resources and the infinite halting tail.

\begin{propositionrep}[Finite precision removes the need for a horizon]\label{prop:eventual-PSPACE}
An eventually robust unit-step system in the sense of Definition~\ref{def:eventual-robust}, with polynomial threshold, decides a language in \(\PSPACE\).
\end{propositionrep}
\begin{proofsketch}
Deterministic rounding turns one permitted execution into an orbit on a finite set of states. Eventual correctness forces all the states of its eventual cycle to carry the correct answer. In other words, a bound on the size of the space of rounded states replaces a bound on the real time of stabilization.
\end{proofsketch}
\begin{proof}
Fix the witnessing system and an input of length \(n\). We set \(s=S(n)\), after enlarging the integer polynomial threshold so that \(s\ge n+1\), and we choose an integer \(r_1\ge r\) and \(b=r_1s+3\). We keep \(x,h\) exact, and we compute \(z_0=c(x,h)\) exactly. At each step, we evaluate the complete rational polynomial update exactly, and we round each coordinate deterministically to the nearest multiple of \(2^{-b}\), with a fixed rule for ties. The error is smaller than \(h^r\), so the hypothesis allows this orbit, and in particular, it remains in the fixed compact box.

We choose an integer \(M\ge1\) such that the box is contained in \([-M,M]^m\). After the first update, the state belongs to
\[
 \mathcal D_b=2^{-b}\mathbb Z^m\cap[-M,M]^m,
 \qquad N_b:=|\mathcal D_b|=(2M2^b+1)^m.
\]
The frozen parameters are the same at every step, so deterministic rounding defines an autonomous map on the states visited by this orbit. Some state repeats among \(z_1,\ldots,z_{N_b+1}\), so \(z_{N_b+1}\) lies on the eventual periodic part, and each state of this part occurs at arbitrarily large indices. Eventual robust decision, applied to this particular sequence of disturbances, forces the output to lie within \(1/4\) of the correct answer at every state of this cycle. Hence reading \(z_{N_b+1}\), and comparing with \(1/2\), decides the input.

The exact initial vector has \(O(n+s)\) bits. Evaluating the polynomial of fixed degree from it, and from every subsequent rounded vector, uses a workspace \(O(n+s+b)=O(s)\): the rational coefficients have a fixed bit length, and every intermediate numerator and denominator has a bit length of this order. Compactness bounds the integer parts of the stored states. Both \(N_b\) and the counter up to \(N_b+1\) have \(O(b)=O(s)\) bits, because \(m,M\) are fixed. Hence the procedure uses space \(O(S(n))\).

Observe that the proof neither computes a time of stabilization of the original real orbit, nor proves a common exponential bound on stabilization for all real sequences of disturbances. It constructs an execution, bounded in space, that decides. We use no effective bound on the eventual index of Definition~\ref{def:eventual-robust}.
\end{proof}

\begin{corollary}[Robust soundness with a supplied horizon]\label{prop:PSPACE-sound}
A robust unit-step scale-controlled system with polynomial threshold, compact active state, noise tolerance \(h^r\), and decision time \(2^{cs}\) decides a language in \(\PSPACE\).
\end{corollary}
\begin{proof}
A common horizon implies eventual robust decision, so Proposition~\ref{prop:eventual-PSPACE} applies. One can also run the same rounded simulation directly up to \(2^{c_1s}\), for an integer \(c_1\ge c\): its state, the exact evaluation of the polynomial and the counter use space \(O(s)\), and the prescribed horizon guarantees the correct readout there.
\end{proof}

\begin{lemmarep}[Exponential accumulation]\label{lem:exponential-accumulation}
Suppose the source computation halts within \(2^{cs}\) transitions. If the fixed exponent satisfies
$$
 K/4>c+2,
$$
then, at all sufficiently large scales, every cycle before the halt of every permitted perturbed trajectory is admissible, and after the halting cycle, the decision is stable at every subsequent micro-step.
\end{lemmarep}
\begin{proofsketch}
The bound on interval products contributes one factor \(1/h\), and the number of transitions contributes \(h^{-c}\). Dividing the resulting error by the radius of the tube contributes a second factor \(1/h\), and the strict inequality on the exponent pays for these three effects. The resetting halt removes any accumulation on the infinite tail.
\end{proofsketch}
\begin{proof}
The unified cycle lemma gives a defect at most \(B\Delta_\eps\), also under the noise \(\nu=\eps^2\), with constants independent of the sequence of disturbances. Applying Lemma~\ref{lem-P-accumulation} with \(T=2^{cs}=h^{-c}\), and \eqref{eq:intrinsic-budget}, we get the relative bound on errors
$$
 \frac{GB}{h^2}\,2^{cs}\Delta_\eps
 \le2GB(1+Ks\log2)h^{K/4-c-2}.
$$
It is at most \(r_*\) for all large \(s\). Hence every hypothesis of the local cycle lemma is obtained by the induction on sections: none of them is assumed along the complete perturbed orbit. After the halt, Proposition~\ref{prop-contracting-halt} applies indefinitely, with the same level of noise, and it gives the assertion of stability at every micro-step.
\end{proof}

\begin{proof}[Proof of Theorem~\ref{thm:main-PSPACE}]
The implication \(2\Rightarrow1\) is Proposition~\ref{prop:eventual-PSPACE}, and \(3\Rightarrow2\) uses the unit mesh. The same proposition proves soundness when a horizon is also prescribed. Conversely, we take the polynomial-space normal form of Lemma~\ref{lem-hard-normal-form} for the source, and we choose one integer \(K\ge256\) with \(K/4>c+2\), where \(c\) is the exponent of the bound \(2^{cs}\) on the number of source configurations. The unified analytic lemmas and the exponential accumulation give a robust simulation for every \(\eta\in(0,1]\), with disturbances bounded by \(\eta h^{2K}\).

The physical time before the completion of the halting cycle is at most
$$
 C_K(1+s)(2^{cs}+1).
$$
A fixed integer \(c_0>c+1\) makes this at most \(2^{c_0s}\), after a final increase of the polynomial threshold, so the number of micro-steps is at most \(\lceil2^{c_0s}/\eta\rceil\). All the trajectories stay in the same compact box, and the accepting control bit remains correct forever after this time. This gives the stated characterization, with the precision exponent \(r=2K\). The proof with scalar noise is uniform in the number of the phase, so we need no additional precision just because the computation, or its halting tail, is long.
\end{proof}

\section{Common polynomial safety certificates}\label{sec:common-certificates}

In this section, we give the two ingredients used in Theorem~\ref{thm:common-certificates}: a converse theorem with a single certificate for every Euler step, and an exact bound on the complexity of checking this certificate. The converse is stated independently of the compiler. It assumes a common invariance under finite updates, with a reserve of admissible disturbance, together with a positive distance from the unsafe set and from the boundary of the box. The two examples below explain why safety in continuous time, and safety without a reserve, are not sufficient for this conclusion.

\begin{propositionrep}[Soundness in both temporal interpretations]\label{prop:common-certificate-soundness}
A common certificate in the sense of Definition~\ref{def:common-certificate} guarantees that \(U\) is avoided from every point of \(I\), for every variable Euler mesh and every measurable continuous forcing bounded by \(\delta\). All the Euler interpolants and all the continuous solutions remain in the compact set \(\{z\in\mathcal K:V(z)\le0\}\subset\operatorname{int}\mathcal K\).
\end{propositionrep}
\begin{proofsketch}
The discrete assertion follows by induction. Each partial step is an allowed update, so interpolation preserves the same sublevel set. Averaging a measurable forcing on shrinking intervals of the mesh, and taking Euler limits, we get the continuous assertion, and compactness gives global existence.
\end{proofsketch}
\begin{proof}
Let \(C=\{z\in\mathcal K:V(z)\le0\}\). Separation gives \(I\subset C\), \(C\cap U=\varnothing\) and \(C\cap\partial\mathcal K=\varnothing\), and \(C\) is compact, by continuity of polynomials. If \(z\in C\), condition~\eqref{eq:certificate-invariance} puts every allowed update back in \(C\), which proves the induction for every finite prefix. A point on the segment from \(z\) to \(z+\eta(f(z)+\xi)\) has the form \(z+\theta\eta(f(z)+\xi)\), with \(0\le\theta\le1\), and it is also in \(C\).

Fix any measurable forcing \(u\) bounded by \(\delta\), and any finite interval \([0,T]\). On a partition of maximal step \(\eta\), we use the averages of \(u\) as the Euler disturbances. The interpolants take values in \(C\), and they have the common bound \(M=\sup_{\mathcal K}\|f\|+\delta\) on their speed. Let \(L_f\) be a Lipschitz constant for \(f\) on the convex box \(\mathcal K\). The integral equations of the interpolants differ from
\[
 z(t)=z(0)+\int_0^t(f(z(v))+u(v))\,dv
\]
by at most \(L_fMT\eta+2\delta\eta\): the first term comes from freezing the state within each cell, and the second one from the last partial cell of the forcing. By compactness and equicontinuity, we get a uniformly convergent subsequence, whose limit solves this equation and remains in the closed set \(C\). The local Lipschitz continuity of \(f\) gives uniqueness for the fixed forcing. Repeating this for every \(T\), or using compactness to continue the solution, we get global existence and the claimed safety. Observe that we used no regularity of the forcing, beyond measurability and boundedness.
\end{proof}

\begin{lemmarep}[A polynomial converse common to all Euler steps]\label{lem:common-converse}
Let \(f:\R^d\to\R^d\) be polynomial, \(\mathcal K\) a compact rational box with nonempty interior, and \(\delta>0\). Suppose a nonempty compact set \(R\subset\operatorname{int}\mathcal K\) satisfies
\begin{equation}\label{eq:inflated-invariance}
 r+\eta(f(r)+u)\in R
 \quad\text{for every }r\in R,\quad
 0\le\eta\le1,\quad\|u\|_\infty\le2\delta.
\end{equation}
If \(I\subset R\) and \(\dist(R,U)>0\) for \(U\subset\mathcal K\), there is a common certificate \(V\in\Q[z]\) for \((f,\mathcal K,I,U,\delta)\), and it can be chosen so that, for every \(0\le\eta\le1\) and \(\|\xi\|_\infty\le\delta\),
\begin{equation}\label{eq:polynomial-boundary-decrease}
 z\in\mathcal K,\quad -\tfrac14\le V(z)\le0
 \quad\Longrightarrow\quad
 V\bigl(z+\eta(f(z)+\xi)\bigr)-V(z)\le-\tfrac12\eta.
\end{equation}
For this assertion of existence, neither \(R\) nor \(f\) needs a rational description. We claim no bound on the degree or on the size of the coefficients.
\end{lemmarep}
\begin{proofsketch}
The reserve of noise makes the distance to \(R\) decrease in a small neighborhood, under every allowed update. We smooth this distance by convolution, and in a thin band around a positive level, the decrease is at least a constant times \(\eta\). We then approximate in \(C^1\) by a rational polynomial: its error on the change over one update is also proportional to \(\eta\), so the same sublevel remains invariant, uniformly as \(\eta\to0\). Notice that a uniform approximation alone would not give this conclusion.
\end{proofsketch}
\begin{proof}
All the distances in this proof use the sup norm. Put \(D(z)=\dist(z,R)\), which is a globally \(1\)-Lipschitz function. We choose \(b>0\) such that every point at distance less than \(b\) from \(R\) is in \(\operatorname{int}\mathcal K\) and outside \(U\) (this follows from compactness and from the two positive distances, and if \(U\) is empty, only the condition on the boundary is needed). Let \(L_f\ge1\) be a Lipschitz constant for \(f\) on the enlarged convex box \(\mathcal K+[-1,1]^d\).

\myparagraph{The distance estimate}
For \(q\) in this enlarged box, we choose a nearest point \(r\in R\). By~\eqref{eq:inflated-invariance}, the closed ball of radius \(2\eta\delta\) around \(r+\eta f(r)\) is contained in \(R\). Consequently, for \(\|v\|\le2\delta\),
\begin{equation}\label{eq:distance-reserve}
 D\bigl(q+\eta(f(q)+v)\bigr)
 \le\max\{0,(1+L_f\eta)D(q)-\eta(2\delta-\|v\|)\}.
\end{equation}
Indeed, the ball of radius \(\eta(2\delta-\|v\|)\) around \(r+\eta(f(r)+v)\) lies in \(R\), and the centers of the two displayed updates differ by at most \((1+L_f\eta)\|q-r\|\). The distance to a ball for the sup norm is the excess of the distance to its center over its radius, or zero if this excess is negative.

Choose a positive rational number \(a_0\) so small that
\[
 a_0<\min\{\tfrac14,\ b/8,\ \delta/[8(L_f+1)]\},
 \qquad \rho=\sigma=a_0/16.
\]
Let \(\psi_\rho\) be a nonnegative smooth mollifier of integral one, supported in \([-\rho,\rho]^d\), and set
\[
 g(z)=\int D(z-w)\psi_\rho(w)\,dw.
\]
Then \(g\) is smooth, and \(|g-D|\le\rho\) everywhere. Consider \(z\in\mathcal K\) with \(g(z)\le a_0+\sigma\), an allowed disturbance \(\|\xi\|\le\delta\), and \(z'=z+\eta(f(z)+\xi)\). First, \(D(z)\le a_0+\sigma+\rho<2a_0\), and applying~\eqref{eq:distance-reserve} with \(q=z,v=\xi\), we get \(D(z')\le D(z)<2a_0<b\), since \(L_fD(z)<\delta\). Hence \(z'\in\operatorname{int}\mathcal K\).

For a vector \(w\) in the support of the mollifier, put \(q=z-w\). Since \(\rho<1\), this point is in the enlarged box on which \(L_f\) is valid, and the identity
\[
 z'-w=q+\eta\bigl(f(q)+v\bigr),\qquad
 v=\xi+f(z)-f(q),\quad \|v\|\le\delta+L_f\rho<2\delta
\]
allows us to use~\eqref{eq:distance-reserve} again. We have \(D(q)\le g(z)+2\rho<2a_0\), and
\(L_fD(q)+L_f\rho<3L_fa_0<\delta/2\). Hence
\begin{equation}\label{eq:shifted-distance-decrease}
 D(z'-w)\le\max\{0,D(z-w)-\eta\delta/2\}.
\end{equation}
By integration, \(g(z')\le g(z)\). If we also have \(g(z)\ge a_0-3\sigma\), then
\(D(z-w)\ge g(z)-2\rho\ge a_0/2\). Since \(\eta\le1\) and \(a_0\le\delta\),
\[
 \min\{D(z-w),\eta\delta/2\}\ge\eta a_0/2.
\]
Integrating~\eqref{eq:shifted-distance-decrease}, we now get the stronger estimate
\begin{equation}\label{eq:smooth-boundary-decrease}
 g(z')-g(z)\le-\eta a_0/2
 \quad\text{if }a_0-3\sigma\le g(z)\le a_0+\sigma.
\end{equation}
The estimates include the case \(\eta=0\), where both changes are zero.

\myparagraph{Rational polynomial approximation}
Let \(M\ge1\) bound \(\|f(z)\|+\delta\) on \(\mathcal K\). Choose \(q\in\Q[z]\) such that, on \(\mathcal K\),
\begin{equation}\label{eq:certificate-c1-approximation}
 |q-g|\le\sigma,\qquad
 \|\nabla q-\nabla g\|_1\le a_0/(4M).
\end{equation}
For completeness, we apply the tensor-product Bernstein approximation to \(\widetilde g=g\circ\mathcal A\), where \(\mathcal A\) is the affine map from \([0,1]^d\) onto the rational box \(\mathcal K\). The Bernstein values converge uniformly, by uniform continuity and by concentration of the binomial weights of sampling. The derivative in coordinate \(i\) is the Bernstein average, with degree lowered by one in this coordinate, of
\[
 N\bigl(\widetilde g((k+e_i)/N)-\widetilde g(k/N)\bigr)
   =\int_0^1\partial_i\widetilde g((k+t e_i)/N)\,dt.
\]
The uniform continuity of the derivatives of \(\widetilde g\) gives the uniform convergence of derivatives as well, since the points of sampling \((k+t e_i)/N\) concentrate at the point of evaluation, uniformly in \(t\in[0,1]\). Going back to \(\mathcal K\) preserves the \(C^1\) convergence, by the chain rule for the fixed affine map. Finally, for a fixed degree of approximation, there are finitely many coefficients, and replacing them by sufficiently close rationals perturbs the values and the derivatives by arbitrarily little on the compact box. Choosing each approximation with a strict slack, we get~\eqref{eq:certificate-c1-approximation}.

\myparagraph{Invariance of the polynomial sublevel}
Put \(V_0=q-a_0\). If \(z\in\mathcal K\) and \(V_0(z)\le0\), then \(g(z)\le a_0+\sigma\), so the preceding arguments already give \(z'\in\mathcal K\) and \(g(z')\le g(z)\). If \(V_0(z)\le-2\sigma\), we get
\[
 V_0(z')\le g(z')+\sigma-a_0
             \le g(z)+\sigma-a_0\le0.
\]
In the band \(-2\sigma\le V_0(z)\le0\), we have \(g(z)\ge a_0-3\sigma\). The segment from \(z\) to \(z'\) lies in the convex box \(\mathcal K\), and its length in the sup norm is at most \(\eta M\), so the bound on the derivative in~\eqref{eq:certificate-c1-approximation} gives
\[
 |(q-g)(z')-(q-g)(z)|\le\eta a_0/4.
\]
With~\eqref{eq:smooth-boundary-decrease}, we get
\(V_0(z')-V_0(z)\le-\eta a_0/4\). This proves the required invariance in both cases, uniformly for every step.

On \(I\subset R\), we have \(V_0\le\rho+\sigma-a_0=-7a_0/8\), and on \(U\cup\partial\mathcal K\), we have \(D\ge b>8a_0\), hence \(V_0>55a_0/8\). Set \(V=2V_0/a_0\), a rational polynomial. Then \(V\le-7/4\) on \(I\) and \(V>55/4\) on \(U\cup\partial\mathcal K\), which proves~\eqref{eq:certificate-separation}, and its sublevel is unchanged. The band \(-2\sigma\le V_0\le0\) becomes \(-1/4\le V\le0\), and the decrease becomes at least \(\eta/2\), which proves~\eqref{eq:polynomial-boundary-decrease}. In particular, dividing by \(\eta>0\) and taking the limit at zero, we get a Lie derivative at most \(-1/2\) on \(V=0\), uniformly over the allowed disturbances.
\end{proof}

\begin{example}[Continuous safety does not imply common Euler safety]\label{ex:flow-not-euler}
Take \(f(z)=-3z\), \(\delta=1/4\), \(I=R=[-1/2,1/2]\), \(\mathcal K=[-2,2]\), and \(U=\{z\in\mathcal K:|z|\ge1\}\). Under any measurable forcing \(|u|\le2\delta\), variation of constants gives
\[
 |z(t)|\le \tfrac12 e^{-3t}+\tfrac16(1-e^{-3t})\le\tfrac12.
\]
Hence the flow is robustly safe, in a compact set separated from \(U\). However, the unforced unit update sends \(z=1/2\) to \(-1\in U\), and no common certificate exists, even at the smaller level \(\delta\) of disturbance.
\end{example}

\begin{example}[The disturbance reserve cannot simply be removed]\label{ex:reserve-needed}
Take \(f(z)=(z^2-1)/4\), \(\delta=1/4\), \(I=[-1,0]\), \(R=[-\sqrt2,0]\), \(\mathcal K=[-2,2]\) and \(U=[1,2]\). For every \(0\le\eta\le1\) and \(|u|\le\delta\), the update \(\Phi_{\eta,u}(z)=z+\eta(f(z)+u)\) is increasing on \(R\), since \(\Phi'_{\eta,u}(z)=1+\eta z/2\ge1-\sqrt2/2>0\). The image of the left endpoint is at least \(-\sqrt2\), and the image of the right endpoint is at most zero. Hence \(R\) is invariant for all these updates, at a positive distance from \(U\) and \(\partial\mathcal K\).

However, there is no common certificate at the level \(\delta\) of disturbance. If one existed, \(V(0)\le-1\) would put some \(z_0\in(0,1)\) in its negative sublevel. Under the permitted constant forcing \(u=\delta\), the solution is \(z(t)=z_0/(1-z_0t/4)\), which reaches \(1\in U\) at time \(4(1/z_0-1)\). Soundness applies from every point of the invariant sublevel, a contradiction. Notice that this example is about the strict initial separation required by Definition~\ref{def:common-certificate}: it does not refute a converse with a smaller certified disturbance.
\end{example}

\begin{propositionrep}[Checking and fixed-degree synthesis]\label{prop:certificate-checking}
Fix the active dimension \(d\) and a rational polynomial field \(F(h,Z)\). For rational \(h,\delta>0\), a rational box \(I\), a fixed rational box \(\mathcal K\) and a fixed rational halfspace defining \(U\cap\mathcal K\), the validity of a proposed common certificate is decidable in time polynomial in the bit length of the input and in the length of the dense list of rational coefficients of the certificate. For each fixed bound \(D\) on the degree, the existence of a common certificate of degree at most \(D\), with arbitrary real coefficients, is decidable in time polynomial in the bit length of the instance.
\end{propositionrep}
\begin{proofsketch}
Validity is a first-order sentence over the reals, with a fixed number of variables for the state, the disturbance and the step. The dense expansion of the update keeps a description of polynomial size, and real quantifier elimination takes polynomial time when the total number of variables is fixed. With a fixed bound on the degree, the unknown vector of coefficients also has a constant dimension.
\end{proofsketch}
\begin{proof}
All the conditions of Definition~\ref{def:common-certificate} are polynomial inequalities with rational coefficients. Membership in either box, in the unsafe halfspace and in the boundary of \(\mathcal K\) has a Boolean description of fixed size in the variables of the state (the boundary is the union of its finitely many faces). The universal implication for updates only uses \(Z\in\R^d\), \(\xi\in\R^d\) and \(\eta\in\R\), and the two conditions of separation can use the same universally quantified variables of the state. Hence validity is one first-order sentence with at most \(2d+1\) variables.

If the proposed polynomial has degree \(D\), substituting the update of fixed degree into it gives a degree \(O(D)\) in this fixed number of variables, and a dense polynomial of this degree has a number of monomials polynomial in \(D\). Successive polynomial multiplications expand it in polynomial time, and the bit lengths of the coefficients remain polynomial in the lengths of the input and of the certificate. In a dense representation, \(D\) is at most the length of the certificate. Clearing rational denominators by positive common multiples has a polynomial cost in bits. The bounds for the decision of the first-order theory of the reals with a fixed number of variables~\cite{Renegar1992II} therefore give the claimed polynomial-time verifier.

For synthesis at a fixed degree \(D\), we introduce one real variable for each of the \(\binom{d+D}{D}\) coefficients, and we quantify these variables existentially before the preceding universal sentence. The total number of variables remains constant, and all the degrees of polynomials are bounded in terms of the fixed field, of \(d\) and of \(D\). The same real-algebraic bound for the decision is polynomial in the bit length of the rational instance, with a polynomial whose exponent may depend on \(D\). Observe that this argument decides existence over \(\R\): it neither assumes nor constructs a short rational vector of coefficients. The first assertion uses dense encodings, and we claim nothing for sparse encodings of polynomials with binary exponents, or for succinct arithmetic circuits.
\end{proof}

\begin{corollaryrep}[The obstruction also applies to ordinary barriers]\label{cor:ordinary-barriers}
For the fixed family of Theorem~\ref{thm:common-certificates}, replace the condition on common updates either by the condition of unit-step invariance alone, or by the condition of a strict barrier in continuous time
\begin{equation}\label{eq:strict-flow-barrier}
 z\in\mathcal K,\ V(z)=0,\ \|\xi\|_\infty\le\delta
 \quad\Longrightarrow\quad
 \nabla V(z)\cdot(F(h,z)+\xi)<0,
\end{equation}
and keep~\eqref{eq:certificate-separation}. For each of these two formats of certificates, every safe instance has a rational polynomial certificate, the validity of a dense rational candidate can be checked in polynomial time, and the same conditional obstructions on the length and on the degree hold.
\end{corollaryrep}
\begin{proofsketch}
The constructed common certificate is strictly decreasing on its zero set, by~\eqref{eq:polynomial-boundary-decrease}. Both alternative formats are sound for the corresponding dynamics, whose safe inputs are still the same \(\PSPACE\)-complete language, and their conditions for checking, and for existence at a fixed degree, again use a fixed number of real variables.
\end{proofsketch}
\begin{proof}
Unit-step invariance follows by taking \(\eta=1\). For the continuous condition, Lemma~\ref{lem:common-converse} gives, whenever \(V(z)=0\),
\[
 V(z+\eta(f(z)+\xi))-V(z)\le-\eta/2
 \quad(0<\eta\le1,\ \|\xi\|\le\delta).
\]
Dividing by \(\eta\) and letting \(\eta\downarrow0\), we see that the derivative is at most \(-1/2\). Hence the same rational polynomial gives both certificates. Notice that this strictness is a property of the construction, and not of every common certificate: cubing any common certificate preserves its sublevel and its normalized separation, but it makes its gradient vanish on its zero set.

The unit-step format is sound by induction. For the format of flows, the compact sublevel set \(C=\{z\in\mathcal K:V(z)\le0\}\) is separated from \(U\) and \(\partial\mathcal K\), and its boundary lies in \(V=0\). This zero set is nonempty, since \(V\le-1\) on the nonempty initial box and \(V\ge1\) on \(\partial\mathcal K\). By continuity, and by compactness of the zero set times the box of disturbances,~\eqref{eq:strict-flow-barrier} gives a uniform negative bound, which persists in a neighborhood of this zero set. Along a solution with a bounded measurable forcing, the chain rule holds almost everywhere. A first crossing of the zero level would have to pass through this neighborhood with \(V\) increasing, whereas its derivative is uniformly negative almost everywhere there. Hence the sublevel is invariant, and compactness gives global existence.

Both formats are therefore sound, and thus exclude certificates on the unsafe members of the same family. Their algebraic conditions use a fixed number of variables, including the variables of disturbance in the Lie derivative, and the arguments of expansion and of quantifier elimination of Proposition~\ref{prop:certificate-checking} apply unchanged, also when the fixed number of unknown coefficients is quantified existentially. The two arguments of collapse of complexity classes of Theorem~\ref{thm:common-certificates} now apply separately to these formats.
\end{proof}

\section{Continuous-time semantics and polynomial length}\label{sec:continuous-bridge}

\subsection{A quantitative limit at a fixed logical scale}

\begin{lemmarep}[Uniform Euler-to-ODE limit]\label{lem:euler-ode}
Fix an input and an allowed scale. Suppose that the exact Euler orbits, for every \(0<\eta\le1\), stay in a common compact box. Then their piecewise-affine interpolants converge on every bounded physical interval to the unique global solution of
$$
 \dot z=F(x,h,z),\qquad z(0)=c(x,h).
$$
If \(M\) and \(A\ge1\) bound the field and its Lipschitz constant on a convex compact neighborhood of the box, the interpolation error up to time \(T\) is at most
$$
 AM\eta\,T e^{AT}
$$
once \(\eta\) is small enough. We do not need this numerical bound on the error to be uniform as \(h\to0\).
\end{lemmarep}
\begin{proofsketch}
The derivative of the interpolant is the field evaluated at the preceding point of the mesh, which differs from the field at the interpolated point by at most \(AM\eta\). Integrating this defect and applying Gr\"onwall, we identify the limit, and compactness prevents the escape of the limiting ODE in finite time.
\end{proofsketch}
\begin{proof}
Let \(z_\eta\) be the affine interpolant. Its values lie in the convex box, and its slope has norm at most \(M\). For almost every \(t\), if \(k\eta\le t<(k+1)\eta\), we have
$$
 \|\dot z_\eta(t)-F(x,h,z_\eta(t))\|
 =\|F(x,h,z_k)-F(x,h,z_\eta(t))\|
 \le AM\eta.
$$
The interpolants are uniformly bounded and equicontinuous on every finite interval. Any uniformly convergent subsequence satisfies the integrated ODE, because the integrated defect tends to zero and \(F\) is uniformly continuous on the box, and local uniqueness for a polynomial field identifies all these limits. The limiting solution remains in the compact box, so it extends beyond every finite horizon. This proves global existence, and the convergence of the whole family. Subtracting the integral equations and using Gr\"onwall, we get the displayed bound. Observe that the argument uses the common compact box to prove existence before using the quantitative estimate: it does not assume the conclusion of global existence.
\end{proof}

The limit is unique because the flow is unique. Under a measurable forcing, Corollary~\ref{cor:forced-ode} keeps uniqueness for each forcing. The statements of this paper quantify over all forcings and all meshes: what is unique is the decision and the observed word, and not the trajectory.

\begin{corollaryrep}[Continuous bounded-forcing semantics]\label{cor:forced-ode}
For the constructed robust field, let \(\xi:[0,\infty)\to\R^m\) be any measurable forcing with \(\|\xi(t)\|_\infty\le h^{2K}\) almost everywhere. The absolutely continuous solution of
$$
 \dot z(t)=F(x,h,z(t))+\xi(t),\qquad z(0)=c(x,h),
$$
exists globally, stays in the same compact active box, and has the same stable decision after the stated physical horizon. In particular, this applies to the witness for polynomial space and to its exponential horizon. The frozen input and scale are not forced.
\end{corollaryrep}
\begin{proofsketch}
We use the average of the measurable forcing on each interval of the mesh as the permitted discrete disturbance. Its integral agrees exactly with that of the original forcing at the times of the grid, and it differs by at most \(2\eta h^{2K}\) between them. Compactness then gives the limit, which is the forced ODE, and the closed bounds of the decision pass to this limit.
\end{proofsketch}
\begin{proof}
Put \(\nu=h^{2K}\) and, for each mesh, choose
$$
 \xi_k^{(\eta)}=\frac1\eta\int_{k\eta}^{(k+1)\eta}\xi(t)\,dt.
$$
These are permitted disturbances. Their piecewise-constant interpolation has an integral that differs from \(\int_0^t\xi(u)\,du\) by at most \(2\eta\nu\), since the integrals agree on complete cells of the grid. The affine interpolation of the state is uniformly Lipschitz, with constant \(M+\nu\), and it remains in the compact box given by the robust discrete theorem. Its integrated defect with respect to the field is at most \(A(M+\nu)\eta T\) on \([0,T]\). Passing to a uniformly convergent subsequence, we therefore get
$$
 z(t)=c(x,h)+\int_0^t F(x,h,z(u))\,du+\int_0^t\xi(u)\,du.
$$
The local Lipschitz continuity of \(F\) gives the uniqueness of this absolutely continuous solution, and compactness gives global existence. Hence the whole family of interpolants converges. The robust discrete bounds of the decision pass to the limit on each fixed finite interval, and hence hold at every time on the infinite tail of the decision. The active length up to a finite time \(T\) is at most \((M+\nu)T\). Notice that we do not infer any converse bound on complexity from continuous robustness alone: the soundness of our characterization still uses the unit-step interpretation.
\end{proof}

\begin{proof}[Proof of Theorem~\ref{thm:main-continuous}]
At a fixed input and scale, Lemma~\ref{lem:euler-ode} applies to the exact mesh-uniform construction. For a fixed \(t>Q(s)\), the points of the mesh around \(t\) eventually lie on the tail of the stable decision. Their output values belong to the closed interval of radius \(1/4\) around the answer, and so does their affine interpolation. Passing to the limit, we get the same bound on the output at \(t\), and continuity gives it at \(t=Q(s)\). Since \(t\) was arbitrary, the decision stays correct forever.

In normalized coordinates, the field is bounded by one constant \(M\), uniformly over all inputs and allowed scales. Therefore, the length of the flow satisfies
$$
 \operatorname{len}_z(0,Q(s))
 =\int_0^{Q(s)}\|F(h,z(t))\|_\infty\,dt\le M Q(s).
$$
This bounds the length of a prefix, and not the length of the oscillating infinite tail. Adding a clock coordinate \(q'=1\), initialized at zero, we get
$$
 t\le\operatorname{len}_{(z,q)}(0,t)\le\max(M,1)t.
$$
The added coordinate is not needed for the discrete characterization: it gives the convention on length used in analog recognition.
\end{proof}

\subsection{Word inputs and a rationally generated initialization}\label{sec:word-inputs}

Let us first assume that the alphabet has two symbols, and let us identify them with the digits \(1,2\). For \(w=d_1\cdots d_n\), put
$$
 \psi_4(w)=(u,n),\qquad u=\sum_{i=1}^n d_i4^{-i}.
$$
The empty word has \(u=0\). This is already a stack code, with exactly the uniform gaps used in the two-stack normal form. The convention from letters to digits of the framework of polynomial length allows these digits in base four, and Section~\ref{sec:hard-normal-form} explains why the stack code uses exactly them. For a larger fixed alphabet, we encode each symbol by a binary block of fixed length before applying this statement, or we use a larger base with an unused largest digit (of course, one cannot assume an injection from an arbitrary finite alphabet into \(\{1,2\}\)).

We initialize the first stack directly to \(u\), and we start the finite control at the state of the simulation reached after the loading of the input. The registers of the readers can be held at zero (their unused tests still have the margins checked earlier), or they may as well be omitted. No arithmetic of readers is performed, so the scale is only used by the clock, by the margins of the unused comparators and by the quantitative estimates on errors, and it need not be dyadic. If the maximal height of stacks on inputs of length \(n\) is \(H(n)\), any scale satisfying
$$
 4^{H(n)}\le1/h
$$
gives the bound on the cumulative gain. All the other tests on stacks and on the control have constant margins, so the preceding proof still applies to a real \(h>0\).

\begin{lemmarep}[A bounded generable scale, including the empty input]\label{lem:generable-scale}
Choose integers \(A,d\ge1\), and set \(S(n)=A(n+1)^{2d}\). Then the function
$$
 h_{\mathrm{gen}}(u,n,\mu)=\exp(-S(n)-\mu)
$$
is generable over the rationals with polynomially bounded generating variables on
$$
 U=(-1,1)\times(-2,\infty)\times(-1,\infty).
$$
Every initial section whose coordinates are rational polynomials in \(u\) and \(h_{\mathrm{gen}}\) is also generable over the rationals, and bounded on \(U\).
\end{lemmarep}
\begin{proofsketch}
We use the rational base point \((u,n,\mu)=(0,-1,0)\), where \(S(n)=0\) and \(h_{\mathrm{gen}}=1\). The generating differential system keeps the coordinates of the input as auxiliary variables, and it has the derivatives \(\partial_nh_{\mathrm{gen}}=-S'(n)h_{\mathrm{gen}}\), \(\partial_\mu h_{\mathrm{gen}}=-h_{\mathrm{gen}}\) and \(\partial_uh_{\mathrm{gen}}=0\). The even exponent bounds \(h_{\mathrm{gen}}\) also on the small extension to negative lengths.
\end{proofsketch}
\begin{proof}
Consider the map \((u,n,\mu)\mapsto(u,n,\mu,h_{\mathrm{gen}})\). Its Jacobian is a polynomial matrix in these four variables, with rational coefficients: the first three rows are the derivatives of the identity coordinates, and the last row is \((0,-S'(n)h_{\mathrm{gen}},-h_{\mathrm{gen}})\). At the base point \((u,n,\mu)=(0,-1,0)\), its value is rational. These identities give the polynomial differential description required by generability. The domain \(U\) is open and connected, and it contains this base point, as well as all encoded inputs and all precisions \(\mu\ge0\).

Since \(S(n)\ge0\) on \(U\) and \(\mu>-1\), we have \(0<h_{\mathrm{gen}}\le e\), so the generating variables have norm at most a fixed linear polynomial in \(\|(u,n,\mu)\|\). Adjoining any fixed polynomial in \(u,h_{\mathrm{gen}}\) preserves a polynomial Jacobian, by the chain rule, and it preserves the bound on growth (the adjoined coordinates are themselves bounded). This proves the claim, in the notation \(\GPVAL_{\mathbb Q}\) of~\cite[Section~3.1]{JournalACM2017}.

Notice that the base point is not an input of the computation, and that we do not require its timers to be small. In particular, \(S(0)=A\) may be arbitrarily large, as needed for the empty word and for all the uniform requirements of small scale. Requiring both \(S(0)=0\) and uniformly small scales for the computation would have been an unnecessary conflict.
\end{proof}

The scale in the next statement is real, not dyadic. The statement therefore leaves the decision semantics of Definition~\ref{def:scale-field}, whose dyadic scale serves the bit counts of the soundness proofs. The estimates of Section~\ref{sec:proof-P} that it uses do not depend on the dyadic form of the scale.

\begin{propositionrep}[Weak computation of the language indicator]\label{prop:weak-bridge}
Let \(L\) be a binary language in \(\Ptime\). There is a rational polynomial core field and a bounded rationally generable initialization at scale
$$
 h=e^{-S(n)-\mu},\qquad \mu\ge0,
$$
whose solution is global and bounded and satisfies
$$
 |o(t)-\mathbf1_L(w)|\le e^{-\mu}
 \quad\text{for all }t\ge Q(n,\mu),
$$
where \(Q\) is a polynomial. Here, \(o\) is an actual output coordinate, and \(S\) may have the form of Lemma~\ref{lem:generable-scale}.
\end{propositionrep}
\begin{proofsketch}
We use the stack machine with the word already loaded. Its source computation takes a number of transitions polynomial in \(n\), independently of the requested accuracy. We increase the exponent of the scale to dominate the gain of the stacks and the accumulated error. When \(\mu\) is added to this exponent, the resetting halt gives an infinite tail, as accurate as we want.
\end{proofsketch}
\begin{proof}
Let \(T_0(n)\) and \(H(n)\) be polynomial bounds on the source time and on the height of stacks, for the directly loaded word. We choose \(K\ge256\), and we put \(a=K/8>2\). Let \(C_{\mathrm{tail}}\) be the constant of Proposition~\ref{prop-contracting-halt}, so that the data stay within \(C_{\mathrm{tail}}\Delta_\eps\) of the halting configuration on the infinite tail. By increasing \(A,d\), the even polynomial \(S(n)=A(n+1)^{2d}\) can ensure simultaneously, for all integers \(n\ge0\),
$$
 S(n)\ge H(n)\log4+C,
 \quad 8GB T_0(n)e^{-(a-2)S(n)}\le r_*,
 \quad 8C_{\mathrm{tail}}e^{-a S(n)}\le1,
$$
and all the conditions of smallness in the analytic lemmas. These inequalities are possible, because an exponential of a sufficiently large polynomial dominates every fixed polynomial, and adding \(\mu\ge0\) preserves them.

We initialize the clock at its exact entry, the timers at \(h^K\), the data at the source state where the word is loaded, the shadows equal to the data, and the comparators equal to their tests. Lemma~\ref{lem:generable-scale} gives a bounded generable initialization, with rational generating data, and the estimate on the gain of stacks is valid at this real scale. The proofs of accumulation and of resetting halt therefore apply, and the limit from Euler schemes to the ODE transfers them to continuous time. On its infinite tail, the output has an error at most
$$
 C_{\mathrm{tail}}\Delta_\eps
 \le8C_{\mathrm{tail}}e^{-a(S(n)+\mu)}\le e^{-\mu}.
$$
Here, the exact scheme has \(\nu=0\), and with \(q=S(n)+\mu\ge0\), its intrinsic error satisfies
$$
 \Delta_\eps=(1+Kq)e^{-Kq/4}\le8e^{-Kq/8}.
$$
Indeed, \(1+Kq\le8e^{Kq/8}\), by \(e^z\ge1+z\), and this takes the logarithmic prefactor explicitly into account. Since \(a>2\), increasing \(\mu\) also preserves the budget on the relative error for accumulation. Each cycle lasts \(O(1+\log(1/h))=O(1+S(n)+\mu)\), so one may take
$$
 Q(n,\mu)=C(1+S(n)+\mu)(T_0(n)+1).
$$
By the same estimates on cycles, the trajectory stays in a fixed compact box, for all times. This verifies all the requirements of weak computation on time, accuracy and growth.
\end{proof}

\begin{propositionrep}[Bridge to strict polynomial-length recognition]\label{prop:bridge}
The weak computation of Proposition~\ref{prop:weak-bridge} gives an analog recognizer of polynomial length, with a polynomial initial condition in the bounded encoding of words, in the strict sense of~\cite{JournalACM2017}. The resulting strict recognizer may have additional variables, and a field different from the scale-controlled core. The core itself keeps the property proved above: the same field works in discrete and in continuous time.
\end{propositionrep}
\begin{proofsketch}
We apply the weak item of the main equivalence and the result comparing polynomial and generable dynamics, to get an indicator computable with polynomial length. The construction for the recognition of languages only needs a computable separator taking the values \(\pm2\) on encoded words, and its mechanism of delayed output gives an irreversible first crossing. The reduction to rational coefficients is a separate cited step: we do not assume that generable constants are rational.
\end{proofsketch}
\begin{proof}
We define \(f\) on the set of encoded words by \(f(\psi_4(w))=\mathbf1_L(w)\). Proposition~\ref{prop:weak-bridge} satisfies the weak computation condition of our main equivalence~\cite[Proposition~3.8, item~(3)]{JournalACM2017}. Its initialization is in \(\GPVAL_{\mathbb Q}\), hence also in the corresponding class over the field of generable coefficients used in this equivalence. By~\cite[Proposition~3.12]{JournalACM2017}, we may use generable dynamics and initializations in that condition. Hence \(f\in\ALP\), and arithmetic closure gives the separator \(g=4f-2\in\ALP\), with \(g(\psi_4(w))\in\{-2,2\}\).

We now apply the construction in our proof of~\cite[Theorem~7.2]{JournalACM2017}. That proof first obtains a separator from a simulation of a machine. Once the separator is available, we use its representation with time and size, a frozen copy of the encoding of the word, a clock and a delayed output variable. The output remains strictly inside \((-1,1)\) until the separator is accurate. Afterwards, if \(z\) denotes the delayed output and \(\chi\in\{-1,1\}\) the correct sign, its equation has the form
\[
 (\chi z)'=\phi(t)(\lambda(t)-\chi z),
 \qquad \phi(t)\ge0,\quad \lambda(t)\ge3/2,
\]
where \(\lambda\) is the signed target. At \(\chi z=1\), the drift is at least \(\phi(t)/2\), so the region \(\chi z\ge1\) is invariant. The positive lower bound on \(\phi\) after a further bounded delay gives a crossing in bounded additional time. This proves stability of the decision without requiring global monotonicity of the output. Polynomial bounds on the state and on time give a polynomial bound on the length, and the added clock gives a length at least the physical time. We apply precisely this construction, from a separator to a recognizer, to the function \(g\) that we just obtained. Finally, we apply our reduction to rational coefficients~\cite[Section~10]{JournalACM2017}. This removes the intermediate field of generable coefficients.

Hence the initial condition and the field of the final recognizer satisfy the strict polynomial convention. Observe that this argument does not try to open an exactly closed polynomial gate after a phase of initialization, and that it does not claim either that an approximate control bit of the original core is monotone during every copy: we need neither claim. The extra compilation is only used for the strict recognizer, and not for the result with the same field.
\end{proof}

\section{Functions and finite readout windows}\label{sec:functions}

For the computation of functions, the output is kept, and not reset, so we use a finite window for the readout. This distinction is essential in the presence of arbitrary persistent perturbations.

\begin{definition}[Scaled function readout]\label{def:function-window}
A scale-controlled field computes \(f:\N^d\to\N\) with threshold \(S\) and time \(Q\) if, for every permitted input, scale and mesh, its active state remains in a fixed compact box for \(0\le k\le2N\), where \(N=\lceil Q(s)/\eta\rceil\), and
$$
 |o_k-hf(x)|\le h/4\qquad(N\le k\le2N).
$$
Robustness requires these properties for all permitted disturbances on this finite prefix. This definition contains no claim on the trajectory after \(2N\).
\end{definition}

\begin{corollaryrep}[Functions with polynomial output length]\label{cor:functions-final}\label{cor-FP-functions}
For functions with an output of polynomially bounded length, the unit-step and mesh-uniform forms of Definition~\ref{def:function-window} characterize \(\FP\) when \(S,Q\) are polynomial. Their robust forms, with a polynomial \(S\), a linear budget of precision and an integer-valued \(Q(s)=2^{O(s)}\) computable in space \(O(s)\), characterize the functions computable in polynomial space with an output of polynomially bounded length.
\end{corollaryrep}
\begin{proofsketch}
We emit the digits of the output, the least significant first, into a bounded pair \((o,\omega')\) initialized to \((0,h)\). Its linear interval gain is bounded by \(1/h\), as for the stacks, so the same proof on cycles and accumulation gives an accuracy smaller than \(h/4\) during the whole finite window of the readout. Conversely, a numerical simulation with absolute precision \(O(h)\), followed by a division by \(h\) and a rounding to an integer, recovers the value of the function.
\end{proofsketch}
\begin{proof}
Let \(m(n)\) bound the length of the output. We increase the polynomial scale threshold so that \(s\ge m(n)+2\). Once the simulated transducer has produced its output on its work tape, the source machine traverses these digits, the least significant first, using the two stacks and the finite control. This costs a polynomial time in the polynomial-time case, and at most the prescribed exponential bound in the polynomial-space case, and it only uses a polynomial additional space.

We adjoin the pair \((o,\omega')\), initialized to \((0,h)\), and for each emitted bit \(b\), we use
$$
 o\leftarrow o+b\omega',\qquad \omega'\leftarrow2\omega'.
$$
After \(m\) bits, \(o=hf(x)\) and \(\omega'=2^m h\le1/4\), and the partial output also stays in \([0,1/4]\). All the emitted bits are determined by the finite control and by the current tests on the stacks. Consequently, the extra operation is affine on each branch, with block matrix
$$
 E_b=\begin{pmatrix}1&b\\0&2\end{pmatrix}.
$$
A product over \(r\) emissions of the output, interspersed with identities, is
$$
 \begin{pmatrix}1&\sum_{j=0}^{r-1}b_j2^j\\0&2^r\end{pmatrix}.
$$
Its maximal row sum is at most \(2^r\le2^m\le1/h\), and it has no coefficient across blocks coming from a register of a stack, because \(b\) is a constant of the branch. Hence the lemma on the cumulative gain still holds for every interval of the augmented source run.

When the emission is finished, we keep \((o,\omega')\) fixed, and we reset the other work registers as before: the linear block of the output is then the identity, and not zero. Let \(T_f(s)\) bound the source computation, emission included. We choose an integer polynomial \(T_f\) in the polynomial-time case, and \(T_f(s)=2^{c_fs}\), for a fixed integer \(c_f\ge1\), in the polynomial-space case. Choose an integer \(C\ge1\) so that \(Q(s)=C(1+s)(T_f(s)+1)\) dominates its physical time of completion. This horizon is integer-valued and computable in space \(O(s)\) in the polynomial-space case. After increasing the threshold, every cycle of the clock has duration at least \(c(1+s)\), for a fixed \(c>0\), so the prefix up to physical time \(2Q(s)+2\), and therefore up to index \(2N\), contains at most \(C'(T_f(s)+1)\) cycles. We apply the argument of accumulation to this number of transitions, counting the holding transitions as well. Choosing \(K\) and the threshold as in the proof for languages makes every error at sections smaller than \(h/16\), with all the local errors on segments smaller than \(h/16\). Once the emission is complete, the output coordinate of both source endpoints is \(hf(x)\), so the estimate on the segments of cycles gives \(|o-hf(x)|\le h/4\) at every intermediate micro-step. The argument of first exit gives compactness over the same prefix. This proves completeness, with or without disturbances, at unit step and uniformly in the mesh, for both resource budgets: polynomial time with a polynomial scale threshold, and polynomial space with the exponential bound on the source time and the linear budget of precision.

For the soundness at unit step in polynomial time, we modify the precision of the truncated simulation of Lemma~\ref{lem-truncated-simulation}, so that the error of approximation at the readout is at most \(h/8\), instead of a constant. This only requires \(s+O(1)\) additional fractional bits, and the simulated output \(\widehat o\) then satisfies
$$
 |\widehat o/h-f(x)|\le 1/4+1/8=3/8<1/2,
$$
so rounding to the nearest integer gives exactly \(f(x)\). The division by the known dyadic scale shifts the binary point, and it has a polynomial cost. For robust soundness in space, a rounded trajectory is already a permitted perturbed trajectory, and storing polynomially many bits (with at least \(s+O(1)\) fractional bits, if necessary) makes the same final readout exact. We compute \(Q(s)\) in space \(O(s)\) and stop after exactly \(Q(s)\) unit steps, at the start of the guaranteed readout window. The counter uses \(O(\log Q(s))=O(s)\) bits, and all the arithmetic uses polynomial space.

Notice that we make no claim on the infinite tail for the kept output of the function. An uncontrolled additive disturbance can accumulate in a held register of output: the finite window avoids this problem, and it is all that soundness needs.
\end{proof}

Polynomial output length is automatic in the \(\FP\) case.

\section{Local algebraic verification}\label{sec:local-certificates}
The proofs above are analytic arguments. AI tools contributed to the development and correction of mathematical drafts, as detailed in the statement on AI methodology below, and the distinction between a written argument and an independent verification is essential. As a supplementary check, the ancillary files of the arXiv version of this paper, \href{https://arxiv.org/abs/2610.40025}{arXiv:2610.40025}, contain a directory \texttt{tests}.
Its file \texttt{tests/local-certificates.json} records exact rational certificates for the
local inequalities on restoration, timers and comparators. They can be checked with
\texttt{tests/verify\_certificates.py}, which only uses the rational arithmetic of the standard library of Python. These tests have been generated by AI tools, under our supervision and through various interactions with them. Several of these tests are actually Python versions of computations and verifications that were initially done using computer algebra software (Maple), so that our readers can check them without a specific proprietary license. For completeness, we discuss them here.

For Proposition~\ref{prop:explicit-policy-certificate} (Appendix~\ref{sec:explicit-certificate}),
\texttt{tests/check\_policy\_certificate.py} checks seven rational inequalities between positive-semidefinite matrices by their principal minors, and 43 scalar inequalities by Bernstein coefficients on whole intervals. These include the margins for the clock box, the gates, the initialization and the disturbances. It also checks the identity of the fixed point and the arithmetic of the comparison of outputs. These are local checks of the displayed algebra, and not an independent validation of the analytic arguments of invariance and of Euler limits.

\texttt{tests/check\_branching\_certificate.py} checks the identity of the quartic, its strict images of intervals, the margins of safety and of the comparators, and the constants of the two-cycle in Theorem~\ref{thm:branching-certificate}. Two checks of Bernstein intervals cover the local inequalities of extinction of the clock used in Proposition~\ref{prop:clock-nonconvexity} (Appendix~\ref{sec:supplement-branching}). The induction on cycles, the obstruction by convexity and the limit for forced flows remain analytic arguments.

\texttt{tests/check\_global\_certificate\_constraints.py} checks the four equilibria of the stalled clock of Proposition~\ref{prop:certificate-coupling-scale}, as polynomial identities in \(\eps\), and the zero gates used in its obstruction for products. These checks neither construct a certificate on the full state, nor independently validate the arguments of compactness and on coefficients. The file \texttt{tests/README.md} lists all the scripts, with the finite discrete cases, the rational margins of the clock switching proof and the rejection of one recorded quartic family, and it states the precise scope of each check.

A Bernstein certificate represents a polynomial on a rational box as
$$
 p(x)=\sum_{\mathbf i} b_{\mathbf i}
       \prod_j {d_j\choose i_j}t_j^{i_j}(1-t_j)^{d_j-i_j},
 \qquad t_j=\frac{x_j-a_j}{b_j-a_j}\in[0,1].
$$
Nonnegative rational coefficients prove nonnegativity on the whole box, and not
only at sampled points. Where the polynomial vanishes in the interior, we instead use an exact
factorization into squares and nonnegative factors. The checker
recomputes the coefficients or the factorization: it does not trust stored decimal
approximations. The checker also verifies the rational margins on exponents used in the
estimates on cuts and on comparators.

These certificates only cover local algebraic obligations. In particular, they
certify neither the induction on first exits, nor the choice of a common threshold for
small parameters, nor the interfaces of cycles, nor the characterizations of complexity: we prove these statements in Sections~\ref{sec:proof-P} and~\ref{sec:proof-PSPACE}. The algebraic checks do not replace those proofs. We use no numerical trajectory as a premise of a theorem.

\section{Discussion and open directions}\label{sec:discussion}

We proved that, for every deterministic decider, one polynomial field can perform the same computation at unit step, at every smaller or variable step and along its flow, with explicit bounds on time and precision. Our model distinguishes logical resolution (the scale \(h\)) from temporal resolution (the mesh): changing the Euler mesh does not change the decoded computation, and the budgets of physical time and of precision remain controlled. This is a property of the constructed witnesses, and not of every polynomial map. In particular, the mesh-uniform assertion is stronger than numerical convergence when the step tends to zero, and weaker than the equality, or the uniform closeness, of all coarse and fine trajectories.

The realization theorem quantifies over all admissible meshes and disturbances, for one fixed field and one fixed initialization.
The proof first controls the discrete executions uniformly, and then gets each continuously forced execution by compactness.
Its two quantitative ingredients are the integrated separation of gates and the bounds on products of transitions over intervals, and the interface of the compiler states the hypotheses on source programs under which they apply.
Of course, these local principles of design are not a sufficient condition for an arbitrary polynomial field to preserve a computation under refinement.

The theorem on common certificates makes this semantic link usable in verification: the same polynomial sublevel set proves safety for every temporal interpretation. Its converse uses an extra tolerance to noise, so we do not claim that every safe polynomial system has a polynomial certificate, and the general proof of existence gives neither a useful estimate on the degree nor an efficient procedure of synthesis. In Appendix~\ref{sec:explicit-certificate}, we give an explicit construction for a structured contractive program (the evaluation of a discounted policy), clock included, but with regularized gates and an exponentially large degree. In Section~\ref{sec:branching-certificate}, we give a quartic invariant of constant size on the source, for a program that branches forever in the unchanged compiler (a scalar controller), together with a complete compositional proof of safety. We also show why the problem on the full state is harder (Appendix~\ref{sec:supplement-branching}): no compact convex robust invariant can contain the prescribed execution. An explicit small nonconvex polynomial on the full state remains to be constructed, even for this example. The question whether general bounds on the degree can be derived from the reserve of disturbance and from geometric margins also remains open. The obstruction on the length in the worst case explains why a uniformly efficient solution cannot follow from compactness and robustness alone, unless \(\mathrm{NP}=\PSPACE\).

\myparagraph{A local lower bound for resolving a small margin}
The logarithmic duration of the modules that amplify signs is explained by Lipschitz dependence. This is not a theorem of optimality for the complete compiler. Let \(F_h\) be \(\Lambda\)-Lipschitz on a convex set containing the trajectories under consideration, and let the readout be \(R\)-Lipschitz there, with \(\Lambda,R>0\) independent of \(h\). Consider two exact Euler executions of \(F_h\) with the same frozen parameters and the same temporal steps \(0<\eta_k\le1\), whose initial active states are at distance at most \(C_0h\). Their distance \(d_k\) satisfies \(d_{k+1}\le(1+\Lambda\eta_k)d_k\), hence \(d_N\le C_0he^{\Lambda t_N}\) at the common physical time \(t_N=\sum_{k<N}\eta_k\). For continuous solutions, Gr\"onwall's inequality gives the same bound at a common observation time. If the readouts are separated by \(\delta>0\) at that time, then \(\delta\le Rd_N\), that is,
$$
 t_N\ge\frac1\Lambda\log\!\left(\frac{\delta}{RC_0h}\right)
 =\Omega(s)\qquad(h=2^{-s}),
$$
for fixed positive \(C_0,R,\delta,\Lambda\). The normalized core field has a uniform Lipschitz bound on each fixed convex compact neighborhood, so this applies to its pairs of trajectories that satisfy the assumption on the initial distance, for instance to the scalar test of signs initialized at two values of order \(h\) and read with a constant separation of outputs. A threshold readout with a Lipschitz constant of order \(1/h\), different frozen parameters, or a time rescaling with an unbounded Lipschitz constant would not satisfy these assumptions. The bound gives no \(\Omega(s)\) lower bound for every operation on stacks (their tests may have constant margins), and it does not prove that the given scale is necessary in other models.

\myparagraph{Extensions that require additional arguments}

A syntactic version would require a discipline of termination and of resources for programs built from affine updates, modules of comparison and clocked composition.
Ranks on the resets of timers, analogous to safe or ramified recurrence, are a possible direction, but such a syntax would need its own compilation theorem, with quantitative bounds.
Indeed, autonomous clocks and affine updates of the data do not impose polynomial time by themselves: our characterization keeps explicit bounds on the scale threshold and on the physical horizon.

The question whether the given scale can be generated from a raw integer, without losing polynomial time, also remains open. The bridge with encodings of words does not answer this question: it starts from a bounded stack code, and it uses a generable initialization before the compilation into a standard analog recognizer.

We proved an active length \(O(1+s)\) per source transition. A bound independent of \(s\) would require an argument on total variation, and not only the estimate with bounded speed (the lower bound on the time of separation above does not exclude such a bound: a slow amplifier can spend a logarithmic time while traveling a bounded distance).

It would be interesting to transfer the construction to learning dynamics in a prescribed family, such as replicator dynamics, but this is a separate, concrete problem. A continuous embedding alone does not transfer the guarantee on Euler schemes. For example, under \(y=x^2\) on \(x>0\), the flow \(x'=x\) becomes \(y'=2y\). Transforming its Euler update gives \(y^+=(1+\eta)^2y\), whereas applying Euler to the transformed field gives \(y^+=(1+2\eta)y\), and their difference is \(\eta^2y\). A nonlinear embedding into a target model would therefore need a new analysis of the error of discretization in the target coordinates, and of the clock, in addition to the continuous simulation. Roughly speaking, this is the step that is missing between our compiler and a theorem on a standard rule of learning.

Finally, small parallel classes would require an explicit account of a growing dimension, of the access to the input, or of parallel composition: these resources are not implicit in our sequential formulation in fixed dimension.

\newcommand\AImethodology{%
\section*{AI methodology}
This paper belongs to a research programme, developed over several years, on programming with ordinary differential equations and on the comparison between classical and analog computation. The question, the model, the decision to unify discrete and continuous time by one field, the theorem stating that this is possible, and the choice among alternative formal statements of it are ours.
Within this frame, we used ChatGPT, including its Codex agent, and Claude to develop and analyze preliminary mathematical arguments. We also used these tools for scripts checking exact arithmetic, for comparisons with related work and for suggestions about the organization of the manuscript. We used Claude to revise the English of the whole manuscript and to condense it by moving the supplementary developments to Appendix~\ref{sec:supplement}.
The idea of applying the construction to shared logical traces, approximate implementations, fixed-field prediction, polynomial space without a supplied horizon, common safety certificates with their converse and complexity, strict source invariants with the branching controller, and function classes with finite readout windows is ours. The developments of Appendix~\ref{sec:supplement}, namely the explicit certificate for clocked policy evaluation and the obstructions to low-degree full-state certificates, were first proposed by the tools and then checked by us.
We also used Google DeepReview for pre-submission feedback on correctness and presentation, as proposed in an email from the STOC 2027 organizers.
Throughout this process, we asked ChatGPT and Claude to review existing proofs and to read drafts as members of different communities would. This changed how several parts are presented and, at times, inspired the development and the presentation of some of these applications.
These uses affected substantive mathematical material and the presentation, and they were not limited to language editing. We have checked the statements, the proofs, the citations and the interpretation ourselves, and we are responsible for their accuracy and their originality. Neither agreement between analyses generated by AI tools, nor the success of a local certificate or of an experiment on trajectories, is an independent verification of the complete proof.
}

\appendixapxproofsubsection
\AtEndDocument{
\section{Supplementary developments on certificates}\label{sec:supplement}
This appendix collects two developments that the main results do not use: an explicit common certificate for a modified clocked implementation of policy evaluation, and obstructions to low-degree certificates on the full state of the unchanged compiler, for the controller of Section~\ref{sec:branching-certificate}. The local algebra of both is checked by the scripts described in Section~\ref{sec:local-certificates}.

\subsection{An explicit certificate for clocked policy evaluation}\label{sec:explicit-certificate}

Let us consider the following program with two registers, with rational rewards \(b\in[0,1/16]^2\):
\begin{equation}\label{eq:policy-program}
 Y^{(0)}=0,\qquad Y^{(j+1)}=\mathsf{M} Y^{(j)}+b,\qquad
 \mathsf{M}=\begin{pmatrix}3/8&1/8\\1/8&3/8\end{pmatrix}.
\end{equation}
This is the evaluation of a discounted policy for the Markov chain with two states, transition matrix \(2\mathsf{M}\) and discount \(1/2\), a standard operation in stochastic dynamic programming~\cite[Chapter~6]{LivrePuterman}. It computes the expected total discounted reward from each starting state:
\begin{equation}\label{eq:policy-value}
 y_*=\sum_{j\ge0}\mathsf{M}^j b
     =(I-\mathsf{M})^{-1}b
     =\tfrac13\begin{pmatrix}5&1\\1&5\end{pmatrix}b
     \in[0,1/8]^2.
\end{equation}
The two coordinates are coupled. The question here is to certify an autonomous implementation, with its clock, its shadows, its error of initialization and its disturbances, under every temporal interpretation.

We use exactly the eight equations of the clock of Section~\ref{sec:polynomial-field}, with \(\eps=h^K=2^{-Ks}\), and the four equations of the data
\begin{equation}\label{eq:policy-field}
 (Y^{\mathrm s})'=g_1(\mathsf{M}Y+b-Y^{\mathrm s}),\qquad
 Y'=g_3(Y^{\mathrm s}-Y),\qquad
 g_i=\eps+(1-2\eps)p_i.
\end{equation}
The rewards and the scale are frozen. Notice that this is an implementation specific to this program, with a positive floor in the gates, and not the unmodified output of the compiler. The floor differs from \(p_i\) by \(\eps(1-2p_i)\), and it ensures progress even at states of the clock outside the intended cycle of phases. For this contractive program, the proof below gives the semantics of the value directly, without using the cycle lemma, and we do not claim equality of finite traces at sections. No field and no budget of the compiler theorem is changed.

Here is a single explicit polynomial for the full state \(Z=(p,\tau,Y,Y^{\mathrm s})\), of dimension twelve. Put
\[
 \delta=\eps^3/4096,\qquad \lambda=\eps^2/16,\qquad
 N=(Ks+6)2^{2Ks},\qquad e=Y-y_*,\quad w=Y^{\mathrm s}-y_*,
\]
\begin{align}
 Q(e,w)&=\|e-w\|_2^2+2w_1^2-2w_1w_2+2w_2^2,\label{eq:policy-quadratic}\\
 \widehat p_i&=\frac{p_i-1/2}{1/2+\eps^2},&
 \widehat\tau_i&=\frac{\tau_i-(1+3\eps^2+\eps/2)/2}
                         {(1+3\eps^2-\eps/2)/2},\notag\\
 H_\eps(p,\tau)&=\sum_{i=0}^3
       \bigl(\widehat p_i^{\,2N}+\widehat\tau_i^{\,2N}\bigr),&
 S(Z)&=4Q(e,w)+H_\eps(p,\tau),\notag\\
 V(Z)&=4S(Z)-2.\label{eq:explicit-policy-certificate}
\end{align}
For each fixed rational \(b\) and scale, this expression is a rational polynomial. Let \(\mathcal K=[-2,2]^{12}\), let \(I\) be the full box of radius \(\delta\) around
\[
 p=(1,0,0,0),\qquad \tau_i=\eps,\qquad Y=Y^{\mathrm s}=0,
\]
and let \(U=\{Z\in\mathcal K:\max(\|Y\|_\infty,\|Y^{\mathrm s}\|_\infty)\ge1\}\).

\begin{proposition}[An explicit common certificate and value semantics]\label{prop:explicit-policy-certificate}
For integers \(K\ge256\), \(s\ge1\), and every rational \(b\in[0,1/16]^2\), the polynomial~\eqref{eq:explicit-policy-certificate} is a common certificate for \((f,\mathcal K,I,U,\delta)\), where \(f\) is the clock together with~\eqref{eq:policy-field}. All twelve active coordinates may be perturbed. More precisely, whenever \(Z\in\mathcal K\) and \(V(Z)\le0\),
\begin{equation}\label{eq:policy-certificate-contract}
 S\bigl(Z+\eta(f(Z)+\xi)\bigr)
 \le(1-\eta\lambda)S(Z)+\eta\lambda/16
 \qquad(0\le\eta\le1,\ \|\xi\|_\infty\le\delta).
\end{equation}
From \(I\), every execution on a variable mesh \(0<\eta_k\le1\), \(\sum_k\eta_k=\infty\), and its interpolant, and every measurably forced flow, satisfy
\begin{equation}\label{eq:policy-value-error}
 \sqrt{Q(e(t),w(t))}
 \le e^{-\lambda t}\sqrt{Q(e(0),w(0))}
            +5\delta/\lambda.
\end{equation}
Without forcing both data vectors converge to \(y_*\). With forcing their coordinate errors have limit superior at most \(\eps/32\). In particular, for \(b=(1/16,0)\), all these interpretations satisfy
\[
 Y_1(t)-Y_2(t)\ge1/48
 \qquad\text{for all }t\ge\lambda^{-1}\log16.
\]
\end{proposition}
\begin{proof}
In this proof, the norms with subscript \(2\) are Euclidean. We write \(q=\eps^2\), so that \(\eps\le1/16\), \(q\le1/256\) and \(\delta\le q/4096\) (these deliberately loose bounds hold for every scale of the statement).

\myparagraph{A common quadratic form for loading and copying}
The matrix \(\mathsf{M}\) is symmetric positive definite, with eigenvalues \(1/2\) and \(1/4\), and
\[
 \mathsf{M}^{-1}=\begin{pmatrix}3&-1\\-1&3\end{pmatrix},\qquad
 Q(e,w)=\|e\|_2^2-2e^\top w+w^\top \mathsf{M}^{-1}w.
\]
Completing the square in either variable gives
\begin{equation}\label{eq:policy-q-bounds}
 Q\ge\tfrac12\|e\|_2^2,\qquad Q\ge\|w\|_2^2,\qquad
 Q\le5\|(e,w)\|_2^2.
\end{equation}
For the first inequality, use
\(Q=e^\top(I-\mathsf{M})e+(w-\mathsf{M}e)^\top \mathsf{M}^{-1}(w-\mathsf{M}e)\).
For the second, use~\eqref{eq:policy-quadratic} and \(\mathsf{M}^{-1}-I\succeq I\).
For the upper bound, \(\mathsf{M}^{-1}\preceq4I\) and
\(2|e^\top w|\le\|e\|_2^2+\|w\|_2^2\) suffice. In particular, \(\sqrt Q\) is a norm, and a vector of disturbances of sup norm \(\delta\) in the four coordinates of the data has this norm at most \(\sqrt{20}\delta<5\delta\).

The unit maps of the data, for the pairs of gates \((g_1,g_3)=(1,0),(0,1),(1,1)\), are respectively
\[
 \mathcal L(e,w)=(e,\mathsf{M}e),\qquad
 \mathcal C(e,w)=(w,w),\qquad
 \mathcal J(e,w)=(w,\mathsf{M}e).
\]
Direct subtraction gives
\[
 Q-Q\circ\mathcal L=(w-\mathsf{M}e)^\top \mathsf{M}^{-1}(w-\mathsf{M}e),\qquad
 Q-Q\circ\mathcal C=\|e-w\|_2^2.
\]
In an orthonormal basis of eigenvectors of \(\mathsf{M}\), the contribution associated with an eigenvalue \(\mu\) is \(e_\mu^2-2e_\mu w_\mu+\mu^{-1}w_\mu^2\), and the map \(\mathcal J\) multiplies this contribution by \(\mu\). Hence \(Q\circ\mathcal J\le Q/2\).
For arbitrary \(g_1,g_3\in[0,1]\), the unit map \(\mathcal A\) of the data is the convex combination of the identity, \(\mathcal L,\mathcal C,\mathcal J\), with respective weights
\[
 (1-g_1)(1-g_3),\quad g_1(1-g_3),\quad
 (1-g_1)g_3,\quad g_1g_3.
\]
By convexity of \(Q\), we get
\begin{equation}\label{eq:policy-gate-contraction}
 Q(\mathcal A(e,w))\le(1-g_1g_3/2)Q(e,w).
\end{equation}

\myparagraph{An explicit invariant clock box}
Consider the auxiliary box
\[
 \mathcal B_\eps=[-q,1+q]^4\times[\eps/2,1+3q]^4
\]
for \((p,\tau)\). We prove that a noisy unit update of the clock maps it strictly inside itself, with each normalized coordinate in \([-1+q,1-q]\).
With the notation of the clock of Section~\ref{sec:polynomial-field},
\[
 -2q\le S_i=p_{i-1}\tau_{i-1}^2\le1+9q,\qquad
 0\le E_i=p_{i+1}^8\le1+9q.
\]
Indeed, \((1+3q)^2\le1+7q\), so that the bounds on \(S_i\) follow by multiplication. For \(E_i\), the binomial expansion gives
\((1+q)^8\le1+8q+247q^2\le1+9q\).
The expression \(a_i=p_i(1-S_i-E_i)+S_i\) is affine in each of its three arguments. At the endpoints \(p_i=-q,1+q\), the preceding bounds give
\[
 -11q\le a_i\le1+11q.
\]
For example, its largest possible value at an endpoint is at most \(1+10q+18q^2\), and its smallest is at least \(-10q-18q^2\). Since \(11q<1/2\), this interval lies in \([-1/2,3/2]\), where \(\restore(a)=3a^2-2a^3\) takes values in \([0,1]\). Hence the noisy unit update satisfies
\[
 -\delta\le p_i^+\le1+\delta.
\]
Its distance from each endpoint of \([-q,1+q]\) is at least \(q-\delta\), and dividing by \(1/2+q\), we get \(|\widehat p_i^+|\le1-q\).

For a timer, the unforced unit update is
\[
 \tau_i^+=\eps+p_i(2\tau_i-\tau_i^2-\eps).
\]
On the stated interval for timers, \(2\tau-\tau^2\in[\eps-q/4,1]\). Evaluating this affine expression at the endpoints, we see that the noisy update of the timer lies in
\[
 [\,\eps-q-\delta,\ 1+q+\delta\,].
\]
Its distances from the lower and upper endpoints of \([\eps/2,1+3q]\) are at least \(\eps/2-q-\delta\ge q\) and \(2q-\delta\ge q\) respectively, and the half-width of the interval for timers is less than one. Hence \(|\widehat\tau_i^+|\le1-q\), as required. The same normalized bounds hold initially on the whole box \(I\): the phase bits are within \(\delta\) of zero or one, and every timer is within \(\delta\) of \(\eps\).

By the explicit choice \(N=(Ks+6)/q\),
\begin{equation}\label{eq:policy-clock-power}
 H_\eps(p^+,\tau^+),\ H_\eps(p(0),\tau(0))
 \le8(1-q)^{2N}
 \le8\,2^{-2(Ks+6)}
 =q/512=\lambda/32.
\end{equation}
Here \((1-q)^{1/q}\le e^{-1}<1/2\). Observe that we use no approximation of the trajectory of the clock.

\myparagraph{Combining the clock and data margins}
On \(\mathcal B_\eps\), the regularized gates satisfy
\[
 \eps/2\le g_i\le1-\eps/2.
\]
Equation~\eqref{eq:policy-gate-contraction} then implies
\[
 \sqrt{Q(\mathcal A(e,w))}
 \le\sqrt{1-\eps^2/8}\sqrt{Q(e,w)}
 \le(1-\lambda)\sqrt{Q(e,w)}.
\]
Including the unit noise on the data, we put \(a=2\sqrt Q\). Then
\[
 a^+\le(1-\lambda)a+10\delta,\qquad
 (a^+)^2\le(1-\lambda)a^2+\frac{100\delta^2}{\lambda}
              \le(1-\lambda)a^2+\lambda/32.
\]
The middle inequality is the convexity of the square, with weights \(1-\lambda\) and \(\lambda\), and the last one follows from \(10\delta/\lambda=5\eps/128\). With~\eqref{eq:policy-clock-power}, we get
\[
 S(Z+f(Z)+\xi)\le(1-\lambda)S(Z)+\lambda/16
\]
whenever the current clock lies in \(\mathcal B_\eps\).

If \(V(Z)\le0\), then \(S(Z)\le1/2\). All the summands of \(S\) are nonnegative, so the clock does lie in \(\mathcal B_\eps\), and the preceding estimate applies. The unit endpoint has \(S\le1/2\), hence \(Q\le1/8\). By~\eqref{eq:policy-q-bounds}, its coordinates of data have modulus at most \(1/8+1/2=5/8<2\), and its clock belongs to \(\mathcal B_\eps\subset(-2,2)^8\). Hence the endpoint belongs to \(\mathcal K\).
The polynomial \(S\) is convex, since it is a positive quadratic form plus even powers of affine functions. The fractional update is the convex combination of \(Z\) and of this unit endpoint, and the convexity of \(\mathcal K\) and of \(S\) proves~\eqref{eq:policy-certificate-contract} and the invariance for all steps. In particular, on \(V=0\), the certificate decreases by at least \(7\eta\lambda/4\).

\myparagraph{Initial and unsafe separation}
At the exact initial center, \(e=w=-y_*\), and
\[
 Q_0=b^\top \mathsf{M}^{-1}(I-\mathsf{M})^{-1}b\le\tfrac{16}{3}\|b\|_2^2
       \le1/24.
\]
The eigenvalues of the displayed matrix are \(4\) and \(16/3\). For an error \(\zeta\) of initialization of the data, of sup norm at most \(\delta\), the version of Young's inequality for quadratic forms gives
\[
 Q((e,w)+\zeta)\le\tfrac98 Q_0+9Q(\zeta)
                  \le3/64+180\delta^2.
\]
Consequently, throughout \(I\),
\[
 S\le3/16+720\delta^2+\lambda/32\le1/4,
\]
using \(\delta\le1/256\) and \(\lambda\le1\). This proves \(V\le-1\).
On \(U\), some coordinate of the data has modulus at least one, so its error from \(y_*\in[0,1/8]^2\) has modulus at least \(7/8\), and~\eqref{eq:policy-q-bounds} gives \(4Q\ge49/32\), hence \(V\ge1\). On a face of \(\partial\mathcal K\) for the clock, some normalized coordinate of the clock has modulus greater than one, so \(H_\eps>1\), and again \(V\ge1\). Every face of \(\partial\mathcal K\) for the data is already in \(U\). This proves both conditions of separation, and \(V\) is a common certificate.

\myparagraph{The computed value, including disturbances}
For a fractional step, the convexity of the norm \(\sqrt Q\) and the unit contraction give
\[
 d_{k+1}\le(1-\eta_k\lambda)d_k+5\eta_k\delta,
 \qquad d_k=\sqrt{Q(e_k,w_k)}.
\]
Iterating, with \(1-\eta_k\lambda\le e^{-\eta_k\lambda}\), we get~\eqref{eq:policy-value-error} at the points of the mesh, and the same argument for a fraction of the next step proves it on interpolants. Without forcing, the additive term vanishes. Proposition~\ref{prop:common-certificate-soundness} gives global existence and safety for measurably forced flows, and its argument by Euler limits passes the same closed bounds on errors to these flows, using averages of the forcing on intervals.
Since \(5\delta/\lambda=5\eps/256\), the inequalities~\eqref{eq:policy-q-bounds} show that the errors on both coordinates have limit superior at most \(\eps/32\).

Finally, the source recurrence has
\(Y^{(j)}=\sum_{l<j}\mathsf{M}^l b\), and \(\|\mathsf{M}\|_\infty=1/2\), which proves~\eqref{eq:policy-value}. For \(b=(1/16,0)\), its value is \((5/48,1/48)\), with gap \(1/12\). The initial bound \(S\le1/4\) gives \(d_0\le1/4\), and for \(t\ge\lambda^{-1}\log16\),
\[
 \|Y(t)-y_*\|_\infty
 \le\sqrt2\left(\frac1{64}+\frac5{4096}\right)<\frac1{32},
\]
where we used \(\eps\le1/16\) in the term of disturbances. Subtracting the errors on the two coordinates from the gap \(1/12\), we get \(Y_1(t)-Y_2(t)\ge1/48\).
\end{proof}

In this example, the reuse of a proof of safety is literal: the same displayed polynomial certifies a full initial box, all the disturbances on the clock and on the data, coarse updates, adaptive refinement and the flow. The quadratic estimate also gives a robust ordering of the computed values. However, this is not an efficient synthesis of certificates: the explicit degree \(2N=2(Ks+6)2^{2Ks}\) is exponential in the scale exponent, and the bound of contraction uses a conservative physical horizon of order \(\eps^{-2}\). The source recurrence is classical, and we do not claim any new result on the algorithmic complexity of policy evaluation. What the example constructs is the certificate for the complete autonomous implementation, together with its common semantics of values.

\subsection{Obstructions to low-degree full-state certificates for the branching controller}\label{sec:supplement-branching}
The compact quartic of Section~\ref{sec:branching-certificate} is an assertion on the source, with a complete argument of lifting. By the converse theorem, a common polynomial certificate on the full state also exists for this example, with a smaller budget of disturbances. However, the geometry of the unchanged clock rules out the most immediate approach with a low degree.

\begin{proposition}[The unchanged clock forces nonconvex robust invariants]\label{prop:clock-nonconvexity}
For the controller of Theorem~\ref{thm:branching-certificate}, at all sufficiently large scales, no compact convex set containing the exact prescribed initial state is invariant under all unit-step updates, for any positive bound on drift disturbances.
Consequently, no polynomial of degree at most three can be a common certificate in the sense of Definition~\ref{def:common-certificate} for an initial set containing this state.

Nevertheless, with \(\mathcal K=[-2,3]^{11}\), the initial box of Theorem~\ref{thm:branching-certificate}, and
\[
 U=\{Z\in\mathcal K:Y\notin(1/5,4/5)
                      \text{ or }Y^{\mathrm s}\notin(1/5,4/5)\},
\]
a rational common polynomial certificate exists for the bound \(\eps^2/2\) on disturbances.
We give no bound on the degree of this certificate on the full state.
\end{proposition}
\begin{proof}
Fix the exact initial center and its unforced unit-step execution, which is global by the preceding theorem. We first select four points on it, one in the waiting part of each phase \(j=0,1,2,3\). Write \(L=\log(1/\eps)\), and choose the grid point at elapsed time \(\lceil L/2\rceil\) after the phase entry \(t_j\) (see Section~\ref{sec:proof-P} for the phase entries \(t_j\) and the timer crossings \(t_j^b\)). For sufficiently large \(s\), this point comes before \(t_j^b\), since \(t_j^b-t_j\ge L-C_{\mathrm{dur}}\).

At this point, the exact scalar box gives \(0\le p_j\le1\). The other three bits tend uniformly to zero as \(\eps\to0\): the bound on the old bit is \(O(\eps^{3/16})\), the bound on the opposite bit is \(O(\eps^{1/4})\), and the bound on the next bit is \(O(\eps^{1/4}+\eps^2)\). These bounds follow from~\eqref{eq:clock-old-far}--\eqref{eq:clock-next}, with zero noise and elapsed time \(L/2+O(1)\).
All four timers tend to zero as well. The active timer is \(O(\eps^{1/2})\), and the next timer is \(O(\eps)\). The opposite timer starts at most at \(B_0\eps^{4/5}\) and stays at rest, which gives \(O(\eps^{4/5}+\eps)\). After a fixed elapsed time, the preceding bit is at most \(1/10\), so the resting estimate of its timer gives \(O(\eps^{2/5}+\eps)\). Hence, at each selected point, all the timers and the nonactive bits tend to zero, while the active bit belongs to \([9/10,1]\).

Let \(z^{(j)}\) denote the four complete states of the implementation. Any invariant set \(D\) containing the initial state contains them, and if \(D\) is convex, it also contains their average \(\bar z\). For sufficiently small \(\eps\), the preceding estimates put the coordinates of the clock of \(\bar z\) in
\begin{equation}\label{eq:extinction-box}
 0\le p_i\le3/10,\qquad 0\le\tau_i\le1/100.
\end{equation}
For instance, we make each nonactive phase bit at most \(1/100\), so that each averaged bit is at most \((1+3/100)/4<3/10\), and we also take \(\eps\le1/1000\).

We now use the exact unit-step equations of the clock, with zero disturbance on the coordinates of the clock. If \(m=\max_i p_i\) and~\eqref{eq:extinction-box} holds, then
\[
 0\le a_i=p_i(1-p_{i+1}^8)+p_{i-1}\tau_{i-1}^2(1-p_i)
       \le(1+10^{-4})m.
\]
Since \(0\le\restore(a)\le3a^2\) in this range,
\[
 m^+\le3(1+10^{-4})^2m^2\le(91/100)m.
\]
The timer update obeys
\[
 0\le\tau_i^+
   =p_i(2\tau_i-\tau_i^2)+(1-p_i)\eps
   \le(3/5)\tau_i+\eps<1/100.
\]
Hence~\eqref{eq:extinction-box} is invariant, and along this execution,
\begin{equation}\label{eq:finite-copy-mass}
 \sum_{n\ge0}p_{3,n}\le\sum_{n\ge0}m_n
       \le\frac{3/10}{1-91/100}=\frac{10}{3}.
\end{equation}

Let \(\delta>0\) be the proposed bound on disturbances. From \(\bar z\), we choose the disturbance \(+\delta\) on the coordinate \(Y\) of the data, and zero on every other coordinate, so that the clock remains the one that we just analyzed. If the resulting execution stayed in a compact set \(D\), the number
\(M_D=\sup_{z\in D}|Y^{\mathrm s}-Y|\) would be finite. The unchanged copy equation and~\eqref{eq:finite-copy-mass} would then give
\[
 Y_n\ge Y_0+\delta n-M_D\sum_{k<n}p_{3,k}
       \ge Y_0+\delta n-\tfrac{10}{3}M_D.
\]
This contradicts the boundedness of \(Y\) on \(D\). Hence no compact convex invariant can contain the initial state.

For the consequence on polynomials, let \(V\) have degree at most three, and be positive on the boundary of its box \(\mathcal K\). We show that \(D=\mathcal K\cap\{V\le0\}\) is convex. Otherwise, a line contains \(x,y\in D\) and a point between them where \(V>0\). We extend this segment to the full chord of \(\mathcal K\), and the restriction of \(V\) is positive at both endpoints of the chord. On the portion before the positive middle point, it has an interior minimum of value at most zero, and on the portion after this point, it has another such minimum. Between these minima, it has an interior maximum of positive value. Hence the derivative of this univariate restriction has at least three distinct zeros. Since its degree is at most two, it would have to vanish identically, a contradiction. Hence \(D\) is convex. If \(V\) were a common certificate, \(D\) would also be compact and invariant, and it would contain the initial state, which contradicts the first assertion.

For existence, let \(R\) be the closure of all the points reached from the prescribed initial box by finitely many Euler updates, with steps in \((0,1]\) and drift disturbances at most \(\eps^2\). Every finite sequence extends to a divergent mesh, so Theorem~\ref{thm:branching-certificate} places \(R\) in a common compact subset of \(\operatorname{int}([-2,3]^{11})\). Its data and shadows remain in \([11/48,37/48]\), at distance \(7/240\) from either unsafe threshold. Hence \(\dist(R,U)\ge7/240\).
By concatenation and continuity, \(R\) is invariant under all allowed finite updates, and Lemma~\ref{lem:common-converse}, with \(2\delta=\eps^2\), gives a rational common certificate for disturbances \(\eps^2/2\). Its invariant sublevel is necessarily nonconvex. Observe that this argument of existence gives neither an explicit list of coefficients nor a compact bound on the degree.
\end{proof}

The positive and the negative statements are about different representations of proofs. The displayed quartic is a certificate of constant size on the source, with a proved transfer to the indefinite perturbed execution of the unchanged compiler. Here, the polynomial on the full state given by the converse is not constructive, and a certificate on the full state of degree at most three is impossible. The positive floor of the gates in Appendix~\ref{sec:explicit-certificate} removes the obstruction due to the finite mass of the copy gate, which explains its role in that construction.

Nonconvexity alone does not solve the problem of construction. An invariant on the full state must also couple the clock to the data. Moreover, the prescribed initialization lies only \(\eps\) away from a state of the clock that stops copying forever. The following necessary conditions use the displayed field directly, without the cycle lemma.

\begin{proposition}[Coupling and scale dependence of full-state certificates]\label{prop:certificate-coupling-scale}
Fix \(0<\eps<1\) and a positive bound \(\delta\) on disturbances for the unchanged controller~\eqref{eq:unchanged-controller-field}. No nonempty compact product
\[
 \Pi=C_{\mathrm{clock}}\times C_{\mathrm{data}}
 \subset\R^8\times\R^3,
 \qquad C_{\mathrm{data}}\text{ on }(v,Y,Y^{\mathrm s}),
\]
containing its exact prescribed initial state is invariant under all disturbed unit updates, even when both factors are nonconvex.

Let \(V(Z)=\sum_\alpha c_\alpha Z^\alpha\) be a common certificate on \([-2,3]^{11}\) whose initial set contains this state, with \(a\) as in Theorem~\ref{thm:branching-certificate}. Then
\begin{equation}\label{eq:certificate-coefficient-lower-bound}
 \sum_{\alpha:\alpha_{\tau_0}>0}|c_\alpha|>\eps^{-1}.
\end{equation}
In particular, if its degree is at most \(D\), then
\[
 \max_\alpha|c_\alpha|>
 \frac{1}{\eps\binom{11+D}{11}}.
\]
For fixed \(K\) and a fixed allowed initial center \(a\), no single polynomial \(W(h,Z)\) satisfies the normalized conditions of a common certificate for all sufficiently small dyadic \(h\), with \(\eps=h^K\) and any positive certified bound on disturbances at each scale.
\end{proposition}
\begin{proof}
\myparagraph{The clock and data cannot be certified independently}
Let \(c_0\) be the initial state of the clock, with \(p=(1,0,0,0)\) and all timers equal to \(\eps\). If the compact product \(\Pi\) contains the initial state, then \(c_0\in C_{\mathrm{clock}}\). By compactness of the nonempty factor of the data, there is a point \(q_*\in C_{\mathrm{data}}\) maximizing its \(Y\)-coordinate, say \(M\). The state \((c_0,q_*)\) belongs to \(\Pi\). At this state, \(p_3=0\), so a unit update with disturbance \(+\delta\) on \(Y\), and zero on the other coordinates, gives \(Y^+=M+\delta\). Its component on the data cannot belong to \(C_{\mathrm{data}}\), which contradicts invariance. Notice that the argument does not assume the convexity of either factor.

\myparagraph{Stalled clock states must be excluded}
We write \(e_j\) for the \(j\)-th unit vector of \(\R^4\). For each \(j\), the state of the clock
\begin{equation}\label{eq:stalled-clock-state}
 p=e_j,\qquad \tau_j=0,\qquad \tau_i=\eps\quad(i\ne j)
\end{equation}
is an exact equilibrium of the eight equations of the clock. Indeed, the only nonzero phase bit has a zero timer, so every activation term \(S_i\) vanishes, and the clearing term of the active bit is zero. Hence \(a_j=1\), \(a_i=0\) for \(i\ne j\), and \(\restore(a_i)=p_i\). The active timer has zero derivative, and each resting timer has derivative \(\eps-\tau_i=0\).

No point with one of these states of the clock can belong to a compact set invariant under all unit updates with disturbances bounded by \(\delta>0\). For \(j\ne3\), we keep a zero disturbance on the clock, and we put \(+\delta\) only on \(Y\). Then \(p_3=0\) forever, and \(Y_n=Y_0+n\delta\). For \(j=3\), we put \(+\delta\) only on \(Y^{\mathrm s}\): now \(p_1=0\) forever, and \(Y^{\mathrm s}_n=Y^{\mathrm s}_0+n\delta\). Both conclusions contradict compactness. This reasoning allows arbitrary initial coordinates for the comparator and the data, since they do not feed back into the clock.

Let \(z_0\) be the exact prescribed initial state, and let \(z_\dagger\) be obtained by replacing only \(\tau_0=\eps\) by zero. Both belong to \([-2,3]^{11}\). If \(V\) is a common certificate, its compact invariant sublevel cannot contain \(z_\dagger\), so
\[
 V(z_\dagger)>0,\qquad V(z_0)\le-1,
 \qquad |V(z_\dagger)-V(z_0)|>1.
\]
Every coordinate of \(z_0\) has absolute value at most one (this includes \(a\), \((a-1/2)/4\) and the timers). The monomials that do not involve \(\tau_0\) cancel in the difference, and any other monomial has value zero at \(z_\dagger\) and absolute value at most \(\eps\) at \(z_0\). Hence
\[
 1<|V(z_\dagger)-V(z_0)|
 \le\eps\sum_{\alpha:\alpha_{\tau_0}>0}|c_\alpha|.
\]
This proves~\eqref{eq:certificate-coefficient-lower-bound}. The number of monomials of total degree at most \(D\) in eleven variables is \(\binom{11+D}{11}\), which gives the displayed bound on the maximal coefficient.

Finally, if \(W(h,Z)=\sum_\alpha c_\alpha(h)Z^\alpha\) is a fixed polynomial, it has finitely many polynomial coefficients \(c_\alpha(h)\), and the sum of their absolute values is bounded for \(0\le h\le1\). The necessary lower bound \(h^{-K}\) diverges along the dyadic scales, a contradiction.
\end{proof}

The bound on the coefficients concerns the fixed margins of separation \(-1\) and \(+1\) of Definition~\ref{def:common-certificate}, in the original eleven coordinates. It is a constraint of conditioning, and not a new obstruction on the degree. It allows a family of fixed degree with coefficients rational in \(\eps\) and poles at zero, and it does not exclude an unnormalized barrier whose margin of separation shrinks with the scale. At fixed degree, it gives a lower bound linear in \(s\) on the bit length of the coefficients, with no upper bound. The compact quartic on the source gives no such formula on the full state. Constructing a nonconvex polynomial of low degree that satisfies the implication at every micro-step remains open for this unchanged controller.
\AImethodology}
\AtEndDocument{\bibliographystyle{alphaurl}\bibliography{\BIBFILES,STOC2027-external}}

\end{document}